\documentclass[9pt]{osa-supplemental-document}
\setboolean{shortarticle}{false}

\usepackage{mismath}
\usepackage{amsmath,amssymb,commath,siunitx,comment}
\usepackage{graphicx} 
\usepackage{mathtools}
\usepackage[version=4]{mhchem}
\usepackage{extarrows}

\newcommand*{\Exp}{\mathrm{e}}
\newcommand*\diff{\mathop{}\!\mathrm{d}}

\newcommand{\etal}[1]{\mbox{{#1} \textit{et al.}}}
\newcommand{\ie}{\textit{i}.\textit{e}.,\ }
\newcommand{\eg}{\textit{e}.\textit{g}., }

\newcommand{\SIadj}[2]{\SI[number-unit-product={\text{-}}]{#1}{#2}}

\DeclareMathOperator{\fftshift}{fftshift}
\DeclareMathOperator{\ifftshift}{ifftshift}
\DeclareMathOperator{\fft}{fft}
\DeclareMathOperator{\ifft}{ifft}
\DeclareMathOperator{\myangle}{angle}
\DeclareMathOperator{\unwrap}{unwrap}

\usepackage{multirow}

\makeatletter
\newcommand{\myBigg}{\bBigg@{3.5}}
\makeatother

\usepackage{stackengine}
\newcommand{\cdddot}{%
\setstackgap{S}{0.4ex}%
\mathrel{\Shortstack{{.} {.} {.}}}}

\usepackage{listings,lstautogobble,verbatim}
\definecolor{mygreen}{RGB}{28,172,0} 
\definecolor{mylilas}{RGB}{170,55,241}
\definecolor{codebg}{RGB}{237,245,207}
\title{Tutorial of Fourier and Hankel transforms for ultrafast optics}
\author{Yi-Hao Chen}
\affil{School of Applied and Engineering Physics, Cornell University, Ithaca NY 14853, USA\\
Corresponding author: yc2368@cornell.edu\\
Revision date: \today}

\begin{abstract}
This tutorial is designed to clarify a few misconceptions in the field of ultrafast optics.\\[0.5em]
(1) \textbf{Analytic signal} that underlies the complex-conjugate decomposition of the field is discussed, which is often ignored in textbooks. The misunderstanding between propagation-constant-offset and frequency-offset analytic signal, as well as \textbf{slowly-varying envelope assumption}, are also discussed.\\[1em]
(2) This article is also written in response to the apparent lack of \textbf{comprehensive introductions to the Fourier transform}. It contains complete derivations of the general formulations of several Fourier-transform relations. It shows the importance of having Fourier-transform constants as parameters, and helps clarify the arbitrary selection of Fourier-transform constants and conventions.\\[1em]
(3) One of the main purposes of this tutorial is to clarify the correct \textbf{Fourier-transform convention} to be employed in ultrafast optics. Most of the time, an implementation of an optical numerical simulation is first verified by checking the smoothness of the generated result (because physical phenomena mostly exhibit some smooth physical features, \eg temporal or spectral profiles) or further by seeing if the simulation reasonably duplicates the ``known physics.'' However, the misuse of Fourier transform, from my observations, cannot be easily detected by such naive verifications.\\[1em]
(4) Moreover, \textbf{multiple Fourier-transform aspects} are discussed, involving convolution, aliasing, phase effect, and short-time Fourier transform.\\[1em]
(5) In addition to the Fourier transform, a tutorial on the \textbf{Hankel transform}, which arises from the two-dimensional spatial Fourier transform of a radially-symmetric function, is provided. Its numerical implementation based on the fast Hankel transform with high accuracy (FHATHA) is also provided, which is the core element of computationally-efficient radially-symmetric full-field optical propagation. Despite being a tutorial, I, for the first time, propose a new numerical scheme for the fast Hankel transform that outperforms both the original FHATHA and the discrete Hankel transform.\\[1em]

Feel free to send me an email if there is any confusion or typo, or if you think that there is more to add to this tutorial. For a deeper understanding into the ultrafast pulse propagation that involves these transforms, please check out our publicly-shared GitHub code at \href{https://github.com/AaHaHaa/MMTools}{https://github.com/AaHaHaa/MMTools} for waveguide and \href{https://github.com/AaHaHaa/3D-UPPE}{https://github.com/AaHaHaa/3D-UPPE} for non-waveguide ultrafast optical propagations. A MATLAB function to compute high-resolution spectrogram for time-frequency analysis is also provided at \href{https://github.com/AaHaHaa/Spectrogram}{https://github.com/AaHaHaa/Spectrogram}.
\end{abstract}

\setboolean{displaycopyright}{false} 

\begin{document}

\maketitle

\begin{titlepage}
{
\centering
{\Huge \textsc{Table of contents}\par}
\vspace{4em}
}
{\large
{\Large\noindent Section \ref{sec:Analytic_signal}: Analytic signal}\\[0.5em]
\indent\ref{subsec:Analytic_signal_intro}: Introduction\\
\indent\ref{subsec:Analytic_signal_f}: Offset frequency $\Omega=\omega-\omega_0$\\
\indent\ref{subsec:Slowly-varying envelope}: Slowly-varying envelope vs.\ propagation-constant-offset analytic signal\\

{\Large\noindent Section \ref{sec:Spectral_Fourier_transform}: Spectral Fourier transform}\\
[0.5em]
\indent\ref{subsec:FT_Definition}: Definition\\
\indent\ref{subsec:Important_use_of_Fourier-transform_convention}: Important correct application of the Fourier-transform convention\\
\indent\ref{subsec:Relations_between_FT_and_DFT}: Conversion of quantities with physical-useful units between FT and DFT\\

{\Large\noindent Section \ref{sec:DFT}: Discrete Fourier transform (DFT)}\\
[0.5em]
\indent\ref{subsec:fftshift}: fftshift/ifftshift and phase unwrapping\\
\indent\ref{subsec:DFT_convolution}: DFT's convolution\\

{\Large\noindent Section \ref{sec:complex_FT}: Complex-valued Fourier transform}\\[0.5em]
\indent\ref{subsec:phenomena_phase}: Phase effect\\
\indent\ref{subsec:time_frequency}: Time-frequency analysis\\[0.2em]
\indent\indent\ref{subsubsec:Wigner_spectrogram}: Wigner distribution and spectrogram\\
\indent\indent\ref{subsubsec:high_resolution_spectrogram}: High-resolution spectrogram\\
\indent\indent\ref{subsubsec:example_spectrogram}: Example cases\\

{\Large\noindent Section \ref{sec:Hankel_transform}: Hankel transform: 2D spatial Fourier transform of $A(r)$}\\
[0.5em]
\indent\ref{subsec:Hankel_intro}: Introduction\\
\indent\ref{subsec:FHATHA}: Numerical computation: Fast Hankel transform of high accuracy (FHATHA)\\[0.2em]
\indent\indent\ref{subsubsec:Magni}: Magni's approach\\
\indent\indent\ref{subsubsec:Improved_FHATHA}: Important steps to improve FHATHA for ultrafast pulse propagation\\[.5em]
\indent\ref{subsec:DHT}: Discrete Hankel transform (DHT)\\
}
\vfill
\end{titlepage}

\pagebreak
\section{Analytic signal}\label{sec:Analytic_signal}
\subsection{Introduction}\label{subsec:Analytic_signal_intro}
When we learn ultrafast optics with textbooks, such as Boyd's Nonlinear Optics \cite{Boyd2008}, they usually start with the field equation for the real-valued field:
\begin{align}
\mathbb{E}(t) & =\frac{1}{2}\left(\mathcal{E}(t)+\mathcal{E}^{\ast}(t)\right)=\frac{1}{2}\left(E(t)\Exp^{-i\omega_0t}+\left(E(t)\right)^*\Exp^{i\omega_0t}\right) \nonumber \\
& =\Re\left[\mathcal{E}(t)\right]\quad,\Re[\cdot]\text{ represents taking the real part of a value}
\label{eq:E(t)}
\end{align}
or simply assume that the complex-valued field has the form:
\begin{equation}
\mathcal{E}(t)=E(t)\Exp^{-i\omega_0t},
\label{eq:mathcalE(t)}
\end{equation}
where the superscript $\ast$ represents the complex-conjugate operation. If the field is a simple sinusoidal wave that follows $\mathbb{E}(t)=\sin(\omega_0 t)=\frac{\Exp^{i\omega_0 t}-\Exp^{-i\omega_0 t}}{2i}$, $\mathcal{E}(t)$ in Eq.~(\ref{eq:E(t)}) can be as simple and trivial as $\frac{-\Exp^{-i\omega_0 t}}{i}=i\Exp^{-i\omega_0 t}$, or as complicated as $\left(i\Exp^{-i\omega_0 t}+i\right)$. Apparently, there are infinite possible options for complex-valued $\mathcal{E}(t)$ that satisfies Eq.~(\ref{eq:E(t)}), even for a simple sinusoidal function. Since $\mathbb{E}(t)=\Re\left[\mathcal{E}(t)\right]$, the imaginary part of $\mathcal{E}(t)$ can be arbitrarily chosen. However, in studies of nonlinear optics, the decomposition following Eq.~(\ref{eq:E(t)}) plays a crucial part in analyzing the evolution of spectral components of involved fields, which are obtained through the Fourier transform. Not correctly defining or understanding the decomposition of the field with Eq.~(\ref{eq:E(t)}) can not only result in a mixed use of Fourier transform of the real-valued $\mathbb{E}(t)$ and complex-valued $\mathcal{E}(t)$ (which will be discussed in detail in Fig.~\ref{fig:DFT_Omega}) but also mislead researchers from obtaining all generated frequencies, leading to the problem with missing negative frequencies \cite{Conforti2013}. Therefore, it is important to understand and apply the correct decomposition. Here I will introduce the ``analytic-signal decomposition,'' which decomposes the real-valued signal into its positive- and negative-frequency components. The positive-frequency part is called the ``analytic signal,'' whose complex conjugate is the negative-frequency part of the real-valued signal.

An analytic signal is a complex-valued function that has no negative-frequency part. If $s(t)$ is a real-valued function with Fourier transform $\mathfrak{F}\left[\mathbb{E}(t)\right]=\mathbb{E}(\nu)$, then it exhibits the Hermitian symmetry about $\nu=0$:
\begin{equation}
\mathbb{E}(-\nu)=\mathbb{E}^*(\nu).
\end{equation}
Thus, there is redundancy if both frequencies are considered; negative-frequency components can be discarded without loss of information.

We define $\mathcal{E}(\nu)$ to represent the positive-frequency part as the following:
\begin{align}
\mathcal{E}(\nu) & =\begin{cases}
2\mathbb{E}(\nu) &, \nu>0 \\
\mathbb{E}(\nu) &, \nu=0 \\
0 &, \nu<0
\end{cases} \nonumber \\
& =\mathbb{E}(\nu)+\sgn(\nu)\mathbb{E}(\nu),
\end{align}
so that
\begin{align}
\mathbb{E}(\nu) & =\begin{cases}
\frac{1}{2}\mathcal{E}(\nu) & ,\nu>0 \\
\mathcal{E}(\nu) & ,\nu=0 \\
\frac{1}{2}\left(\mathcal{E}(-\nu)\right)^* & ,\nu<0
\end{cases} \nonumber \\
& =\frac{1}{2}\left[\mathcal{E}(\nu)+\left(\mathcal{E}(-\nu)\right)^*\right]
\end{align}
and thus, after inverse Fourier transform,
\begin{equation}
\mathbb{E}(t)=\frac{1}{2}\left[\mathcal{E}(t)+\left(\mathcal{E}(t)\right)^*\right].
\label{eq:s(t)}
\end{equation}
From Eq.~(\ref{eq:s(t)}), we can see that $\mathbb{E}(t)$ is a combination of its analytic signal $\mathcal{E}(t)$ and the corresponding complex conjugate, which is the negative-frequency part of the $\mathbb{E}(t)$.

Since $s_a(t)$ is the inverse Fourier transform of $S_a(\nu)$, we can derive the relation with its real-valued signal also as follows:
\begin{align}
\mathcal{E}(t) & =\mathfrak{F}^{-1}\left[\mathcal{E}(\nu)\right] \nonumber \\
& =\mathfrak{F}^{-1}\left[\mathbb{E}(\nu)+\sgn(\nu)\mathbb{E}(\nu)\right] \nonumber \\
& =\mathfrak{F}^{-1}\left[\mathbb{E}(\nu)\right]+\mathfrak{F}^{-1}\left[\sgn(\nu)\mathbb{E}(\nu)\right] \nonumber \\
& =\mathbb{E}(t)+C_{\mathfrak{F}}\left\{\mathfrak{F}^{-1}\left[\sgn(\nu)\right]*\mathfrak{F}^{-1}\left[\mathbb{E}(\nu)\right]\right\} \quad\text{with the convolution theorem [Eq.~(\ref{eq:cc1})]} \nonumber \\
& =\mathbb{E}(t)+C_{\mathfrak{F}}\left[\left(C_{\mathfrak{IF}}c_s\frac{2}{it}\right)*\mathbb{E}(t)\right] \quad\because\mathfrak{F}^{-1}\left[\sgn(\nu)\right]=C_{\mathfrak{IF}}c_s\frac{2}{it} \nonumber \\
& =\mathbb{E}(t)+i\left[\left(-\frac{c_s}{\pi t}\right)*\mathbb{E}(t)\right],\quad\because C_{\mathfrak{F}}C_{\mathfrak{IF}}=\frac{1}{2\pi}
\label{eq:sa_uniquely}
\end{align}
As previously discussed, while the complex-conjugate relation of an analytic signal and its real-valued signal indicates $\mathbb{E}(t)=\Re[\mathcal{E}(t)]$ [Eq.~(\ref{eq:E(t)})], it opens up infinite possible options of $\mathcal{E}(t)$ that satisfies this relation  a seemingly arbitrarily-varying imaginary part. \textbf{By imposing the ``positive-frequency'' requirement of analytic signal, the imaginary part becomes uniquely defined}, which corresponds to the Hilbert transform of $\mathbb{E}(t)$ following $\left[\left(-\frac{c_s}{\pi t}\right)*\mathbb{E}(t)\right]$[Eq.~(\ref{eq:sa_uniquely})].

Any real-valued signal can be decomposed into its positive-frequency (analytic-signal) and negative-frequency parts, which underlies the decomposition of Eq.~(\ref{eq:E(t)}). In addition, this tells us that the Fourier transform of the real-valued field is different from that of its analytic signal; they exhibit different spectral components. Hence, it is crucial to define clearly what is being used, especially in studies of, for example, four-wave mixing and Raman scattering that involves nonlinear evolutions of different frequencies. In principle, any derivations of nonlinear optics should start with the real-valued signal, followed by its decomposition into the ``analytic signal.'' If a derivation starts directly with a single complex-valued signal (either the positive- or negative-frequency part) [Eq.~(\ref{eq:mathcalE(t)})], then readers need to be cautious about two things: (1) whether there is any missing frequency component due to ignoring the complex-conjugate part, and (2) whether there is a deviation of a factor of \num{2}. As an example, \cite{Chen2007} starts with the complex-valued field for deriving the Raman-induced index change, which eventually results in a reported deviation of a factor of \num{2} between their theory and experiments. (Correct derivation with analytic signal, as well as generalization to an arbitrary polarization, is in \cite{Chen2024}.) As another example, the third-order nonlinear polarization $\vec{\mathbb{P}}(t)=\int^{\infty}_{-\infty}\chi^{(3)}(t_1,t_2,t_3)\cdddot\vec{\mathbb{E}}(t-t_1)\vec{\mathbb{E}}(t-t_2)\vec{\mathbb{E}}(t-t_3)\diff t_1\diff t_2\diff t_3$ is apparently different from $\vec{\mathcal{P}}'(t)=\int^{\infty}_{-\infty}\chi^{(3)}(t_1,t_2,t_3)\cdddot\vec{\mathcal{E}}(t-t_1)\vec{\mathcal{E}}(t-t_2)\vec{\mathcal{E}}(t-t_3)\diff t_1\diff t_2\diff t_3$, as $\vec{\mathcal{P}}'(t)$ lacks all frequency components that result from combinations including the negative-frequency components of three $\vec{\mathbb{E}}(t)$'s. Therefore, $\vec{\mathcal{P}}'(t)$ is not the correct calculation for the analytic signal $\vec{\mathcal{P}}(t)$ of $\vec{\mathbb{P}}(t)$. A correct $\vec{\mathcal{P}}(t)$ should be calculated by identifying the positive-frequency part of $\vec{\mathbb{P}}(t)=\frac{1}{2}\left(\vec{\mathcal{P}}(t)+\vec{\mathcal{P}}^{\ast}(t)\right)=\int^{\infty}_{-\infty}\chi^{(3)}(t_1,t_2,t_3)\cdddot\vec{\mathbb{E}}(t-t_1)\vec{\mathbb{E}}(t-t_2)\vec{\mathbb{E}}(t-t_3)\diff t_1\diff t_2\diff t_3$.

\subsection{Offset frequency \texorpdfstring{$\Omega=\omega-\omega_0$}{Ω=ω-ω0}}\label{subsec:Analytic_signal_f}
\begin{figure}[!ht]
\centering
\includegraphics[width=0.6\linewidth]{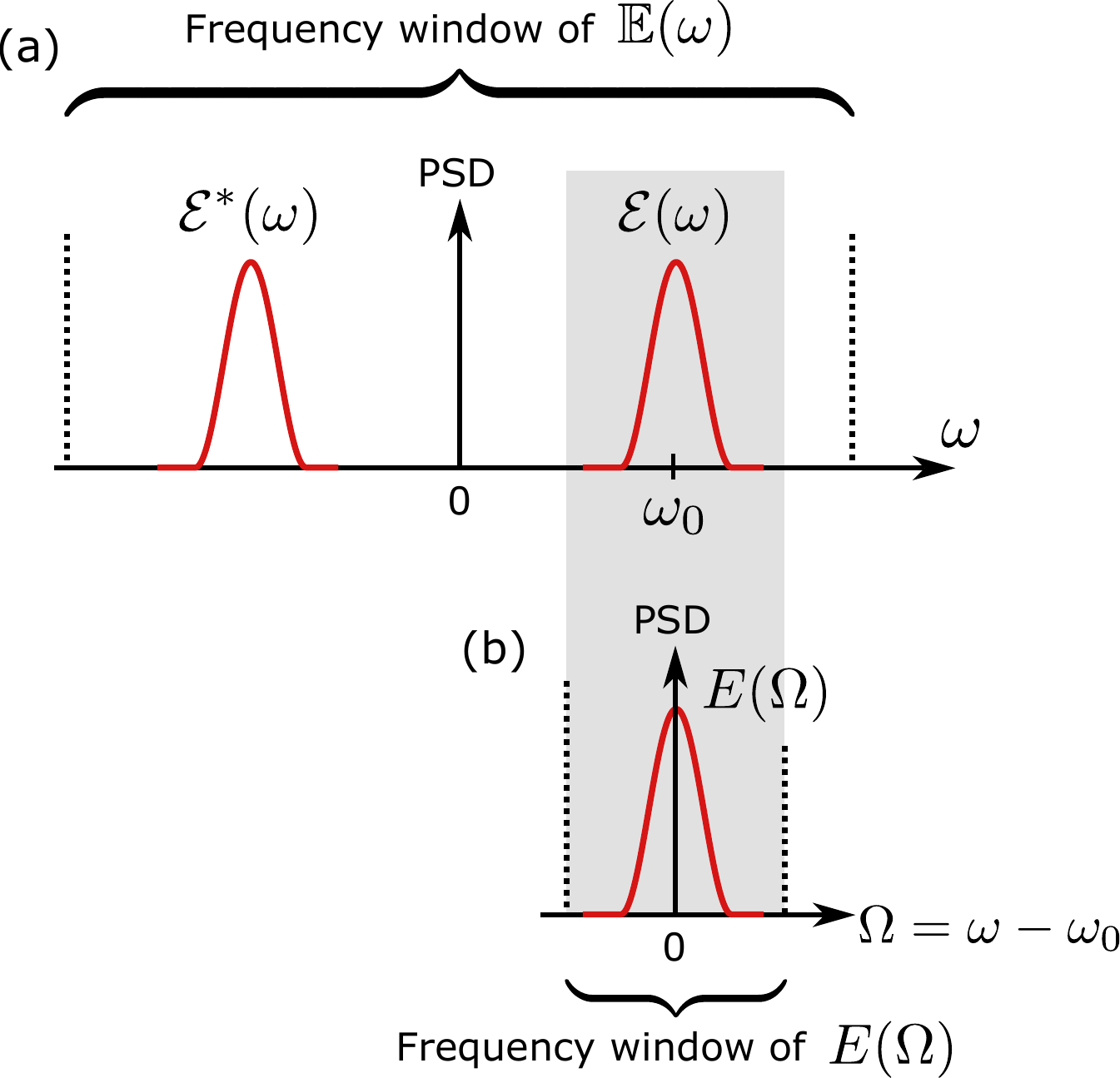}
\caption{Spectral domain of the Fourier-transform components of (a) the real-valued signal and (b) the envelope of its analytic signal [Eq.~(\ref{eq:E(t)})]. PSD: power spectral density $\sim\abs{\mathfrak{F}\left[\cdot\right]}^2$}
\label{fig:DFT_Omega}
\end{figure}

Applying analytic signal provides benefits in numerical computations. Because real-valued signal contains both positive and negative frequencies, its frequency window should cover both signs of frequency [Fig.~\ref{fig:DFT_Omega}(a)]. By extracting $\omega_0$ out as in Eq.~(\ref{eq:E(t)}), the Fourier transform of $E(t)$ (to obtain $E(\Omega)=\mathfrak{F}\left[E\right]=C_{\mathfrak{F}}\int^{\infty}_{-\infty}E(t)\Exp^{ic_s\Omega t}\diff t$ [Eq.~(\ref{eq:Fourier_transform})]) is equivalent to applying it with respect to the offset frequency $\Omega=\omega-\omega_0$ [Fig.~\ref{fig:DFT_Omega}(b)]. The offset center of the frequency window, from $\omega=0$ to $\omega_0$, enables a small window covering only around the signal's spectrum, free from the redundant negative-frequency components. As an example, a broadband simulation of a \ce{Yb}-doped fiber laser requires a \SIadj{100}{\THz} window to cover signals from \SIrange[range-phrase=--,range-units=single]{1}{1.2}{\micro\m} without inducing aliasing. \footnote{A rule of thumb for the size of frequency window is that it needs to be around \numrange[range-phrase=--]{5}{10} times as the pulse's bandwidth.} On the other hand, the frequency window of the real-valued $\mathbb{E}(t)$ should be $\SIadj{\sim700}{\THz}$ wide for a \SIadj{1}{\micro\m} ($\approx\SI{300}{\THz}$) pulse [if we assume the same \SIadj{100}{\THz} coverage around the signal: $700=2\times\left(300+100/2\right)$]. If $\omega_0$ happens to be the pulse's center frequency, $E(t)$ is called the ``slowly-varying'' envelope of $\mathcal{E}(t)$, as the highly-oscillatory component is extracted out. $E(t)$ can be real-valued if the analytic signal $\mathcal{E}(t)$ exhibits no extra phase variation other than sinusoidal waves, \ie it is a ``transform-limited'' signal if it is symmetric in time (Sec.~\ref{subsec:phenomena_phase}). Thus, the frequency window of the analytic signal's envelope is \num{7} times smaller than that of the real-valued signal, allowing a numerical simulation with a larger temporal sampling period and thus less sampling points. For narrowband simulations, the computational improvement can be more significant (an improvement factor of $\frac{600+x}{x}$ with a small $x$, the size of the narrowband-signal's frequency window). In general, $\omega_0$ does not need to be at the pulse's center frequency. For broadband nonlinear processes such as vibrational-Raman generation in \ce{H2} that creates frequency \SI{125}{\THz} apart from the pump frequency \cite{Konyashchenko2008,Chen2023}, or \SI{\geq250}{\THz} for cascaded processes \cite{Loree1976,Grasyuk1984}, $\omega_0$ should be placed such that the frequency window can cover all generated frequency components.

\subsection{Slowly-varying envelope vs.\ propagation-constant-offset analytic signal}\label{subsec:Slowly-varying envelope}
In this section, I would like to clarify a few misconceptions of the slowly-varying envelope assumption (SVEA). Conclusion first: \textbf{Slowly-varying envelope assumption shouldn't be confused with propagation-constant-offset or frequency-offset analytic signal.}

First, what is SVEA? To simplify the discussion, we consider only linear propagation (\ie no nonlinearity) with a single spatial mode (\ie one constant $\beta_0$). Derivation usually leads to the famous Helmholtz equation:
\begin{equation}
\triangledown^2\vec{\mathcal{E}}(\vec{r},t)+\beta^2\vec{\mathcal{E}}(\vec{r},t)=0,
\label{eq:Helmholtz}
\end{equation}
where $\beta=n\omega/c$ is assumed to be a constant (for narrowband fields) and $\vec{\mathcal{E}}$ is the analytic signal of $\vec{\mathbb{E}}$ [Eq.~(\ref{eq:E(t)})]. By assuming that
\begin{equation}
\vec{\mathcal{E}}=\vec{E}(z,t)\Exp^{i\left(\beta_0z-\omega_0t\right)},
\label{eq:E_tmp}
\end{equation}
it results in
\begin{equation}
\triangledown_{\perp}^2\vec{E}(\vec{r},t)+\partial_z^2\vec{E}(\vec{r},t)+2i\beta_0\partial_z\vec{E}(\vec{r},t)+\left(\beta^2-\beta_0^2\right)\vec{E}(\vec{r},t)=0.
\label{eq:SVEA0}
\end{equation}
Under the SVEA in the $z$ dimension,
\begin{equation}
\abs{\partial_z^2\vec{E}}\ll\abs{2i\beta_0\partial_z\vec{E}},
\label{eq:SVEA}
\end{equation}
making the equation a first-order derivative with respect to $z$ and thus solvable:
\begin{align}
\partial_z\vec{E}(\vec{r},t) & =i\frac{\beta^2-\beta_0^2}{2\beta_0}\vec{E}(\vec{r},t)+\frac{i}{2\beta_0}\triangledown_{\perp}^2\vec{E}(\vec{r},t) \nonumber \\
&  \approx i\triangle\beta_0\vec{E}(\vec{r},t)+\frac{i}{2\beta_0}\triangledown_{\perp}^2\vec{E}(\vec{r},t).
\label{eq:SVEA1}
\end{align}
by having $\beta=\beta_0+\triangle\beta$ such that $\beta^2-\beta_0^2\approx2\beta_0\triangle\beta$. That is to say, SVEA slows down the phase variation of the target $\vec{E}$ in $z$ by picking $\beta_0$ in Eq.~(\ref{eq:E_tmp}).

It is worth noting that \textbf{(traditional) SVEA automatically involves unidirectional optical propagation} when assuming that $\vec{\mathcal{E}}$ is dominated only by the effect of $\beta_0$ in the $z$ dimension and thus $\vec{E}\sim1$ [Eq.~(\ref{eq:E_tmp})]. If there is a backward-propagating field, the beating leads to spatial oscillation at a rate of $\left(2\beta_0\right)$, having $\vec{\mathcal{E}}\sim\Exp^{i2\beta_0z}$ and then $\vec{E}\sim\Exp^{i\beta_0z}$. This fast oscillation invalidates the SVEA [Eq.~(\ref{eq:SVEA})]. On the other hand, unidirectional optical propagation can be directly applied to Eq.~(\ref{eq:Helmholtz}):
\begin{align}
& \left(\partial_z^2+\beta^2\right)\vec{\mathcal{E}}(\vec{r},t)+\triangledown_{\perp}^2\vec{\mathcal{E}}(\vec{r},t)=0 \nonumber \\
& \Rightarrow\left(\partial_z+i\beta\right)\left(\partial_z-i\beta\right)\vec{\mathcal{E}}(\vec{r},t)+\triangledown_{\perp}^2\vec{\mathcal{E}}(\vec{r},t)=0.
\label{eq:UPPE_derivation0}
\end{align}
Unidirectionality makes $\partial_z$ operator approximately $\left(i\beta_0\right)$, and thus we can approximate Eq.~(\ref{eq:UPPE_derivation0}) with
\begin{align}
& i\left(\beta_0+\beta\right)\left(\partial_z-i\beta\right)\vec{\mathcal{E}}(\vec{r},t)+\triangledown_{\perp}^2\vec{\mathcal{E}}(\vec{r},t)=0 \nonumber \\
& \Rightarrow\partial_z\vec{\mathcal{E}}(\vec{r},t)=i\beta\vec{\mathcal{E}}(\vec{r},t)+\frac{i}{\beta_0+\beta}\triangledown_{\perp}^2\vec{\mathcal{E}}(\vec{r},t).
\end{align}
Recovering to the previously-derived SEVA formula [Eq.~(\ref{eq:SVEA1})] can be achieved by introducing Eq.~(\ref{eq:E_tmp}). Because $\beta_0$ is picked close to $\beta$, the denominator of transverse component $\beta_0+\beta\approx2\beta_0$, as shown in Eq.~(\ref{eq:SVEA1}).

Before I begin to clarify SVEA, let me introduce the most general formulation, whose real-valued electric field follows
\begin{equation}
\vec{\mathbb{E}}(\vec{r},t)=\frac{1}{2}\left[\vec{\mathcal{E}}(\vec{r},t)+\text{c.c.}\right]=\int\diff\omega\frac{1}{2}\left[\vec{E}(\vec{r},\omega)\Exp^{i\left[\beta(\omega)z-\omega t\right]}+\text{c.c.}\right].
\label{eq:E_freq_slow}
\end{equation}
Also, it can be written into
\begin{equation}
\vec{\mathbb{E}}(\vec{r},t)=\frac{1}{2}\left[\vec{E}(\vec{r},t)\Exp^{i\left(\beta_0z-\omega_0t\right)}+\text{c.c.}\right].
\label{eq:E_time_offset}
\end{equation}

Eq.~(\ref{eq:E_freq_slow}) represents the SVEA applied in the frequency domain. By extracting the $\beta(\omega)$ out of the field, $\vec{E}(\vec{r},\omega)$ exhibits the \emph{slowly-varying} feature in propagation distance $z$. Due to the extraction of $\omega$, it also exhibits the \emph{slowly-varying} feature in time (if $\vec{E}(\vec{r},\omega)$ is inverse-Fourier-transformed). In this general formulation, the field can be broadband or consist of multiple frequency components (\eg two colors of pump and Stokes frequencies in Raman generation, or three colors in optical parametric generation), so $\vec{E}(\vec{r},\omega)$ itself can still oscillate fast in $z$ due to the beating between different frequency components with different propagation constants $\beta(\omega)$, or in time due to the beating between different frequency components.

Eq.~(\ref{eq:E_time_offset}) represents the propagation-constant-offset (by $\beta_0$) and frequency-offset (by $\omega_0$) operation to the analytic signal of $\vec{\mathbb{E}}$, obtaining $\vec{E}$ (Sec.~\ref{sec:Analytic_signal}\ref{subsec:Analytic_signal_f}). If two offsets are picked correctly, the oscillations in $\vec{E}$ is minimized, which aligns with the traditional concept of SVEA. However, in principle, they can be picked arbitrarily. During numerical computation with the unidirectional pulse propagation equation, the goal of picking these two values is to minimize the variation of propagation constant and frequency of the field during evolution to allow the largest step size $\triangle z$ and the smallest frequency window possible. If they are picked with other values, the simulation can still be solved but is just slower.

Fundamentally, Eq.~(\ref{eq:SVEA}) is the core equation of SVEA and is why it is called ``slowly varying.'' Eq.~(\ref{eq:SVEA}) is satisfied only when the field is single-frequency, narrowband, and unidirectional. Under these conditions, Eqs.~(\ref{eq:E_freq_slow}) and (\ref{eq:E_time_offset}) become the same equation. In the general case where the field can involve broadband or multi-frequency components, \textbf{SVEA should correspond only to $\boldsymbol{\vec{E}(\vec{r},\omega)}$ in Eq.~(\ref{eq:E_freq_slow}) while $\boldsymbol{\vec{E}(\vec{r},t)}$ in Eq.~(\ref{eq:E_time_offset}) corresponds to propagation-constant-offset and frequency-offset operations. Fundamentally, SVEA represents the slowly-varying nature in the coefficient of the plane-wave expansion.}

\noindent\rule{\textwidth}{1pt}

Here, I elaborate more on the propagation-constant-offset operation. Typical single-spatial-mode pulse propagation follows
\begin{equation}
\partial_zA(z,\Omega)=i\left[\beta(\omega)-\left(\beta_{(0)}+\beta_{(1)}\Omega\right)\right]A(z,\Omega)+\hat{\mathcal{N}}A(z,\Omega),
\label{eq:partial_zA}
\end{equation}
where $A$ is the pulse's field following Eq.~(\ref{eq:mathbbE_Ap}) such that $A(z,\Omega)\sim\vec{E}(\vec{r},\omega)$ in Eq.~(\ref{eq:E_freq_slow}), $\beta(\omega)$ is the propagation constant, $\beta_{(0)}$ and $\beta_{(1)}$ are to facilitate the simulation speed by reducing the linear phase variation. $\beta_{(0)}$ reduces the global phase that does not affect the overall physical phenomenon; $\beta_{(1)}$ creates a temporally-moving time window that follows the pulse such that the pulse is always at $t=0$ in the window, for example. Apparently, the choice of $\beta_{(1)}$ does not affect the physical phenomenon as well. $\hat{\mathcal{N}}$ is the nonlinear operation that, for example in a $\chi^{(3)}$ system, is related to $\left[A(t)\right]^3$. Where to pick $\beta_{(0)}$ and $\beta_{(1)}$ is the core of the discussion in this section. Because a numerical simulation must correctly resolve the phase evolution, especially when it can further affect nonlinear evolutions in $\hat{\mathcal{N}}A$ (\eg in a multimode system, linear phase evolution modulates the spatial speckles, further affecting the nonlinear evolutions), effective reduction of the accumulated propagation constant $\beta_{\text{sim}}=\max_{\omega\in\text{pulse's spectrum}}\abs{\beta(\omega)-\left(\beta_{(0)}+\beta_{(1)}\Omega\right)}$ decreases the phase increment per numerical step size $\left(\beta_{\text{sim}}\triangle z\right)$. This allows the simulation to run with a large step size and thus improves the simulation speed. Ultimately, the simulation speed is limited by the minimum achievable $\beta_{\text{sim}}$.

It is important to understand when to compute a component, at the center of the frequency window $\omega_0$ or at the pulse's center frequency $\omega_{\text{pulse}}$. Since SVEA places the pulse at the center of the frequency window, these two frequencies are the same ($\omega_0=\omega_{\text{pulse}}$). In addition, it naturally picks $\beta_{(0)}$ and $\beta_{(1)}$ from Taylor-series coefficients of $\beta(\omega)$ at the center of the frequency window. In general, $\omega_0$ and $\omega_{\text{pulse}}$ do not need to overlap. In such a situation, to still have the fastest simulation speed, $\beta_{(0)}$ and $\beta_{(1)}$ should be computed from $\beta(\omega)$ at pulse's frequency $\omega_{\text{pulse}}$, not $\omega_0$, because it is the pulse's evolution that determines the phase increment. Understanding the important difference between $\omega_0$ and $\omega_{\text{pulse}}$ enables us to freely establish a frequency window that meets our simulation needs. For example, if the pulse will be significantly nonlinearly redshifted during propagation, the frequency window should be centered on the long-wavelength side of the pulse (rather than at the pulse's center frequency), which allows us to use a frequency window with the minimum size and the least number of sampling points; otherwise, the short-wavelength side of the window is wasted in this redshifting process if it is centered at the pulse's frequency.

\pagebreak
\section{Spectral Fourier transform}\label{sec:Spectral_Fourier_transform}
In this section, we explain the Fourier transform in its continuous and discrete formats. In addition, all constants are represented as parameters to be compatible with various conventions people in different fields use. I would also like to advocate people to derive their equations based on the parametrized Fourier transform, as in Eq.~(\ref{eq:Fourier_transform}). As I will show later, many relations depend on the Fourier-transform convention and its constants. Non-uniform conventions (\eg $C_{\mathfrak{F}}=1$ or $C_{\mathfrak{F}}=\frac{1}{\sqrt{2\pi}}$ [Eq.~(\ref{eq:Fourier_transform})]), arbitrarily employed by different people, can create misleading equations. For example, if an equation has $\pi$, we do not know whether it is dependent on Fourier-transform constants or not; it can come from anywhere, such as the frequency relation $\omega=2\pi\nu$ that is irrelevant to the Fourier transform. Furthermore, losing the information of Fourier-transform constants prevents people from correctly transforming the equation from continuous to discrete Fourier transform for numerical computations, as I will show later in Sec.~\ref{sec:Spectral_Fourier_transform}\ref{subsec:Relations_between_FT_and_DFT}.

Here in this tutorial, the overall trend of notation follows the physical convention whose inverse spectral Fourier transform follows $A(t)\sim\int A(\omega)\Exp^{i\left(k_zz-\omega t\right)}\diff\omega$, which is the spectral Fourier transform in mathematics or engineering. Note that in contrast to the spectral Fourier transform, the physical convention ($k_zz-\omega t$) potentially shows that the spatial Fourier transform $\mathfrak{F}_{k_z}$ is consistent with the mathematical convention; however, we typically do not calculate based on the $k_z$-space. Here, we will focus only on the spectral Fourier transform.

\subsection{Definition}\label{subsec:FT_Definition}
In general, the Fourier transform is defined as
\begin{equation}
\begin{aligned}
A(\omega)=\mathfrak{F}[A(t)] & \equiv C_{\mathfrak{F}}\int^{\infty}_{-\infty}A(t)\Exp^{ic_s\omega t}\diff t\\
A(t)=\mathfrak{F}^{-1}[A(\omega)] & \equiv C_{\mathfrak{IF}}\int^{\infty}_{-\infty}A(\omega)\Exp^{-ic_s\omega t}\diff\omega.
\end{aligned} \label{eq:Fourier_transform}
\end{equation}
If $c_s=1$, the Fourier transform follows the physical convention; whereas it follows the mathematical convention if it is $-1$. Its constants satisfy $C_{\mathfrak{F}}C_{\mathfrak{IF}}=\frac{1}{2\pi}$, which is found with 
\begin{align}
A(t) & =C_{\mathfrak{IF}}\int^{\infty}_{-\infty}A(\omega)\Exp^{-ic_s\omega t}\diff\omega \nonumber \\
& =C_{\mathfrak{IF}}\int^{\infty}_{-\infty}\left(C_{\mathfrak{F}}\int^{\infty}_{-\infty}A(\tau)\Exp^{ic_s\omega\tau}\diff\tau\right)\Exp^{-ic_s\omega t}\diff\omega \nonumber \\
& =C_{\mathfrak{F}}C_{\mathfrak{IF}}\int^{\infty}_{-\infty}\int^{\infty}_{-\infty}A(\tau)\Exp^{ic_s\omega\left(\tau-t\right)}\diff\omega\diff\tau \nonumber \\
& =C_{\mathfrak{F}}C_{\mathfrak{IF}}\int^{\infty}_{-\infty}A(\tau)\frac{2\pi}{\abs{c_s}}\delta(\tau-t)\diff\tau \nonumber \\
& =C_{\mathfrak{F}}C_{\mathfrak{IF}}A(t)2\pi \quad\because\abs{c_s}=1 \nonumber \\
\Rightarrow C_{\mathfrak{F}}C_{\mathfrak{IF}} & =\frac{1}{2\pi}.
\label{eq:CFCIF_derivation}
\end{align}

Since the convolution theorem is a commonly-used relation, we show it below:
\begin{subequations}
\begin{align}
\mathfrak{F}[A*B] & =\frac{1}{C_{\mathfrak{F}}}\mathfrak{F}[A]\mathfrak{F}[B] \label{eq:cc1} \\
\mathfrak{F}^{-1}[A*B] & =\frac{1}{C_{\mathfrak{IF}}}\mathfrak{F}^{-1}[A]\mathfrak{F}^{-1}[B], \label{eq:cc2}
\end{align} \label{eq:continuous_convolution_thm}
\end{subequations}
where the convolution (denoted with $*$) follows
\begin{equation}
\left(A*B\right)(t)\equiv\int_{-\infty}^{\infty}A(\tau)B(t-\tau)\diff\tau.
\label{eq:convolution}
\end{equation}
The discrete counterpart (also called ``discrete-time Fourier transform'') is
\begin{equation}
\begin{aligned}
\mathfrak{F}_{D_c}[A*B] & =\frac{1}{C_{\mathfrak{F}}}\mathfrak{F}_{D_c}[A]\mathfrak{F}_{D_c}[B] \\
\mathfrak{F}_{D_c}^{-1}[A*B] & =\frac{1}{C_{\mathfrak{IF}}}\mathfrak{F}_{D_c}^{-1}[A]\mathfrak{F}_{D_c}^{-1}[B]
\end{aligned}
\quad\text{ if }\quad
\begin{aligned}
A(\omega)=\mathfrak{F}_{D_c}[A(t_n)] & \equiv C_{\mathfrak{F}}\sum_{n=1}^{\mathfrak{N}}A(t_n)\Exp^{ic_s\omega t_n}\triangle t\\
A(t)=\mathfrak{F}_{D_c}^{-1}[A(\omega_m)] & \equiv C_{\mathfrak{IF}}\sum_{m=1}^{\mathfrak{N}}A(\omega_m)\Exp^{-ic_s\omega_mt}\triangle\omega
\end{aligned}.
\label{eq:discrete_Fourier_transform_c}
\end{equation}
$\mathfrak{F}$ and $\mathfrak{F}_{D_c}$ are the continuous and discrete versions of Fourier transform, respectively. $\mathfrak{N}$ is the number of discrete points. In the discrete manner, $t_n=n\triangle t$ and $\omega_m=m\triangle\omega$. The time window is $T^w=\mathfrak{N}\triangle t$, and the frequency window $\nu^w=\frac{1}{\triangle t}=\frac{\mathfrak{N}}{T^w}=\mathfrak{N}\triangle\nu$. The angular-frequency spacing $\triangle w=2\pi\triangle\nu=\frac{2\pi}{T^w}$. If the sampling frequency is infinitely high in an infinitely-wide time window (which then leads to an infinitely-wide frequency window), $\mathfrak{F}_{D_c}\approx\mathfrak{F}$ and $\mathfrak{F}_{D_c}^{-1}\approx\mathfrak{F}^{-1}$.

In practice, during numerical computations, discrete Fourier transform (DFT) is used to compute Fourier-transform components:
\begin{equation}
\begin{aligned}
A_D(\omega)=\mathfrak{F}_D[A(t_n)] & \equiv C_{\mathfrak{F}_D}\sum_{n=1}^{\mathfrak{N}}A(t_n)\Exp^{ic_s\omega t_n}\\
A(t)=\mathfrak{F}_D^{-1}[A_D(\omega_m)] & \equiv C_{\mathfrak{IF}_D}\sum_{m=1}^{\mathfrak{N}}A_D(\omega_m)\Exp^{-ic_s\omega_m t},
\label{eq:discrete_Fourier_transform}
\end{aligned}
\end{equation}
which, however, differs from the actual Fourier transform [Eq.~(\ref{eq:discrete_Fourier_transform_c})] in constants and units:
\begin{subequations}
\begin{alignat}{3}
\mathfrak{F} & =\frac{C_{\mathfrak{F}}}{C_{\mathfrak{F}_D}} && \triangle t && \cdot\mathfrak{F}_D \label{eq:F_op_conversion} \\
\mathfrak{F}^{-1} & =\frac{C_{\mathfrak{IF}}}{C_{\mathfrak{IF}_D}} && \triangle\omega && \cdot\mathfrak{F}_D^{-1}. \label{eq:Finv_op_conversion}
\end{alignat} \label{eq:F_FD_conversion}
\end{subequations}
If we replace variables following $\omega_m=m\triangle\omega=m\frac{2\pi}{T^w}=m\frac{2\pi}{\mathfrak{N}\triangle t}$ and $t_n=n\triangle t$, Eq.~(\ref{eq:discrete_Fourier_transform}) becomes
\begin{equation}
\begin{aligned}
A_D(\omega_m)=\mathfrak{F}_D[A(t_n)] & \equiv C_{\mathfrak{F}_D}\sum_{n=1}^{\mathfrak{N}}A(t_n)\Exp^{ic_s\frac{2\pi}{\mathfrak{N}}mn} \\
A(t_n)=\mathfrak{F}_D^{-1}[A_D(\omega_m)] & \equiv C_{\mathfrak{IF}_D}\sum_{m=1}^{\mathfrak{N}}A_D(\omega_m)\Exp^{-ic_s\frac{2\pi}{\mathfrak{N}}nm},
\end{aligned}
\label{eq:discrete_Fourier_transform0}
\end{equation}
where $C_{\mathfrak{F}_D}C_{\mathfrak{IF}_D}=\frac{1}{\mathfrak{N}}$, found with the similar process to Eq.~(\ref{eq:CFCIF_derivation}).

The DFT $A_D(\omega)$ [Eq.~(\ref{eq:discrete_Fourier_transform})] is denoted with an extra ``$D$'' subscript to distinguish it from $A(\omega)$ of Eqs.~(\ref{eq:Fourier_transform}) and (\ref{eq:discrete_Fourier_transform_c}). Therefore, it is important to derive a relationship between $A(\omega)$ and $A_D(\omega)$, which follows
\begin{equation}
\frac{1}{C_{\mathfrak{F}_D}}A_D(\omega)\triangle t=\frac{1}{C_{\mathfrak{F}}}A(\omega)
\label{eq:AD-A_transform}
\end{equation}
so that they obtain the same $A(t)$. For the commonly-used Fourier-transform convention in the laser field (and the one we emphasize in this article), $c_s=1$ so that the inverse Fourier transform is consistent with the use of $A(t)\sim\int A(\omega)\Exp^{i\left(k_zz-\omega t\right)}\diff\omega$ in physical representation. Since this convention in physics is different from that in mathematics ($c_s=-1$), it is customary to define the DFT constant in physics with $C_{\mathfrak{F}_D}=\frac{1}{\mathfrak{N}}$, the inverse-DFT constant in mathematics, such that Eq.~(\ref{eq:AD-A_transform}) becomes $A_D(\omega)=A(\omega)\triangle\nu/C_{\mathfrak{F}}$ under this convention.

With Eqs.~(\ref{eq:continuous_convolution_thm}) and (\ref{eq:F_FD_conversion}), DFT's convolution theorem follows
\begin{subequations}
\begin{align}
\mathfrak{F}_D[A*B] & =\frac{\triangle t}{C_{\mathfrak{F}_D}}\mathfrak{F}_D[A]\mathfrak{F}_D[B] \label{eq:cc1D} \\
\mathfrak{F}_D^{-1}[A*B] & =\frac{\triangle\omega}{C_{\mathfrak{IF}_D}}\mathfrak{F}_D^{-1}[A]\mathfrak{F}_D^{-1}[B]. \label{eq:cc2D}
\end{align} \label{eq:DFT_convolution_thm}
\end{subequations}

In addition to the convolution [Eq.~(\ref{eq:convolution})], ``cross correlation'' is another formula that exhibits a convolution-theorem-like relation [Eq.~(\ref{eq:continuous_cross_convolution_thm})]. Although it is generally less-frequently used, it is the core element of the FFT-based Hankel transform (Sec.~\ref{sec:Hankel_transform}\ref{subsec:FHATHA}). It is denoted with $\star$ and follows
\begin{align}
\left(A\star B\right)(t) & \equiv\int_{-\infty}^{\infty}\overline{A(\tau)}B(t+\tau)\diff\tau \nonumber \\
& =\int_{-\infty}^{\infty}\overline{A(\tau-t)}B(\tau)\diff\tau.
\label{eq:cross_convolution}
\end{align}
It satisfies
\begin{subequations}
\begin{align}
\mathfrak{F}[A\star B] & =\frac{1}{C_{\mathfrak{F}}}\overline{\mathfrak{F}[A]}\mathfrak{F}[B] \label{eq:ccc1} \\
\mathfrak{F}^{-1}[A\star B] & =\frac{1}{C_{\mathfrak{IF}}}\overline{\mathfrak{F}^{-1}[A]}\mathfrak{F}^{-1}[B]. \label{eq:ccc2}
\end{align} \label{eq:continuous_cross_convolution_thm}
\end{subequations}
Similarly, its discrete counterpart is
\begin{subequations}
\begin{align}
\mathfrak{F}_D[A\star B] & =\frac{\triangle t}{C_{\mathfrak{F}_D}}\overline{\mathfrak{F}_D[A]}\mathfrak{F}_D[B]\xlongequal{\text{also}}\frac{\triangle t}{C_{\mathfrak{IF}_D}}\mathfrak{F}^{-1}_D[\overline{A}]\mathfrak{F}_D[B] \label{eq:ccc1D} \\
\mathfrak{F}_D^{-1}[A\star B] & =\frac{\triangle\omega}{C_{\mathfrak{IF}_D}}\overline{\mathfrak{F}_D^{-1}[A]}\mathfrak{F}_D^{-1}[B]\xlongequal{\text{also}}\frac{\triangle\omega}{C_{\mathfrak{F}_D}}\mathfrak{F}_D[\overline{A}]\mathfrak{F}_D^{-1}[B]. \label{eq:ccc2D}
\end{align} \label{eq:DFT_cross_convolution_thm}
\end{subequations}

Autocorrelation is defined as the cross correlation of itself:
\begin{align}
\mathcal{A}\equiv\left(A\star A\right)(t) & =\int_{-\infty}^{\infty}\overline{A(\tau)}A(t+\tau)\diff\tau \nonumber \\
& =\int_{-\infty}^{\infty}\overline{A(\tau-t)}A(\tau)\diff\tau.
\label{eq:acc}
\end{align}
It is a technique to measure the ultrafast-pulse duration (about a few femtoseconds or picoseconds) that is too short to be directly measured by any optical sensor. The pulse width of $\mathcal{A}$, a measurable ``time delay'' between pulses with a sensor, provides information for the pulse duration. However, it has been considered outdated due to a lack of accurate pulse information (\eg actual pulse temporal profile and phase), compared to recent more-advanced frequency-resolved optical gating (FROG) \cite{Kane1993} and dispersion-scan (d-scan) \cite{Miranda2012,Miranda2012a}. It satisfies [from Eq.~(\ref{eq:continuous_cross_convolution_thm})]
\begin{subequations}
\begin{align}
\mathfrak{F}[\mathcal{A}] & =\frac{1}{C_{\mathfrak{F}}}\abs{\mathfrak{F}[A]}^2 \label{eq:acc1} \\
\mathfrak{F}^{-1}[\mathcal{A}] & =\frac{1}{C_{\mathfrak{IF}}}\abs{\mathfrak{F}^{-1}[A]}^2, \label{eq:acc2}
\end{align} \label{eq:continuous_autoconvolution_thm}
\end{subequations}
Eq.~(\ref{eq:acc1}) is the basis of the Fourier-transform spectroscopy that utilizes a Michelson interferometer to measure the spectrum $\abs{\mathfrak{F}[A]}^2$ of an optical signal.

\pagebreak
\subsection{Important correct application of the Fourier-transform convention}\label{subsec:Important_use_of_Fourier-transform_convention}
Many physics equations are derived by assuming that the phase follows $\left(k_zz-\omega t\right)$, which implies that the Fourier transform in physics ($c_s=1$) has a different convention from that in mathematics ($c_s=-1$). For example in the single-spatial-mode unidirectional pulse propagation equation of an electronic-only $\chi^{(3)}$ nonlinear process \cite{Chen2024a}:
\begin{align}
\partial_zA(z,\Omega) & =i\left[\beta(\omega)-\left(\beta_{(0)}+\beta_{(1)}\Omega\right)\right]A(z,\Omega)+i\frac{\omega n_2}{cA_{\text{eff}}}\mathfrak{F}\left[\abs{A(z,T)}^2A(z,T)\right], \label{eq:UPPE}
\end{align}
whose derivation can be found in the supplement of \cite{Chen2024}. In the derivation, we can see that the field $A$ results from
\begin{equation}
\vec{\mathbb{E}}(\vec{r},t)=\int\diff\omega\frac{1}{2}\left[\frac{\vec{F}(\vec{r}_{\perp},\omega)}{N(\omega)}A(z,\omega)\Exp^{i\left[\beta(\omega)z-\omega t\right]}+\text{c.c.}\right],
\label{eq:mathbbE_Ap}
\end{equation}
where $A$ represents the frequency-offset part of the analytic signal of the real-valued electric field $\vec{\mathbb{E}}(\vec{r},t)$. $\Omega=\omega-\omega_0$. Some might rewrite it in the time domain (with a narrowband assumption with Taylor-series expansion around center frequency $\omega_0$) as
\begin{align}
\partial_zA(z,T) & =\left\{ i\left[\beta^{(0)}(\omega_0)-\beta_{(0)}\right]-\left[\beta^{(1)}(\omega_0)-\beta_{(1)}\right]\partial_T\right\}A(z,T)+i\sum_{m\geq2}\frac{\left(i\partial_T\right)^m}{m!}\beta^{(m)}(\omega_0)A(z,T) \nonumber \\
&\hspace{1em} +\frac{i\omega_0n_2}{cA_{\text{eff}}}\left[1+\tau_{1111}\left(i\partial_T\right)\right]\abs{A(z,T)}^2A(z,T),
\label{eq:GMMNLSE}
\end{align}
intending to be free from spectral computations. Still, terms with $i\partial_TA(z,T)$ that results from $\Omega A(z,\Omega)$ are related to the convention of Fourier transform with $c_s=1$:
\begin{equation}
\partial_TA(z,T) \xlongrightarrow{\mathfrak{F}} -ic_s\Omega A(z,\Omega).
\end{equation}

\begin{figure}[b]
\centering
\includegraphics[width=0.4\linewidth]{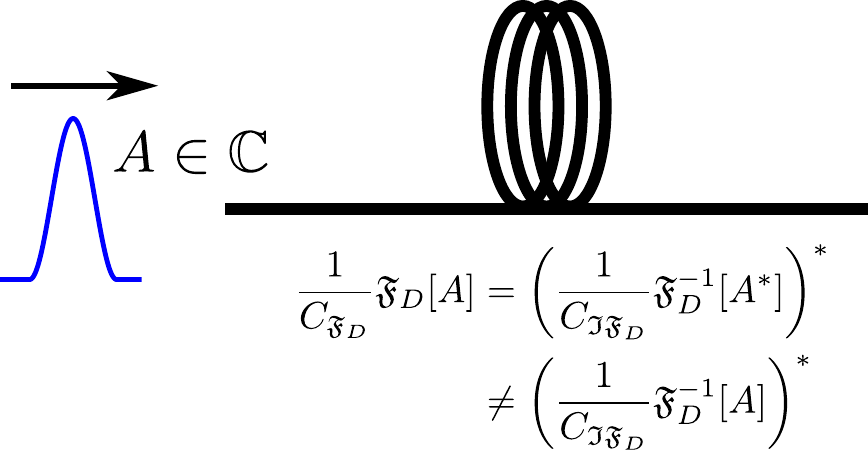}
\caption{Importance of applying the correct (same) Fourier-transform convention for the signal as the propagation equation. If the wrong convention is employed such that a signal in one convention is propagated with an equation of a different convention, the equation essentially sees a different signal. For example, the pulse, designed to be up-chirped (spectral components temporally vary from low at the leading edge to high frequencies at the trailing edge of a pulse), the equation with a different convention will treat it as down-chirped. Then the pulse propagation will be a propagation of a down-chirped pulse, which behaves differently from an up-chirped pulse. As an example, if the fiber exhibits normal dispersion, a down-chirped pulse will get dechirped and becomes shorter in time; however, an up-chirped pulse will be temporally stretched and up-chirped more.}
\label{fig:FT_convention_importance}
\end{figure}

As a result, it is crucial to apply the correct Fourier-transform convention. Since DFT depends only on the numerical sampling, but not on the actual time or frequency, the DFT with $c_s=-1$ is equivalent to the inverse DFT with $c_s=1$, if we ignore their Fourier-transform constants here. In nonlinear optics, it follows the physical convention, so numerically it relies on the mathematics-convention inverse DFT (inverse DFT with $c_s=-1$, or MATLAB's $\ifft$) for Fourier-transform into the spectral domain and mathematics-convention DFT (MATLAB's $\fft$) for inverse-Fourier-transform into the temporal domain. Some might think that, in numerical computations, the wrong use of Fourier-transform convention simply creates a signal that is a complex conjugate of the correct one, ``both following the same pulse propagation equation.'' Since the analytic signal, or its envelope, is generally complex-valued, the spectral signal transformed with $\fft$ is not a complex conjugate of the spectral signal transformed with $\ifft$: $\frac{1}{C_{\mathfrak{F}_D}}\mathfrak{F}_D[\cdot]\neq\left(\frac{1}{C_{\mathfrak{IF}_D}}\mathfrak{F}_D^{-1}[\cdot]\right)^{\ast}$ (Fig.~\ref{fig:FT_convention_importance}). Note that although a real-valued signal satisfies the complex-conjugate relation: $\frac{1}{C_{\mathfrak{F}_D}}\mathfrak{F}_D[A^{\mathbb{R}}]=\left(\frac{1}{C_{\mathfrak{IF}_D}}\mathfrak{F}_D^{-1}[A^{\mathbb{R}}]\right)^{\ast}$ [$A(z=0)=A^{\mathbb{R}}\in\mathbb{R}$], it becomes complex-valued during pulse propagation and violates the relation. Since $A(z,\Omega)$ is the envelope centering around $\omega_0$ [Fig.~\ref{fig:DFT_Omega}(b)], the wrong convention represents the field of the reversed frequency sign $A(z,-\Omega)$ that should not follow the same pulse propagation equation as $A(z,\Omega)$. For example, the dispersion term $\beta(\omega)$ [Eqs.~(\ref{eq:UPPE} and (\ref{eq:GMMNLSE}))] should be reversed around $\omega_0$ as well:
\begin{equation}
\beta(\omega=\omega_0+\Omega)\rightarrow\beta(\omega_0-\Omega).
\end{equation}
In general, a wrong convention of Fourier transform applied to an evolution/propagation equation derived with a different convention simply creates a wrong pulse-propagation result, unless the equation is revised accordingly.

A signal should experience the propagation equation in the same Fourier-transform convention. Since global phase does not affect physical phenomena, the signals in different conventions are related by complex conjugate: $\frac{1}{C_{\mathfrak{F}_D}}\mathfrak{F}_D[A]=\left(\frac{1}{C_{\mathfrak{IF}_D}}\mathfrak{F}_D^{-1}[A^{\ast}]\right)^{\ast}$. If a signal will be applied to a propagation equation with a different convention, complex-conjugate to transform the signal to the same convention should be applied before undergoing the propagation. In conclusion, a signal from one Fourier-transform convention cannot be sent into a propagation equation with a different convention, unless a complex-conjugate transformation is employed to ensure that the signal has the same convention as the propagation equation.

\begin{figure}[!ht]
\centering
\includegraphics[width=0.6\linewidth]{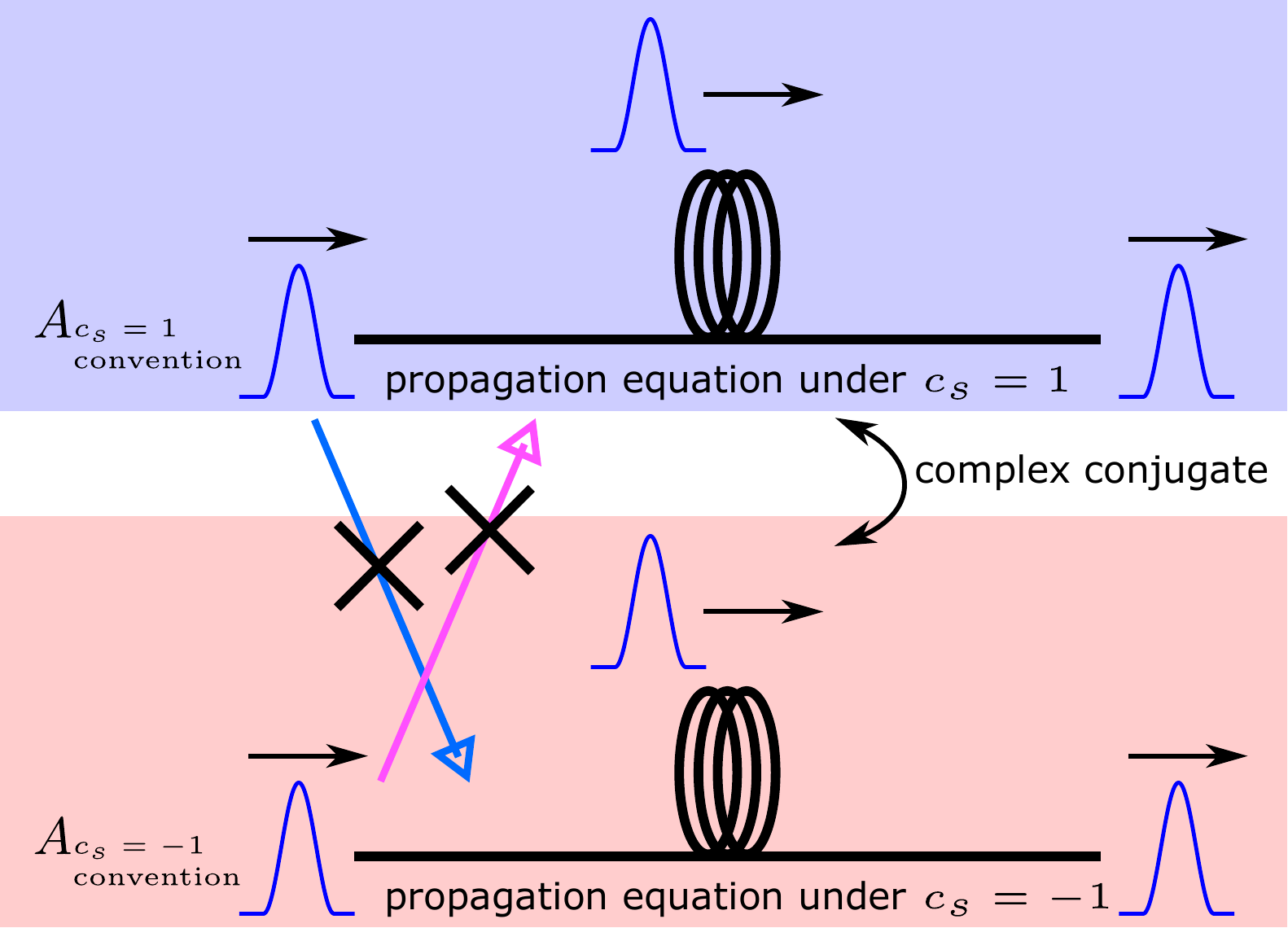}
\caption{Signals in different conventions differ by complex conjugate. However, a mixed use between two conventions is forbidden (Fig.~\ref{fig:FT_convention_importance}).}
\label{fig:FT_convention_importance2}
\end{figure}
\clearpage

\subsection{Conversion of quantities with physical-useful units between FT and DFT}\label{subsec:Relations_between_FT_and_DFT}
In this section, we derive several formulae for conversion of physical quantities between Fourier transform and discrete Fourier transform.
\begin{align}
\int_{-\infty}^{\infty}\abs{A(t)}^2\diff t & =\int_{-\infty}^{\infty}C_{\mathfrak{IF}}^2\int_{-\infty}^{\infty}\int_{-\infty}^{\infty}A(\omega)A^{\ast}(\omega')\Exp^{-ic_s\left(\omega-\omega'\right)t}\diff\omega\diff\omega'\diff t \nonumber \\
& =C_{\mathfrak{IF}}^2\int_{-\infty}^{\infty}\int_{-\infty}^{\infty}A(\omega)A^{\ast}(\omega')\left[\int_{-\infty}^{\infty}\Exp^{-ic_s\left(\omega-\omega'\right)t}\diff t\right]\diff\omega\diff\omega' \nonumber \\
& =C_{\mathfrak{IF}}^2\int_{-\infty}^{\infty}\int_{-\infty}^{\infty}A(\omega)A^{\ast}(\omega')\left[2\pi\delta\left(c_s(\omega-\omega')\right)\right]\diff\omega\diff\omega' \nonumber \\
& =2\pi C_{\mathfrak{IF}}^2\int_{-\infty}^{\infty}\abs{A(\omega)}^2\diff\omega,\quad\delta\left(c_s(\omega-\omega')\right)=\frac{\delta(\omega-\omega')}{\abs{c_s}}=\delta(\omega-\omega') \nonumber \\
& =\frac{C_{\mathfrak{IF}}}{C_{\mathfrak{F}}}\int_{-\infty}^{\infty}\abs{A(\omega)}^2\diff\omega.\quad\because C_{\mathfrak{IF}}=\frac{1}{2\pi C_{\mathfrak{F}}} \label{eq:Parseval_derivation}
\end{align}
Eq.~(\ref{eq:Parseval_derivation}) leads to the general formulation of the Parseval's theorem:
\begin{equation}
\frac{1}{C_{\mathfrak{IF}}}\int_{-\infty}^{\infty}\abs{A(t)}^2\diff t=\frac{1}{C_{\mathfrak{F}}}\int_{-\infty}^{\infty}\abs{A(\omega)}^2\diff\omega.
\label{eq:Parseval}
\end{equation}
With Eq.~(\ref{eq:AD-A_transform}) and $\triangle\omega=\frac{2\pi}{\mathfrak{N}\triangle t}$, we can derive the discrete version of the Parseval's theorem:
\begin{equation}
\frac{1}{C_{\mathfrak{IF}_D}}\sum_{n=1}^{\mathfrak{N}}\abs{A(t_n)}^2=\frac{1}{C_{\mathfrak{F}_D}}\sum_{m=1}^{\mathfrak{N}}\abs{A_D(\omega_m)}^2.
\label{eq:discrete_Parseval}
\end{equation}

Rewriting the Parseval's theorem in powers leads to $\int_{-\infty}^{\infty}P(t)\diff t=\int_{-\infty}^{\infty}P(\omega)\diff\omega$, where $P(t)=\abs{A(t)}^2$. Since $P(t)$ has the unit of ``\si{\W}=\si{\joule/\s},'' $P(\omega)=\frac{C_{\mathfrak{IF}}}{C_{\mathfrak{F}}}\abs{A(\omega)}^2=\frac{1}{2\pi C_{\mathfrak{F}}^2}\abs{A(\omega)}^2$ has the unit of ``\si{\joule/(\radian\cdot\Hz)}'' [$\omega$ has a unit of ``$\si{\Hz}/(2\pi)=\si{\radian\cdot\Hz}$'']. To calculate the spectrum with the unit of ``\si{\joule/\Hz}'' numerically,
\begin{equation}
P(\nu)=2\pi P(\omega)=\frac{1}{C_{\mathfrak{F}}^2}\abs{A(\omega)}^2=\left(\frac{\triangle t}{C_{\mathfrak{F}_D}}\right)^2\abs{A_D(\omega)}^2,
\label{eq:Pnu}
\end{equation}
by applying Eq.~(\ref{eq:AD-A_transform}) and $C_{\mathfrak{F}}C_{\mathfrak{IF}}=\frac{1}{2\pi}$. $\int P(\omega)\diff\omega=\int P(\nu)\diff\nu$ leads to $P(\nu)=2\pi P(\omega)$. With the DFT convention we use here ($C_{\mathfrak{F}_D}=\frac{1}{\mathfrak{N}}$), it becomes $P(\nu)=\left(T^w\right)^2\abs{A_D(\omega)}^2$.

The Parseval's theorem assumes the unit of energy (\si{\joule}) after the integral. However, for continuous waves, a unit in terms of \emph{power} makes more sense in the frequency domain. With a known time window $T^w=\mathfrak{N}\triangle t$, the continuous-wave spectral energy in this time window is $P_{\text{CW}}(\omega)T^w=\frac{1}{2\pi C_{\mathfrak{F}}^2}\abs{A_{\text{CW}}(\omega)}^2$, where $P_{\text{CW}}(\omega)$ is in ``\si{\W/(\radian\cdot\Hz)}.'' Hence,
\begin{equation}
\abs{A_{\text{CW}}(\omega)}=C_{\mathfrak{F}}\sqrt{2\pi P_{\text{CW}}(\omega)T^w}=C_{\mathfrak{F}}\sqrt{P_{\text{CW}}(\nu)T^w},
\label{eq:ACW_P}
\end{equation}
by use of the relation $P_{\text{CW}}(\nu)=2\pi P_{\text{CW}}(\omega)$. $P_{\text{CW}}(\nu)$ is in \si{\watt/\Hz}. This leads to, with Eq.~(\ref{eq:AD-A_transform}),
\begin{equation}
\abs{A_{D,\text{CW}}(\omega)}=\frac{C_{\mathfrak{F}_D}}{\triangle t}\sqrt{2\pi P_{\text{CW}}(\omega)T^w}=\frac{C_{\mathfrak{F}_D}}{\triangle t}\sqrt{P_{\text{CW}}(\nu)T^w},
\label{eq:ADCW_P}
\end{equation}
which results in $\abs{A_{D,\text{CW}}(\omega)}=\sqrt{\frac{P_{\text{CW}}(\nu)}{T^w}}=\sqrt{P_{\text{CW}}(\nu)\triangle\nu}$ with the DFT convention we use here (note that $T^w=1/\triangle\nu$).

In the common model of adding noise photon (\eg shot noise), the noise is added as a CW background with one noise photon per frequency mode/bin, or equivalently, with an one-noise-photon spectral distribution ($\si{\joule}=\si{\watt/\Hz}$) $P_{\text{noise}}(\nu)=h\nu$ \cite{Eggleston1980,Dudley2006,Frosz2010,Genier2019}. Eq.~(\ref{eq:ACW_P}) leads to
\begin{subequations}
\begin{align}
\abs{A_{\text{noise photon}}(\omega)} & =C_{\mathfrak{F}}\sqrt{T^wh\nu} \\
\abs{A_{D,\text{noise photon}}(\omega)} & =\frac{C_{\mathfrak{F}_D}}{\triangle t}\sqrt{T^wh\nu}, \label{eq:AD_CWnoise}
\end{align}
\end{subequations}
Eq.~(\ref{eq:AD_CWnoise}) gives $\abs{A_{D,\text{noise photon}}(\omega)}=\sqrt{h\nu/T^w}=\sqrt{h\nu\triangle\nu}$ with the DFT convention we use here.

The power spectral density $P(\omega)$ or $P(\nu)$ can also be represented in the wavelength domain. First, we derive the relation
\begin{align}
& c=\nu\lambda \nonumber \\
&\Rightarrow 0=\lambda\diff\nu+\nu\diff\lambda \nonumber \\
&\Rightarrow \diff\nu=-\frac{\nu}{\lambda}\diff\lambda=-\frac{c}{\lambda^2}\diff\lambda
\end{align}
which leads to
\begin{align}
\int P(\lambda)\abs{\diff\lambda} & =\int P(\nu)\diff\nu \nonumber \\
& =\int P(\nu)\abs{-\frac{c}{\lambda^2}\diff\lambda}=\int\left(\frac{c}{\lambda^2}P(\nu)\right)\abs{\diff\lambda} \nonumber \\
&\Rightarrow P(\lambda)=\frac{c}{\lambda^2}P(\nu)
\label{eq:P_lambda}
\end{align}
Because wavelength and frequency have an inverse relation, and the power is always positive, an absolute value is taken in derivation. $P(\lambda)$ has the unit ``\si{\joule/\m},'' whose CW version [Eqs.~(\ref{eq:ACW_P}) and (\ref{eq:ADCW_P})] is in ``\si{\watt/\m}.''

\pagebreak
\section{Discrete Fourier transform (DFT)}\label{sec:DFT}
This section focuses on the numerical aspect of DFT, so the DFT discussion here can be applied to both spatial/$\vec{k}_{\perp}$-space and time/frequency versions. For simplicity, we employ the time/frequency version.

In practice, DFT is computed with finite sampling in a finite window. If a signal is temporally-confined [Fig.~\ref{fig:DFT_periodic_explain_f}(a)], the integration range of Fourier transform can be changed from $(-\infty,\infty)$ to a small window that is just sufficiently large to cover the signal, which is practically realizable with DFT. However, in general, DFT does not necessarily duplicate the Fourier transform because finite-sampling DFT discards the signal information beyond the window. For example, the DFT of a known temporal signal remains the same for $\omega\rightarrow\omega+2\pi\nu^w$:
\begin{align}
A(\omega+2\pi\nu^w) & =\mathfrak{F}_D[A(t_n)]\equiv C_{\mathfrak{F}_D}\sum_{n=1}^{\mathfrak{N}}A(t_n)\Exp^{ic_s\left(\omega+2\pi\nu^w\right)t_n} \nonumber \\
& =A(\omega),\quad\because t_n\nu^w=n,
\end{align}
where $\nu^w$ is the frequency window. The formulation of DFT establishes the implicit periodicity of the transformed signal [Fig.~\ref{fig:DFT_periodic_explain_f}(b)]. Due to the symmetry in formulation of DFT and inverse DFT, the implicit periodicity is established for both the temporal and spectral signals, or the spatial and $\vec{k}_{\perp}$-space signals.

\begin{figure}[!ht]
\centering
\includegraphics[width=0.7\linewidth]{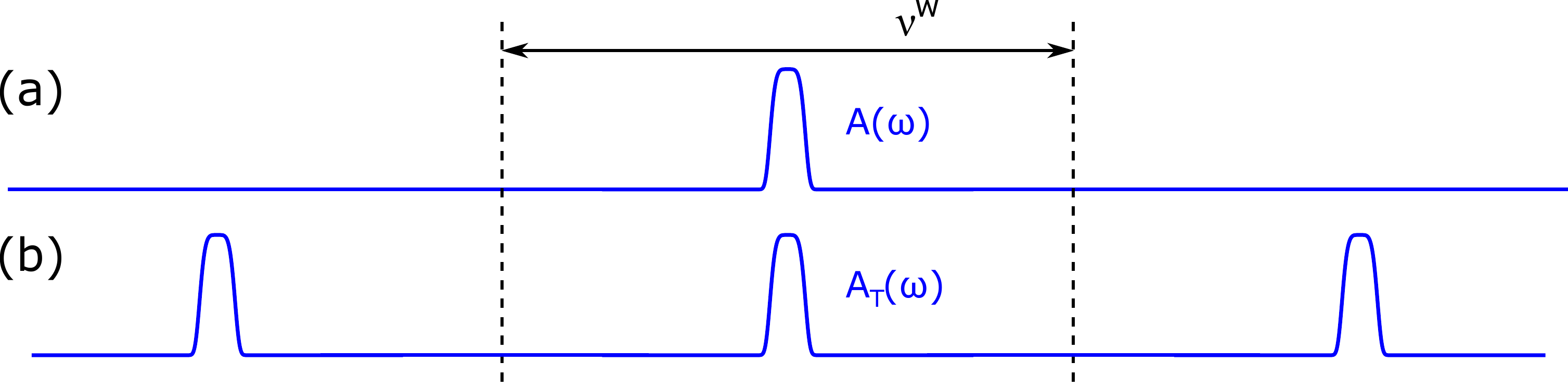}
\caption{(a) Spectrally-confined signal. A frequency window $\nu^w$ covers the signal. (b) Spectrally-periodic signal with a period $\nu^w$.}
\label{fig:DFT_periodic_explain_f}
\end{figure}

\subsection{fftshift/ifftshift and phase unwrapping}\label{subsec:fftshift}
During DFT operations in numerical computations, it is sometimes necessary to apply (MATLAB's) ``$\fftshift$'' and ``$\ifftshift$,'' the offset of the signal. For example, because the frequency coordinate of the spectral field calculated from DFT has zero frequency at the first sampling point (the left edge of the frequency window), the spectrum appears split into half, one at the left edge and the other at the right edge (Fig.~\ref{fig:DFT}). This occurs because of the assumed periodicity of the DFT spectral field. It is then desirable to ``shift'' the signal to the center of the window for visualization or other computations. Despite various Fourier-transform conventions, $\fftshift$ is used to shift the signal to the center of the window (move $t_x=0$ or $\nu_x=0$ to window's center), whether it is in the temporal or spectral domain. On the other hand, $\ifftshift$ cancels the $\fftshift$ effect and shifts the signal to center at $t_x=0$ or $\nu_x=0$ (left edge of the window). $t_x$ and $\nu_x$ represent the sampling coordinates in the temporal and spectral domains, respectively. They should not be used simply as a pair of ($\fft$,$\fftshift$) and ($\ifft$,$\ifftshift$) due to different Fourier-transform conventions. For example, to compute the spectrum under the physical Fourier-transform convention ($c_s=1$), we should follow (in the MATLAB syntax below)
\begin{lstlisting}[language=MATLAB]
Nt = size(field,1); % the number of sampling points
t = (-Nt/2:Nt/2-1)'*dt; % ps; time coordinates; dt = temporal sampling spacing
c = 299792.458; % nm/ps; speed of light
f = f0+(-Nt/2:Nt/2-1)'/(Nt*dt); % THz; frequency coordinate; f0 = center frequency of the freuqnency window
wavelength = c./f; % nm
correct_unit = (Nt*dt)^2/1e3; % to make the spectrum of the correct unit "nJ/THz" [Eq.(32)]
                              % "/1e3" is to make pJ into nJ
                              % "field" unit: sqrt(W)
spectrum = abs(fftshift(ifft(field),1)).^2*correct_unit; % nJ/THz; centered at the frequency window [Fig.5(c)]
\end{lstlisting}

In practice, the time coordinate is placed such that the pulse at the center of the numerical time window locates at $t=0$ in real-time ($t$) coordinates [Fig.~\ref{fig:DFT}(b)]. Do not confuse it with the pulse locating at $t_x=0$ [Fig.~\ref{fig:DFT}(a)]. In principle, the time coordinate can be placed arbitrarily because numerical computations see only the sampling-point coordinate $t_x$ or $\nu_x$.

\begin{figure}[!ht]
\centering
\includegraphics[width=.9\linewidth]{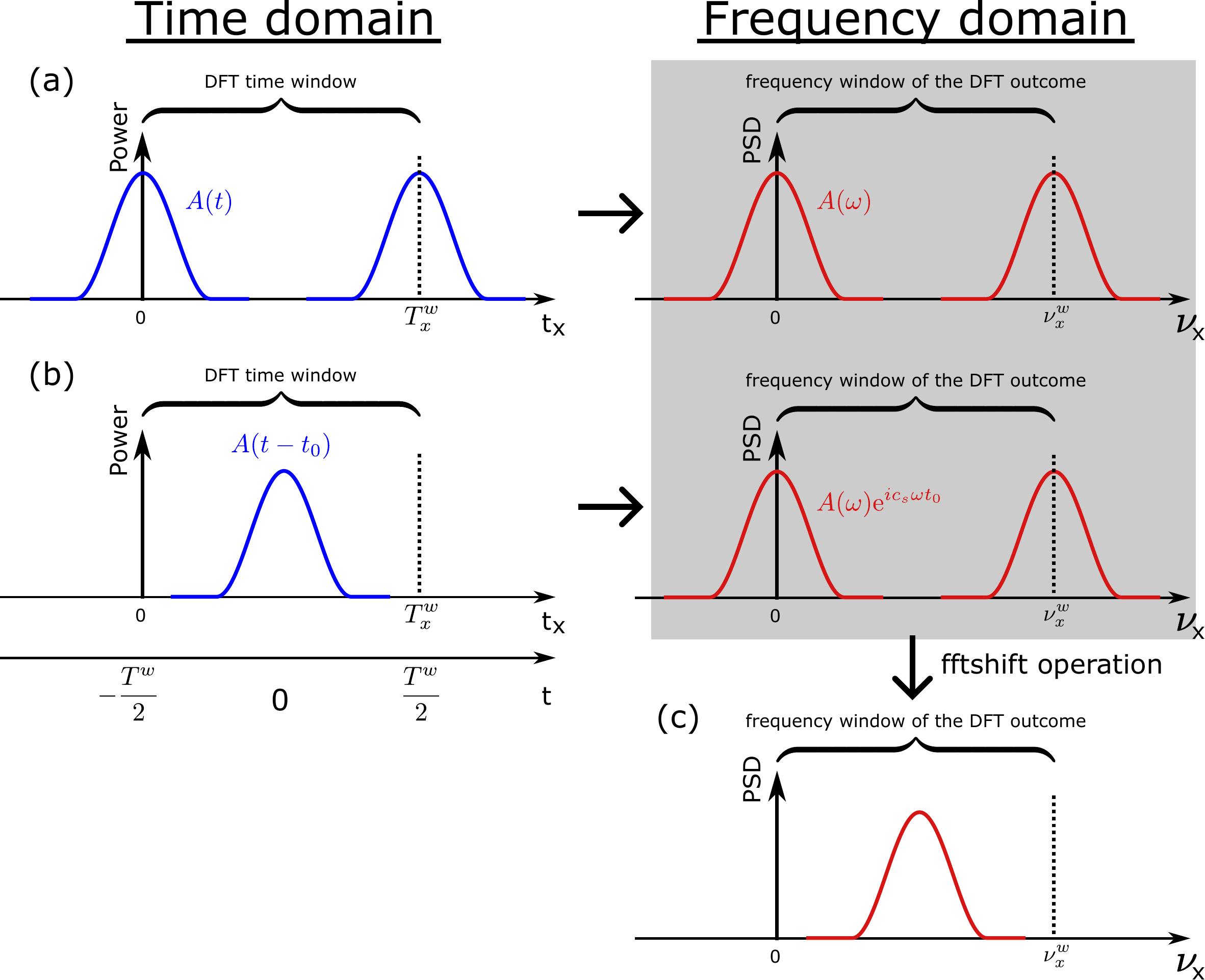}
\caption{DFT conversion. (a) is the ``formal'' use of DFT when the temporal profile $\abs{A(t)}^2$ is centered at $t=0$. However, in numerical simulations, it is common to place the pulse at the center of the time window for visualization purpose (b), resulting in a spectral phase shift that doesn't affect the spectral shape. (c) is the result after $\fftshift$ centers the spectrum with respect to the frequency window. PSD: power spectral density $\sim\abs{A(\omega)}^2$. Here, the subscript ``$x$'' represents the coordinate of sampling points, rather than the actual time and frequency coordinates.}
\label{fig:DFT}
\end{figure}
\clearpage

Due to the numerical phase-unwrapping process, the temporal position of the pulse can affect the acquisition of a smooth phase relation $\phi(\omega)$. Numerically, phase of a complex number is found within the range of either $(0,2\pi]$ or $(-\pi,\pi]$. Phase unwrapping is a process that removes the phase jump by adding multiples of $\pm2\pi$ when the phase change between two data points is larger than a threshold value, typically $\pi$ [Fig.~\ref{fig:unwrap}(a)]. This operation is crucial in phase computations, such as characterizing the second-order spectral phase $\od[2]{\phi}{\omega}$ in a pulse to determine the pulse-dechirping strategy (see Sec.~\ref{subsec:phenomena_phase}). Therefore, it is important to unwrap the phase correctly such that it does not arbitrarily add multiples of $\pm2\pi$ that distorts the phase relation $\phi(\omega)$. Phase unwrapping can go wrong when the phase relation $\phi(\omega)$ changes too drastically beyond the threshold value between two sampling points. The unwrapping operation will attempt to reduce the variation, leading to a kink, followed by a slope of reverse sign due to the continuous reduction of the phase change by $\pm2\pi$ [Fig.~\ref{fig:unwrap}(b)]. Since a temporal offset of a pulse adds a linear spectral phase $(c_s\omega t_0)$ [Fig.~\ref{fig:DFT}(b)], this extra addition could increase the rate of phase change and trigger a wrong unwrapping operation. As a result, in ultrafast optics, the fundamental principle is to first remove the temporal offset and place the pulse at $t_x=0$ (left edge of the time window), as in Fig.~\ref{fig:DFT}(a), before applying any spectra-phase computations. The value of temporal offset is remembered to recover the pulse position after the operation if necessary. However, if the phase relation inherently shows a rapid phase change, as illustrated in Fig.~\ref{fig:unwrap}(b), the only viable solution to accurately unwrap the phase is to enhance the spectral resolution, which necessitates extending the numerical time window (because $T^w=1/\triangle\nu$).

\begin{figure}[!ht]
\centering
\includegraphics[width=0.7\linewidth]{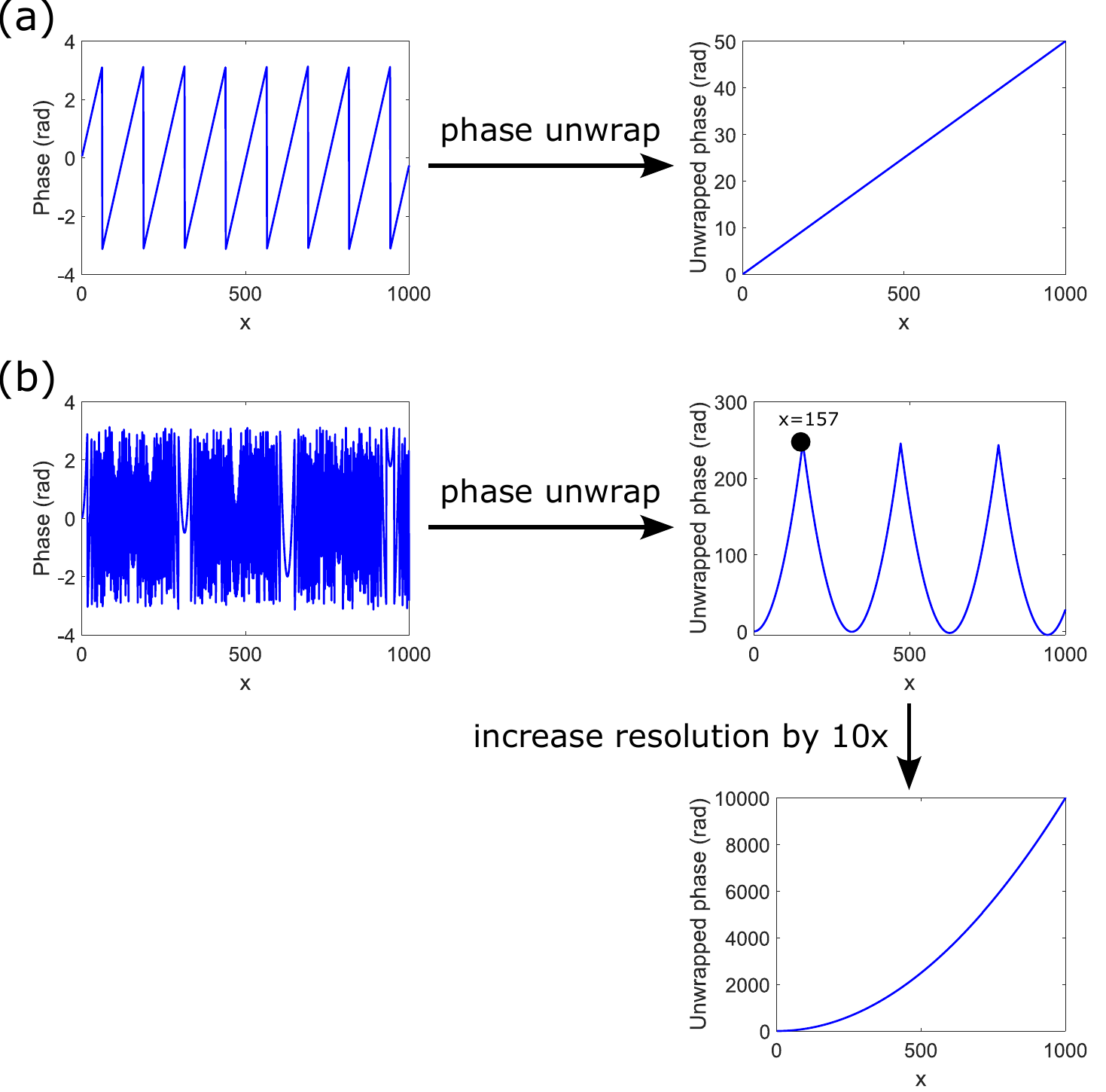}
\caption{Phase-unwrapping process: $\phi(x)=\unwrap\left(\myangle\left(y(x)\right)\right)$ ($x=0:1:1000$ here, an integer array from $0$ to $1000$ in the MATLAB syntax). In MATLAB, phase from $\myangle(\cdot)$ is in $(-\pi,\pi]$. (a) Linear spectral phase: $y=\Exp^{ix/20}$. Its phase does not vary dramatically between data points, so phase is correctly unwrapped. (b) Quadratic spectral phase: $y=\Exp^{ix^2/100}$. This phase rapidly varies for larger values of $x$. Beyond $x=157$, the phase change between two consecutive points always exceeds $\pi$, constantly triggering phase-unwrapping operation to add another $-2\pi$. This results in decreasing phase values, ultimately leading to the formation of multiple parabolic curves. To correctly unwrap the phase, the resolution is increased to \num{10} times: $x=0:0.1:1000$ (from $0$ to $1000$ with $0.1$ spacing).}
\label{fig:unwrap}
\end{figure}

\subsection{DFT's convolution}\label{subsec:DFT_convolution}
In ultrafast optics, convolution is commonly used. There are two types of convolutions:
\begin{subequations}
\begin{alignat}{2}
\text{linear convolution: } && \left(A*B\right)(t) & \equiv\int_{-\infty}^{\infty}A(\tau)B(t-\tau)\diff\tau \label{eq:linear_convolution} \\
\text{circular convolution: } && \left(A_T*B_T\right)_{\text{circular}}(t) & \equiv\int_{t_0}^{t_0+T}A_T(\tau)B_T(t-\tau)\diff\tau. \label{eq:circular_convolution}
\end{alignat}
\end{subequations}
with $A_T$ and $B_T$ being periodic with period $T$, and $t_0$ is an arbitrary value. Both satisfy similar convolution theorems:
\begin{subequations}
\begin{alignat}{2}
\text{linear convolution theorem: } && \mathfrak{F}[A*B] & =\frac{1}{C_{\mathfrak{F}}}\mathfrak{F}[A]\mathfrak{F}[B] \label{eq:linear_convolution_thm} \\
\text{circular convolution theorem: } && \mathfrak{F}_T[\left(A_T*B_T\right)_{\text{circular}}] & =\frac{1}{C_{\mathfrak{F}}}\mathfrak{F}_T[A_T]\mathfrak{F}_T[B_T], \label{eq:circular_convolution_thm}
\end{alignat}
\end{subequations}
where $\mathfrak{F}_T[A_T]=C_{\mathfrak{F}}\int_{t_0}^{t_0+T}A_T(t)\Exp^{ic_s\omega t}\diff t$ is the ``modified'' Fourier transform with a small integration range $T$. In practical numerical computations, it is required to define a window that covers the signal and decide the sampling conditions (uniform vs.\ non-uniform sampling, sampling spacing, and the number of sampling points), which lead to finite number of sampling points within a finite window. Therefore, the linear convolution theorem in Eq.~(\ref{eq:linear_convolution_thm}), which needs an infinite integration range $(-\infty,\infty)$ and thus an infinite number of sampling points, is never numerically useful. On the other hand, the circular convolution theorem is useful, which can be easily computed with the (finite-sampling) fast-Fourier-transform (FFT)-based DFT that offers significant computational efficiency. However, in physics equations, the convolution is always the ``linear'' convolution. Only under specific conditions do the two convolutions become equivalent, allowing us to solve the ``linear'' convolution with the computationally-efficient circular convolution theorem.

Circular (or cyclic) convolution arises from the linear convolution involving a (or two) periodic signal(s) [Fig.~\ref{fig:convolution_periodic}(b)]:
\begin{align}
\left(A*B_T\right)(t) & \equiv\int_{-\infty}^{\infty}A(\tau)B_T(t-\tau)\diff\tau,\quad B_T\text{ is periodic with period }T \nonumber \\
& =\sum_{k=-\infty}^{\infty}\int_{t_0}^{t_0+T}A(\tau'+kT)B_T(t-\tau'-kT)\diff\tau'=\sum_{k=-\infty}^{\infty}\int_{t_0}^{t_0+T}A(\tau'+kT)B_T(t-\tau')\diff\tau' \nonumber \\
& =\int_{t_0}^{t_0+T}\left(\sum_{k=-\infty}^{\infty}A(\tau'+kT)\right)B_T(t-\tau')\diff\tau' \nonumber \\
& =\int_{t_0}^{t_0+T}A_T(\tau')B_T(t-\tau')\diff\tau',\quad A_T=\sum_{k=-\infty}^{\infty}A(\tau'+kT)\text{ is periodic with period }T \nonumber \\
& \equiv \left(A_T*B_T\right)_{\text{circular}}(t).
\label{eq:circular_convolution_derivation}
\end{align}
Furthermore, if $t\in[t_0,t_0+T]$ which $\mathfrak{F}_T^{-1}$ is defined to recover to, it can be calculated with
\begin{equation}
\left(A_T*B_T\right)_{\text{circular}}=\mathfrak{F}_T^{-1}\left[\frac{1}{C_{\mathfrak{F}}}\mathfrak{F}_T[A_T]\mathfrak{F}_T[B_T]\right] \label{eq:circular_convolution_thm2}
\end{equation}
due to the circular convolution theorem, which can be computed with the finite-sampling DFT.

Only under limited conditions is discrete linear convolution theorem [Eq.~(\ref{eq:DFT_convolution_thm}); discrete version of Eq.~(\ref{eq:linear_convolution_thm})] practically numerically available, \ie with finite sampling points in a limited (temporal/spectral/spatial) range. For signals that are confined in time [Fig.~\ref{fig:convolution_periodic}(a)], the integration range of linear convolution [Eq.~(\ref{eq:linear_convolution})] can be significantly reduced from $({-\infty},{\infty})$ to a finite range $T^w$ that covers the signals. Because linear convolution can be visualized with integration of $A$ and a sliding $t$-inverted $B$ (whose amount of sliding is determined by $C$'s coordinate), the approximate width of the convoluted signal $C$ is the summation of widths of $A$ and $B$: $\triangle t_C=\triangle t_A+\triangle t_B$. Therefore, the condition to satisfy to avoid a distorted convolution is $T^w\geq\triangle t_C$, which, in the discrete manner, becomes $T^w\geq\triangle t_C=\mathfrak{N}_C=\mathfrak{N}_A+\mathfrak{N}_B-1$ [$\mathfrak{N}_i$ is the number of points of signal $i$ ($i=A,B,C$)]. With finite discrete sampling, the continuous Fourier transform and inverse Fourier transform within the reduced range $T^w$ can be, respectively, numerically computed with FFT-based DFT and inverse DFT, which leads to efficient computation of the discrete linear convolution theorem.

\begin{figure}[!ht]
\centering
\includegraphics[width=\linewidth]{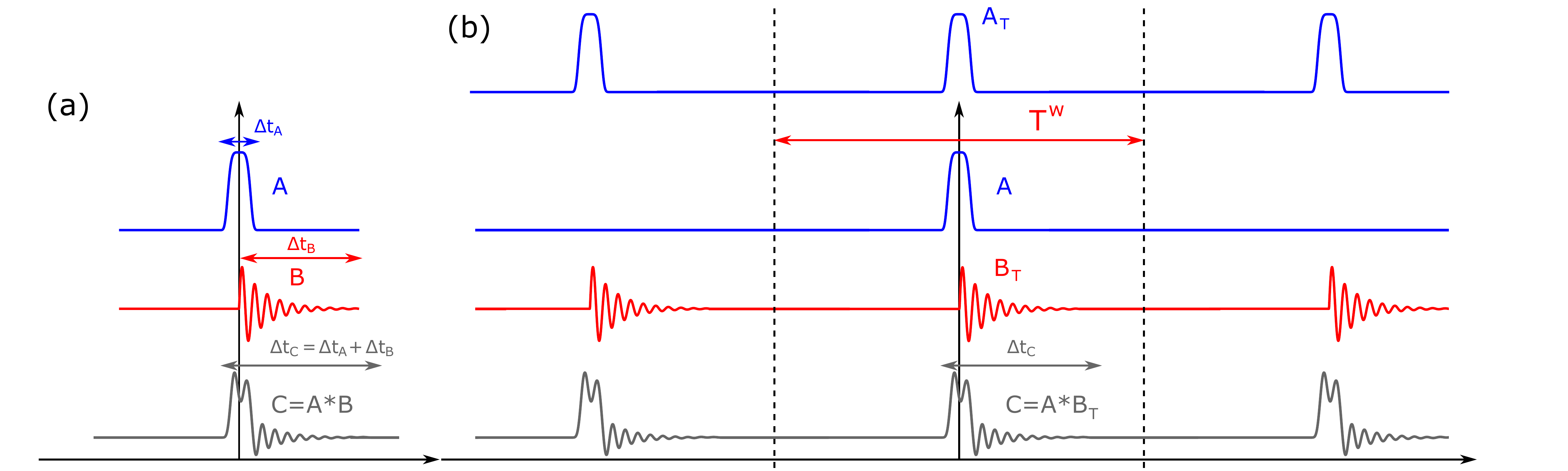}
\caption{Linear convolution $C=A*B$. (a) is the convolution scheme when signals $A$ and $B$ are confined in time. (b) is the scheme when $B$ becomes a periodic signal with period $T^w$.}
\label{fig:convolution_periodic}
\end{figure}

The DFT-based convolution theorem [Eq.~(\ref{eq:cc1D})]
\begin{equation}
A*B=\mathfrak{F}_D^{-1}\left[\frac{\triangle t}{C_{\mathfrak{F}}}\mathfrak{F}_D[A]\mathfrak{F}_D[B]\right] \label{eq:discrete_convolution_thm}
\end{equation}
implicitly assumes periodicity for the convoluted signal $\left(A*B\right)$ due to $\mathfrak{F}_D^{-1}$. The periodicity in convolution implies periodicity in either signal $A$ or signal $B$, or both. This establishes equivalence between the DFT-based convolution theorem [Eq.~(\ref{eq:discrete_convolution_thm})] and the circular convolution theorem [Eq.~(\ref{eq:circular_convolution_thm})]. In conclusion, the practical finite-sampling (with a finite range) DFT-based convolution theorem is applicable to confined signals or periodic signals.

It is important to note that the discussion here applies to inverse-Fourier-transform-based convolution theorems (with $\mathfrak{F}\leftrightarrow\mathfrak{F}^{-1}$or $\mathfrak{F}_D\leftrightarrow\mathfrak{F}_D^{-1}$) [Eqs.~(\ref{eq:cc2}) and (\ref{eq:cc2D})], or the spectral (or $\vec{k}_{\perp}$) domain (with $t\leftrightarrow\nu$ or $\vec{r}_{\perp}\leftrightarrow\vec{k}_{\perp}$), due to the symmetric formulation of the Fourier-transform pair.

In ultrafast optics, there are situations when the pulse is short but the Raman response to be convoluted with extends in time [Fig.~\ref{fig:convolution_zeropadding}(a) with an extended signal $B$]. Since only the physical phenomena of the pulse matters, it is desirable to compute the convolution with a window that is comparable to the pulse width so that only a small number of sampling points is considered [Fig.~\ref{fig:convolution_zeropadding}(b)]. However, the small time window clips the extended signal, resulting in a distorted convoluted outcome. Because the width of the convoluted signal $\triangle t_C$ is larger than the clipped extended signal $\triangle t_B$, there is a width of $\triangle t_C-\triangle t_B=\triangle t_A$ of distortion. If $\triangle t_B$ is too small due to significant clipping, the distortion happens inside the time window. By temporally shifting the extended signal $B$ such that it is clipped less (it is still the same function $B(t)$ but is shifted with respect to numerical sampling points), the valid part of the convolution increases so that the entire convoluted signal in the time window is valid [Fig.~\ref{fig:convolution_zeropadding}(c)]. Due to the assumed periodicity resulting from the DFT-based convolution theorem, it is important to have a window that covers the convoluted signal; otherwise, there will be distortion due to overlapped convoluted signals [Fig.~\ref{fig:convolution_zeropadding}(d)]. The distortion usually creates or modulates the frequency components at the low- or high-frequency sides of the frequency window, resulting from other periodic signals. This phenomenon is also called ``aliasing.'' Zero-padding is an efficient solution to extend the window and avoid aliasing [see MATLAB code below; Fig.~\ref{fig:convolution_zeropadding}(e)]. Since the convoluted signal has a width of $\triangle t_C$, both signals are extended to a width larger than $\triangle t_C=\mathfrak{N}_A+\mathfrak{N}_B-1$ from their original widths ($\triangle t_A=\mathfrak{N}_A$ or $\triangle t_B=\mathfrak{N}_B$). After convolution (with the DFT-based convolution theorem) with zero-padded signals, only those in the original time window are kept, which is the valid part of the convoluted signal.
\begin{lstlisting}[language=MATLAB]
% For example,
A = (1:5)';
B = rand(10,1);

Na = length(A);
Nb = length(B);
Nc = Na + Nb - 1; % length of the convoluted signal

% A and B are column vectors here
A = [A;zeros(Nc-Na,1)]; % zero-padding
B = [B;zeros(Nc-Nb,1)]; % zero-padding
C = Nc*fft(ifft(A).*ifft(B)); % Eq.(43) but dt is ignored here

% C equals conv(A,B).
\end{lstlisting}
Lastly, it is important to note that aliasing is a fundamental issue in DFT, beyond the convolution operation. When any signal to be Fourier-transformed with extends to the edges of the time or frequency window, it results in temporal or spectral aliasing that artificially generates temporal or spectral components, respectively.

\begin{figure}[!ht]
\centering
\includegraphics[width=\linewidth]{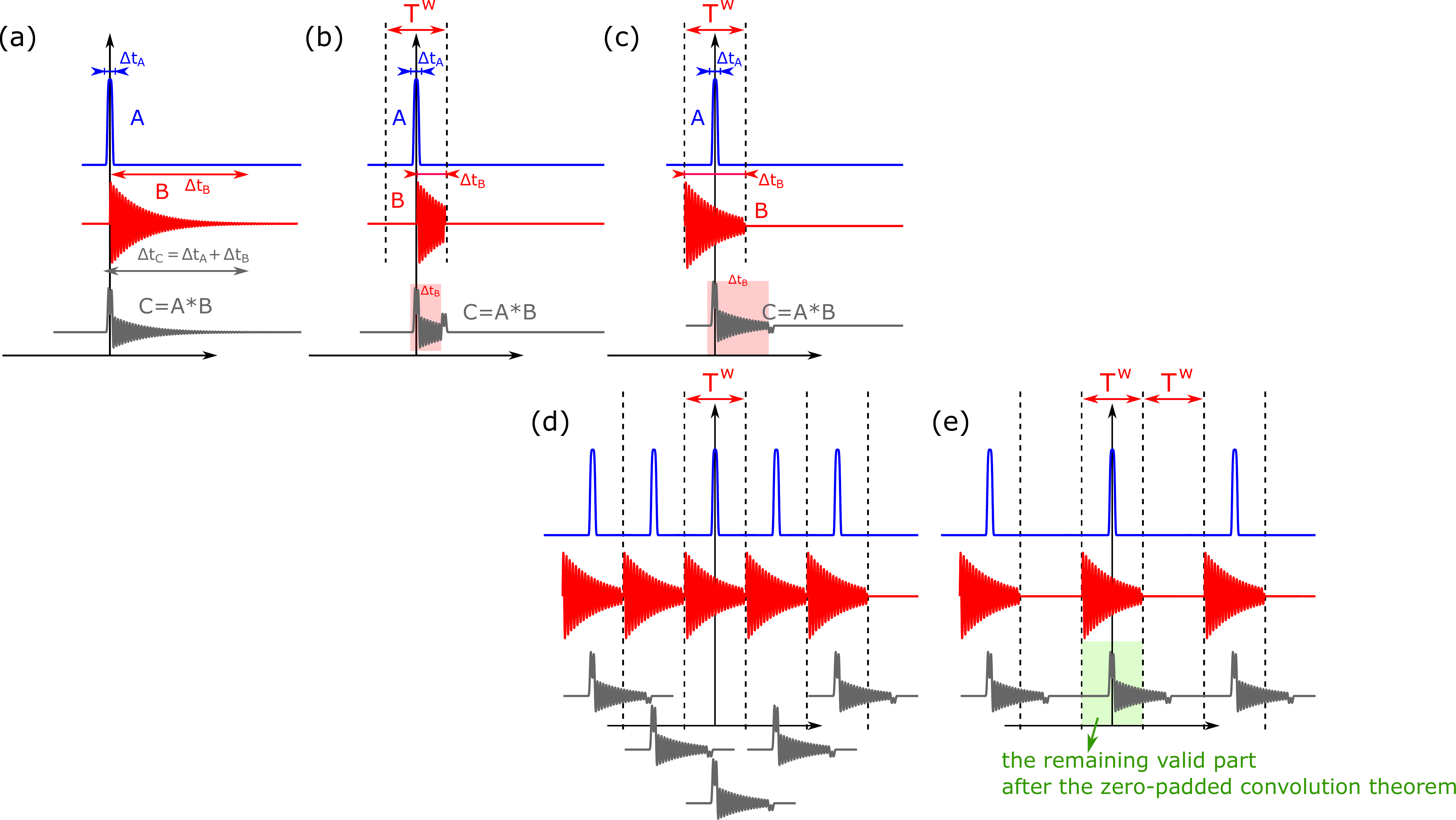}
\caption{(a) Linear convolution of $A$ with a temporally-extended $B$. (b) and (c) are the convolutions with $B$ clipped beyond the window $T^w$, with a shifted $B$ in (c). (d) and (e) are convolution results without and with zero-padded signals, respectively. Convoluted signals in (d) from each window/period overlap, distorting the outcome, while zero-padding avoids the distortion from overlapping.}
\label{fig:convolution_zeropadding}
\end{figure}
\clearpage

\pagebreak
\section{Complex-valued Fourier transform}\label{sec:complex_FT}
\subsection{Phase effect}\label{subsec:phenomena_phase}
Figs.~\ref{fig:DFT_chirp} and \ref{fig:DFT_chirp3} show how the phase affects the signal. When the pulse has a flat phase in the spectral domain, it is called ``a transform-limited pulse.'' Its temporal and spectral widths satisfy a fixed time-bandwidth product: a more-broadband pulse has a smaller duration. Adding a spectrally-varying phase to the pulse's spectral profile modulates the signal, making it longer in time. Therefore, a transform-limited pulse is usually considered the ``shortest'' pulse for a specific spectrum. Similarly due to the Fourier-transform symmetry, if a temporally-varying phase is applied to the temporal profile, its spectrum is broadened. For example, adding a parabolic (second-order) phase to the temporal profile of a transform-limited signal creates a chirp, \ie varying frequency at different temporal slices, broadening the spectrum [Fig.~\ref{fig:DFT_chirp}(b)]. The temporal frequency change follows $\triangle\omega(t)=\frac{1}{-c_s}\od{\phi}{t}(t)$, so the phase effect on the signal is also dependent on the convention of the Fourier transform. Similarly, adding a parabolic phase to the spectral profile increases the pulse duration, which is the physical ``dispersion'' effect: different frequencies moves at different speeds for a certain distance, widening the pulse's temporal profile [Figs.~\ref{fig:DFT_chirp}(c) and \ref{fig:spectrogram_dispersion}]. Fig.~\ref{fig:DFT_chirp3} shows the effect of a cubic phase, which broadens the spectral [Fig.~\ref{fig:DFT_chirp3}(b)] or temporal [Fig.~\ref{fig:DFT_chirp3}(c)] profiles with pedestals. Adding a linear chirp to the pulse to increase the pulse duration [Fig.~\ref{fig:DFT_chirp}(c)], reducing the peak power, is the basis of chirp-pulse amplification \cite{Strickland1985} that was awarded the Nobel Prize in Physics in 2018. It enables amplification of an ultrashort pulse without suffering from significant nonlinear-phase accumulations due to a reduced peak power.

\begin{figure}[!ht]
\centering
\includegraphics[width=.8\linewidth]{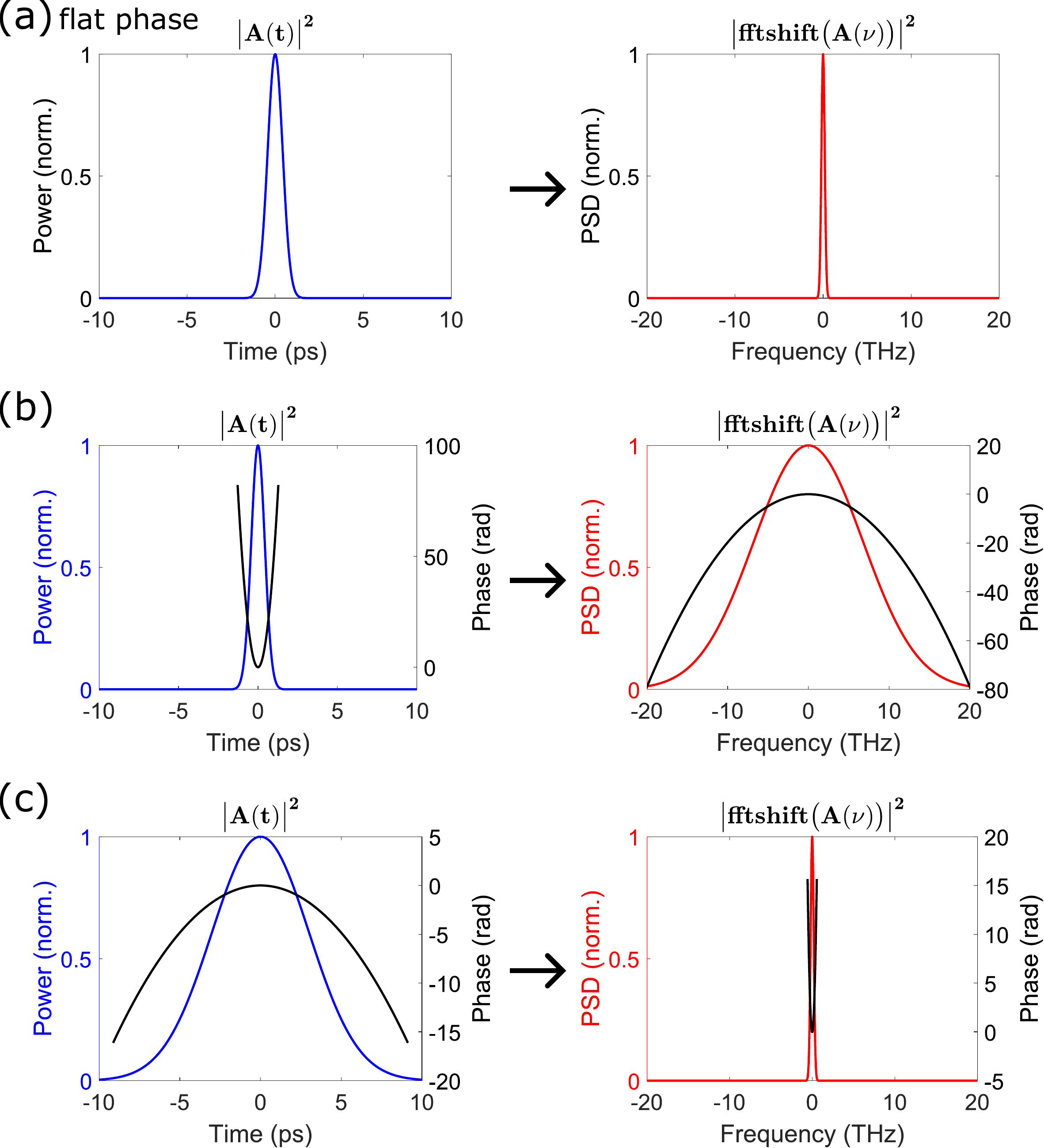}
\caption{DFT conversion of a second-order chirped signal. (a) transform-limited pulse that has only flat phase in time domain. (b) and (c) add a parabolic phase to the temporal and spectral profiles, respectively.}
\label{fig:DFT_chirp}
\end{figure}

\begin{figure}[!ht]
\centering
\includegraphics[width=.8\linewidth]{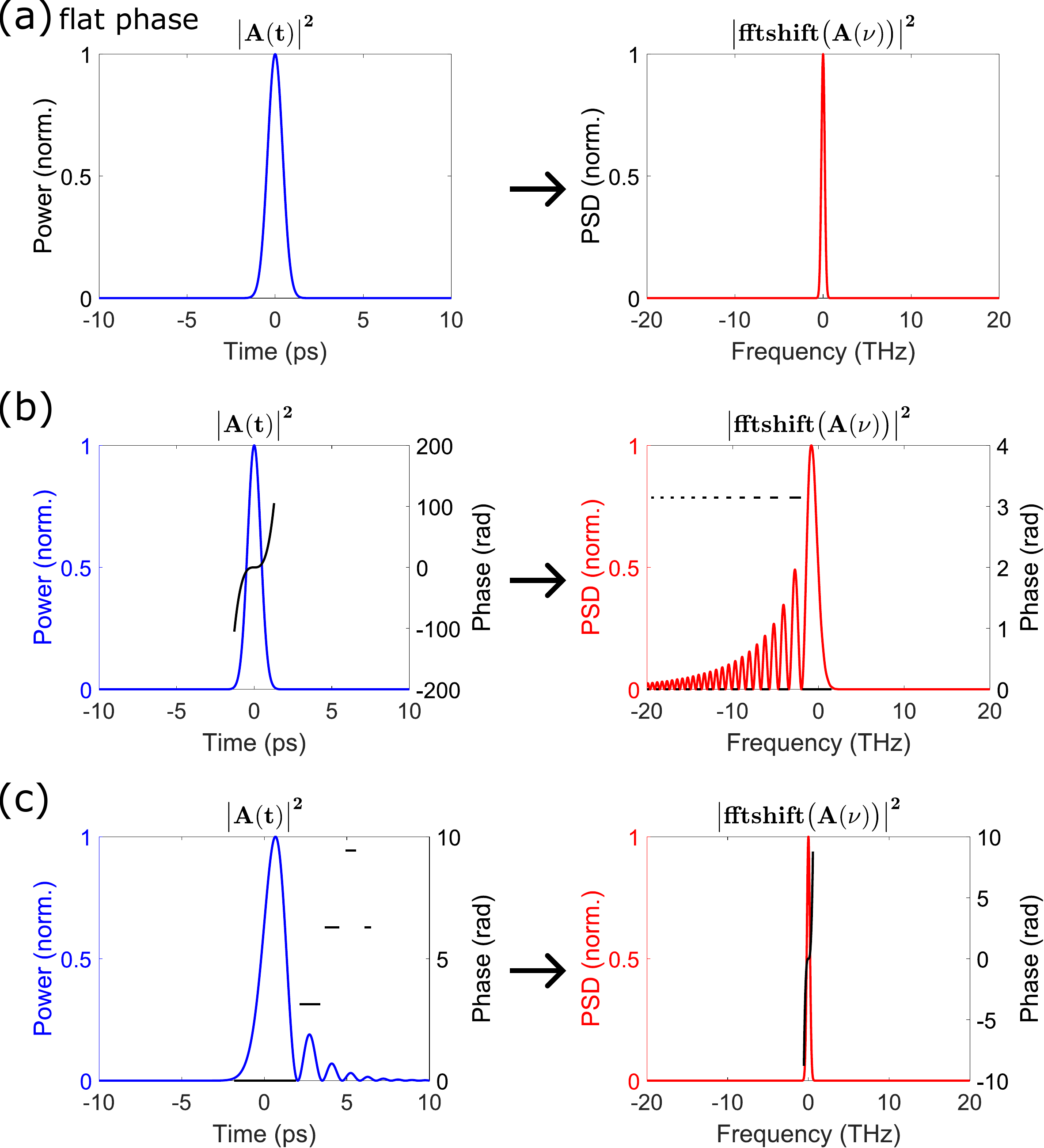}
\caption{DFT conversion of a third-order chirped signal. (a) transform-limited pulse that has only flat phase in time domain. (b) and (c) add a cubic phase to the temporal and spectral profiles, respectively.}
\label{fig:DFT_chirp3}
\end{figure}
\clearpage

\subsection{Time-frequency analysis}\label{subsec:time_frequency}
As discussed previously, the phase of a field is correlated with the frequency distribution in time, \ie different temporal positions of a field have different frequencies. This results in the necessity of time-frequency analysis of a field, which can be achieved by calculating
\begin{equation}
P_K(t,\omega)=\frac{C_{\mathfrak{F}}}{2\pi}\int_{-\infty}^{\infty}\int_{-\infty}^{\infty}\int_{-\infty}^{\infty}s_a(u+\frac{\tau}{2})\left[s_a(u-\frac{\tau}{2})\right]^{\ast}K(v,\tau)\Exp^{ic_s\left[v(u-t)+\omega\tau\right]}\diff u\diff v\diff\tau,
\end{equation}
where $s_a$ is the analytic signal of a field, and $K(v,\tau)$ is the kernel defining different time-frequency representations \cite{Najmi1994}.

\subsubsection{Wigner distribution and spectrogram}\label{subsubsec:Wigner_spectrogram}
If $K(v,\tau)=1$, it leads to the Wigner distribution:
\begin{align}
P_{\text{Wigner}}(t,\omega) & =\frac{C_{\mathfrak{F}}}{2\pi}\int_{-\infty}^{\infty}\int_{-\infty}^{\infty}s_a(u+\frac{\tau}{2})\left[s_a(u-\frac{\tau}{2})\right]^{\ast}\left[\int_{-\infty}^{\infty}\Exp^{ic_sv(u-t)}\diff v\right]\Exp^{ic_s\omega\tau}\diff u\diff\tau \nonumber \\
& =\frac{C_{\mathfrak{F}}}{2\pi}\int_{-\infty}^{\infty}\int_{-\infty}^{\infty}s_a(u+\frac{\tau}{2})\left[s_a(u-\frac{\tau}{2})\right]^{\ast}\left[2\pi\delta(u-t)\right]\Exp^{ic_s\omega\tau}\diff u\diff\tau,\quad\because \abs{c_s}=1 \nonumber \\
& =C_{\mathfrak{F}}\int_{-\infty}^{\infty}s_a(t+\frac{\tau}{2})\left[s_a(t-\frac{\tau}{2})\right]^{\ast}\Exp^{ic_s\omega\tau}\diff\tau.
\end{align}
It satisfies the ``marginal'' conditions:
\begin{subequations}
\begin{align}
\int_{-\infty}^{\infty}P_{\text{Wigner}}(t,\omega)\diff t & =\frac{1}{C_{\mathfrak{F}}}\abs{S_a(\omega)}^2 \label{eq:marginal_t} \\
\int_{-\infty}^{\infty}P_{\text{Wigner}}(t,\omega)\diff\omega & =\frac{1}{C_{\mathfrak{IF}}}\abs{s_a(t)}^2, \label{eq:marginal_omega}
\end{align} \label{eq:marginal_Wigner}
\end{subequations}
where $S_a(\omega)=\mathfrak{F}[s_a(t)]$. Again, this leads to the Parseval's theorem previously derived [Eq.~(\ref{eq:Parseval})]:
\begin{align}
\int_{-\infty}^{\infty}\int_{-\infty}^{\infty}P_{\text{Wigner}}(t,\omega)\diff t\diff\omega & \xlongequal{\text{Eq.~(\ref{eq:marginal_t})}}\int_{-\infty}^{\infty}\frac{1}{C_{\mathfrak{F}}}\abs{S_a(\omega)}^2\diff\omega \nonumber \\
& \xlongequal{\text{Eq.~(\ref{eq:marginal_omega})}}\int_{-\infty}^{\infty}\frac{1}{C_{\mathfrak{IF}}}\abs{s_a(t)}^2\diff t.
\end{align}

If $K(v,\tau)=\int_{-\infty}^{\infty}w(u'-\frac{\tau}{2})w(u'+\frac{\tau}{2})\Exp^{ic_svu'}\diff u'$, where $w(t)$ is the window function, it leads to the spectrogram, which results from the absolute squared of a short-time Fourier transform of a field. With the use of coordinate transformation and the Jacobian matrix
\begin{align}
& \because
\begin{cases}
x=u+\frac{\tau}{2} \\
y=u-\frac{\tau}{2}
\end{cases}\quad\Rightarrow\quad
\begin{cases}
u=\frac{1}{2}\left(x+y\right) \\
\tau=x-y
\end{cases} \nonumber \\
& \therefore
\int f(u,\tau)\diff u\diff\tau=\int f(u(x,y),\tau(x,y))
\abs{\begin{vmatrix}
\frac{1}{2} & \frac{1}{2} \\
1 & -1
\end{vmatrix}}\diff x\diff y=\int f(u(x,y),\tau(x,y))\diff x\diff y,
\end{align}
we obtain the spectrogram, the absolute squared of the short-time Fourier transform $S_{a,\text{STFT}}$ with a $1/C_{\mathfrak{F}}$ prefactor:
\begin{align}
& P_{\text{spectrogram}}(t,\omega) \nonumber \\
& =\frac{C_{\mathfrak{F}}}{2\pi}\int_{-\infty}^{\infty}\int_{-\infty}^{\infty}\int_{-\infty}^{\infty}s_a(u+\frac{\tau}{2})\left[s_a(u-\frac{\tau}{2})\right]^{\ast}w(u'-\frac{\tau}{2})w(u'+\frac{\tau}{2})\Exp^{ic_s\omega\tau}\left[\int_{-\infty}^{\infty}\Exp^{ic_s\left[v(u'+u-t)\right]}\diff v\right]\diff u\diff u'\diff\tau \nonumber \\
& =C_{\mathfrak{F}}\int_{-\infty}^{\infty}\int_{-\infty}^{\infty}\int_{-\infty}^{\infty}s_a(u+\frac{\tau}{2})\left[s_a(u-\frac{\tau}{2})\right]^{\ast}w(u'-\frac{\tau}{2})w(u'+\frac{\tau}{2})\Exp^{ic_s\omega\tau}\delta(u'+u-t)\diff u\diff u'\diff\tau \nonumber \\
& =C_{\mathfrak{F}}\int_{-\infty}^{\infty}\int_{-\infty}^{\infty}s_a(u+\frac{\tau}{2})\left[s_a(u-\frac{\tau}{2})\right]^{\ast}w(t-u-\frac{\tau}{2})w(t-u+\frac{\tau}{2})\Exp^{ic_s\omega\tau}\diff u\diff\tau \nonumber \\
& =C_{\mathfrak{F}}\int_{-\infty}^{\infty}\int_{-\infty}^{\infty}s_a(u+\frac{\tau}{2})w(t-\left(u+\frac{\tau}{2}\right))\left[s_a(u-\frac{\tau}{2})w(t-\left(u-\frac{\tau}{2}\right))\right]^{\ast}\Exp^{ic_s\omega\tau}\diff u\diff\tau \nonumber \\
& =C_{\mathfrak{F}}\int_{-\infty}^{\infty}\int_{-\infty}^{\infty}s_a(x)w(t-x)\left[s_a(y)w(t-y)\right]^{\ast}\Exp^{ic_s\omega(x-y)}\diff x\diff y \nonumber \\
& =\frac{1}{C_{\mathfrak{F}}}\abs{C_{\mathfrak{F}}\int_{-\infty}^{\infty}s_a(x)w(t-x)\Exp^{ic_s\omega x}\diff x}^2=\frac{1}{C_{\mathfrak{F}}}\abs{S_{a,\text{STFT}}}^2
\end{align}
The short-time Fourier transform satisfies a different marginal condition:
\begin{align}
\int_{-\infty}^{\infty}S_{a,\text{STFT}}\diff t=S_a(\omega),
\end{align}
where the window function satisfies
\begin{equation}
\int_{-\infty}^{\infty}w(t)\diff t=1.
\end{equation}
For a sharp window [Fig.~\ref{fig:stft_w}(a)], we assume that
\begin{equation}
w(t)=\delta(t)\approx\frac{1}{a}\left(H(t+\frac{a}{2})-H(t-\frac{a}{2})\right),\quad\text{with }a\rightarrow0,
\end{equation}
where $H(t)$ is the Heaviside step function that is one when $t>0$ or zero otherwise. We can find that its autocorrelation has a unity area [Fig.~\ref{fig:stft_w}(b)], so it is also a Dirac delta function. Therefore, we can derive that
\begin{equation}
\mathcal{A}_w(x_1,x_2)=\int_{-\infty}^{\infty}w(t-x_1)w(t-x_2)\diff t=\int_{-\infty}^{\infty}w(t')w(t'+x_1-x_2)\diff t'=\delta(x_1-x_2)\quad\text{for a sharp window}.
\label{eq:stft_w_auto}
\end{equation}
\begin{figure}[ht!]
\centering
\includegraphics[width=0.5\linewidth]{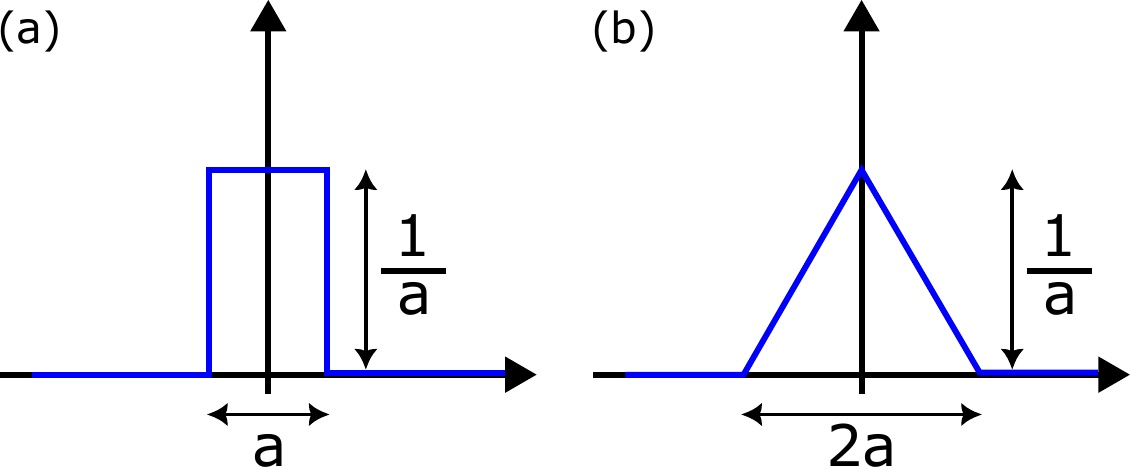}
\caption{(a) Square-wave window function and (b) its autocorrelation $\mathcal{A}_w(t)=\int_{-\infty}^{\infty}w(\tau)w(\tau+t)\diff\tau$.}
\label{fig:stft_w}
\end{figure}\\
With Eq.~(\ref{eq:stft_w_auto}), we can derive that, under the condition of a sharp sliding window, the spectrogram also satisfies similar marginal conditions to the Wigner distribution [Eq.~(\ref{eq:marginal_Wigner})]:
\begin{subequations}
\begin{align}
\int_{-\infty}^{\infty}P_{\text{spectrogram}}(t,\omega)\diff t & =C_{\mathfrak{F}}\int_{-\infty}^{\infty}\int_{-\infty}^{\infty}s_a(x_1)\left[s_a(x_2)\right]^{\ast}\mathcal{A}_w(x_1-x_2)\Exp^{ic_s\omega\left( x_1-x_2\right)}\diff x_1\diff x_2\approx\frac{1}{C_{\mathfrak{F}}}\abs{S_a(\omega)}^2 \label{eq:marginal_t2} \\
\int_{-\infty}^{\infty}P_{\text{spectrogram}}(t,\omega)\diff\omega & =\frac{1}{C_{\mathfrak{IF}}}\abs{s_a(t)}^2\ast\left[w(t)\right]^2\approx\frac{1}{a}\frac{1}{C_{\mathfrak{IF}}}\abs{s_a(t)}^2,\quad\text{with }a\rightarrow0. \label{eq:marginal_omega2}
\end{align} \label{eq:marginal_spectrogram}
\end{subequations}
Compared to those of the Wigner distribution, the window function applies smoothing to the temporal and spectral profiles.

There are two widely-used tools for time-frequency analysis, spectrogram and Wigner distribution. The idea of spectrogram, which is based on the short-time Fourier transform, is to take the Fourier transform but only to each temporal slice and obtain the corresponding spectrum (Fig.~\ref{fig:MATLAB_stft}). This requires the use of a sliding time window to select a small temporal slice of a field. However, this method fails to provide simultaneously-high temporal and spectral resolutions (Fig.~\ref{fig:stft_comparison}). A small time window to localize spectral components in time leads to a poor spectral resolution; a huge time window for high spectral resolution inevitably lacks good temporal resolution. Thus, it is required to select an optimal window size depending on the signal investigated. Wigner distribution, on the other hand, is free from the resolution issue, as it does not rely on a sliding time window. Nevertheless, it suffers from the presence of cross terms when applied to multi-component signals due to its quadratic nature \cite{Cohen1989}.

\begin{figure}[ht!]
\centering
\includegraphics[width=0.9\linewidth]{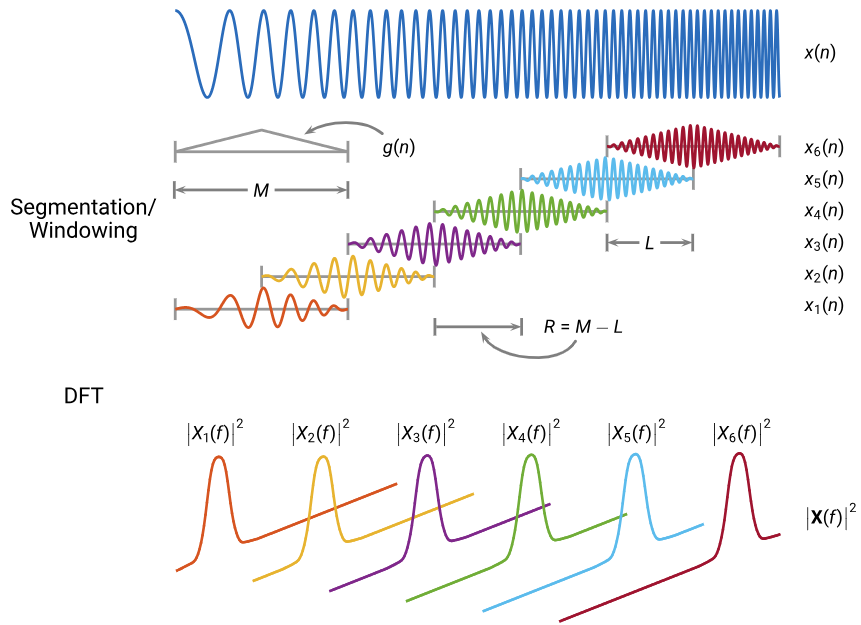}
\caption{Illustration of the short-time Fourier transform (figure extracted from \cite{MATLAB_stft}). The signal $x$ is scanned by a sliding temporal window $g$ to generate the frequency distribution $X$ of each temporal slice.}
\label{fig:MATLAB_stft}
\end{figure}

\begin{figure}[ht!]
\centering
\includegraphics[width=\linewidth]{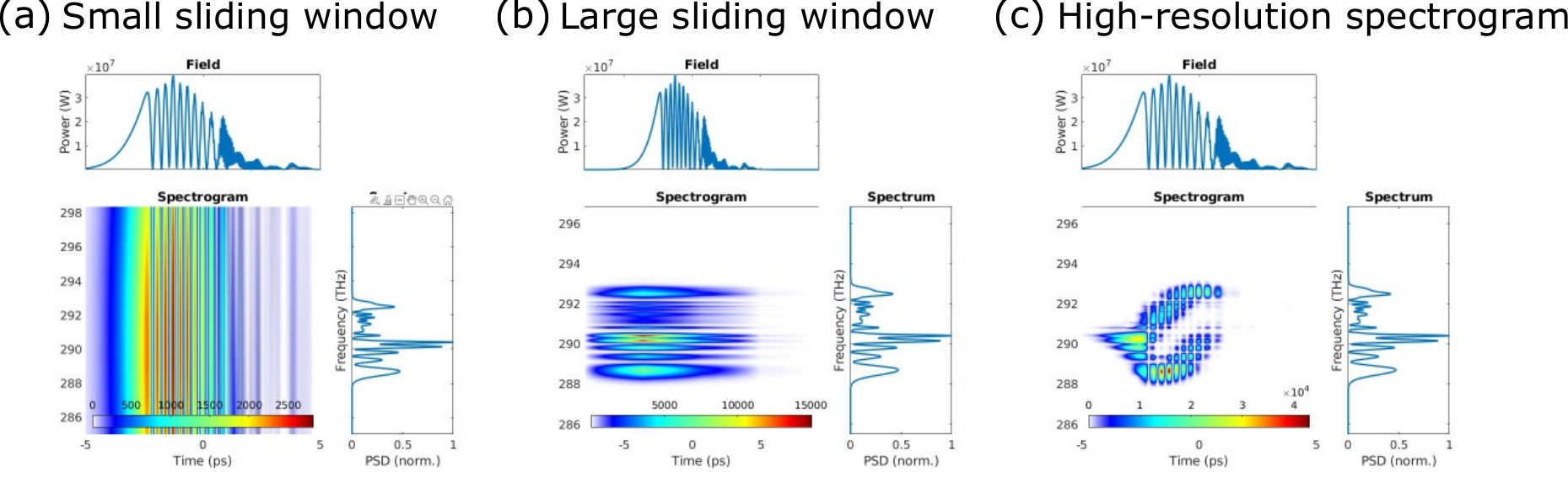}
\caption{Spectrograms with (a) a small sliding time window and (b) a large window. (c) Spectrogram with high time-frequency resolution generated by combining multiple spectrograms \cite{Khan2011}. Field used here is generated from the residual pump after nonlinear transient-Raman generation where both self-phase modulation and Raman transfer occur \cite{Chen2024}. The central hole results from energy transfer to the Raman-Stokes light not shown here.}
\label{fig:stft_comparison}
\end{figure}

\subsubsection{High-resolution spectrogram}\label{subsubsec:high_resolution_spectrogram}
In this section, we will focus only on spectrogram and its underlying short-time Fourier transform. Due to the linear nature of its mathematical formulation, it does not suffer from cross terms as in Wigner distribution, but from the time-frequency resolution. As the uncertainty principle in quantum mechanics, a single mathematical tool cannot obtain simultaneously-high temporal and spectral resolutions. This issue can be resolved by combining multiple spectrograms under different sizes of sliding window and overcome the restriction set by the uncertainty principle \cite{Khan2011}. If the field exists at a certain time-frequency point, spectrogram should show the signal under any time-window size. If the field does not exist there, spectrogram showing signal at that point results from the spread of energy from signals nearby due to an insufficient time-frequency resolution. Hence, a high-resolution spectrogram should be achievable by identifying the maximum and minimum energies at each time-frequency point under various window sizes. As it is difficult to know whether there should be a signal and thus whether a minimum or maximum should be picked beforehand, multiplication of obtained maximum and minimum values is employed. Numerically, as the spectrograms under different window sizes have different frequency spacings ($\triangle\nu=1/T_{\text{sliding window}}$), the energies within different frequency spacings are different, leading to different absolute values of the generated spectrograms. Furthermore, as previously mentioned, the sliding window applies a smoothing operation to the signal. As a result, a simple comparison of each grid value in spectrograms of different window sizes to find the minimum and maximum is not meaningful. Meaningful comparison is achieved with the normalized values by their temporal or spectral profiles to remove these ambiguities from scaling and smoothing.

\noindent\rule{\textwidth}{1pt}

Below I lay out the process of how to obtain a high-resolution spectrogram through combining multiple spectrograms under different sliding window sizes:
\begin{enumerate}
\item Compute spectrograms ($P_i$) under different sliding window sizes.

\item Compute each of their temporal and spectral profiles.
\begin{subequations}
\begin{align}
\abs{s_{a,i}(t_n)}^2 & =C_{\mathfrak{IF}}\sum_{\nu_n}P_i(t_n,\omega_n)\triangle\omega_{\text{sliding window},i}=C_{\mathfrak{IF}}\sum_{\nu_n}P_i(t_n,\omega_n)\frac{2\pi}{T_{\text{sliding window},i}} \\
\abs{S_{a,i}(\omega_n)}^2 & =C_{\mathfrak{F}}\sum_{t_n}P_i(t_n,\nu_n)\triangle t.
\end{align}
\end{subequations}

\item Normalize spectrograms ($\bar{P}_i$) with respect to their temporal and spectral profiles. Normalization to weak signals needs to be avoided due to the numerical precision error.
\begin{subequations}
\begin{align}
\bar{P}_{t,i}(t_n,\omega_n) & =\frac{P_i(t_n,\omega_n)}{\abs{s_{a,i}(t_n)}^2} \\
\bar{P}_{\nu,i}(t_n,\omega_n) & =\frac{P_i(t_n,\omega_n)}{\abs{S_{a,i}(\omega_n)}^2}.
\end{align}
\end{subequations}

\item Apply interpolation so that all normalized spectrograms have the same time-frequency coordinates.

\item Find minimum and maximum of two sets of normalized spectrograms over different window sizes.
\begin{subequations}
\begin{align}
& \bar{P}_{t,\max}(t_n,\omega_n)=\max_i\bar{P}_{t,i}(t_n,\omega_n),\quad && \bar{P}_{t,\min}(t_n,\omega_n)=\min_i\bar{P}_{t,i}(t_n,\omega_n) \\
& \bar{P}_{\nu,\max}(t_n,\omega_n)=\max_i\bar{P}_{\nu,i}(t_n,\omega_n),\quad && \bar{P}_{\nu,\min}(t_n,\omega_n)=\min_i\bar{P}_{\nu,i}(t_n,\omega_n).
\end{align}
\end{subequations}

\item Multiply maximum and minimum values of each set of spectrograms.
\begin{subequations}
\begin{align}
\bar{P}_t(t_n,\omega_n) & =\bar{P}_{t,\max}(t_n,\omega_n)\bar{P}_{t,\min}(t_n,\omega_n) \\
\bar{P}_{\nu}(t_n,\omega_n) & =\bar{P}_{\nu,\max}(t_n,\omega_n)\bar{P}_{\nu,\min}(t_n,\omega_n).
\end{align}
\end{subequations}

\item Recover from normalization by multiplying by the original temporal or spectral profiles. Interpolation is employed if necessary so that the signal's original time and frequency coordinates are consistent with those of spectrograms.
\begin{subequations}
\begin{align}
P_t(t_n,\omega_n) & =\bar{P}_t(t_n,\omega_n)\abs{s_a(t_n)}^2 \\
P_{\nu}(t_n,\omega_n) & =\bar{P}_{\nu}(t_n,\omega_n)\abs{S_a(\omega_n)}^2.
\end{align}
\end{subequations}

\item Find minimum between the two obtained spectrograms at each time-frequency point.
\begin{equation}
P(t_n,\omega_n)=\min_{j=t,\nu}P_j(t_n,\omega_n).
\end{equation}

\item Since spectrogram is affected by the smoothing operation of a sliding window [Eqs.~(\ref{eq:marginal_spectrogram})], it is important to isolate the high-resolution spectrogram from this effect. A natural option is to normalize it such that it satisfies the same marginal conditions as the Wigner distribution that is free from sliding windows [Eqs.~(\ref{eq:marginal_Wigner})].
\begin{equation}
P_{\text{final}}(t_n,\omega_n)=P(t_n,\omega_n)\frac{E/C_{\mathfrak{IF}}}{\int_{-\infty}^{\infty}\int_{-\infty}^{\infty}P(t_n,\omega_n)\diff t\diff\omega},
\end{equation}
where $E=\int_{-\infty}^{\infty}\abs{s_a(t)}^2\diff t$ is the pulse energy.
\end{enumerate}
\noindent\rule{\textwidth}{1pt}

Moreover, below I show the calculation for obtaining a spectrogram with a physical-useful unit:
\begin{align}
& \begin{aligned}
\because P(t,\omega)\xlongrightarrow{\text{unit analysis}}\frac{1}{\diff\omega}\frac{1}{C_{\mathfrak{IF}}}\abs{s_a(t)}^2\text{ from Eq.~(\ref{eq:marginal_omega}),} & \text{ so }C_{\mathfrak{IF}}P(t,\omega)\text{ is in ``\si{\joule/\s/\radian}''} \\
\Rightarrow & \frac{1}{C_{\mathfrak{F}}}P(t,\nu)\text{ is in ``\si{\joule/\s/\Hz}'' by using }
\begin{cases}
C_{\mathfrak{F}}=\frac{1}{2\pi C_{\mathfrak{IF}}} \\
P(\omega)=2\pi P(\nu)\text{ [Eq.~(\ref{eq:Pnu})]}
\end{cases}
\end{aligned} \nonumber \\
& \therefore \frac{1}{C_{\mathfrak{F}}}P_{\text{spectrogram}}(t,\nu)=\frac{1}{C_{\mathfrak{F}}^2}\abs{S_{a,\text{STFT}}(t,\nu)}^2\text{ is in ``\si{\joule/\s/\Hz}''}.
\end{align}

\subsubsection{Example cases}\label{subsubsec:example_spectrogram}
As an example, Fig.~\ref{fig:spectrogram_dispersion} shows the spectrogram of a positively-chirped pulse where lower-frequency components are in the temporal leading edge. Another example in Fig.~\ref{fig:stft_comparison}(c) demonstrates the necessity of applying the high-resolution spectrogram to analyze the time-frequency relation of a complicated field. Fig.~\ref{fig:optical_wave_breaking} displays the nonlinear process during optical wave breaking \cite{Anderson1992,Heidt2016}. Self-phase modulation in a $\chi^{(3)}$ nonlinear process can broaden a pulse with spectral modulations \cite{Finot2018}. In environments with normal dispersion, low-frequency light travels faster than high-frequency light, and multiple frequencies during the broadening process can temporally overlap, leading to sharp temporal edges with tiny structures due to the interference [Fig.~\ref{fig:optical_wave_breaking}(b)]. In the meantime, four-wave mixing occurs, generating new frequencies at the far spectral edges. Hence, typical features of optical wave breaking include tiny and sharp temporal structures, and low spectral edges. From these examples, we can see that spectrogram is a nice tool to straightforwardly visualize the time-frequency relation, compared to mathematical phase relation previously shown in Figs.~\ref{fig:DFT_chirp} and \ref{fig:DFT_chirp3}.
\begin{figure}[!ht]
\centering
\includegraphics[width=.55\linewidth]{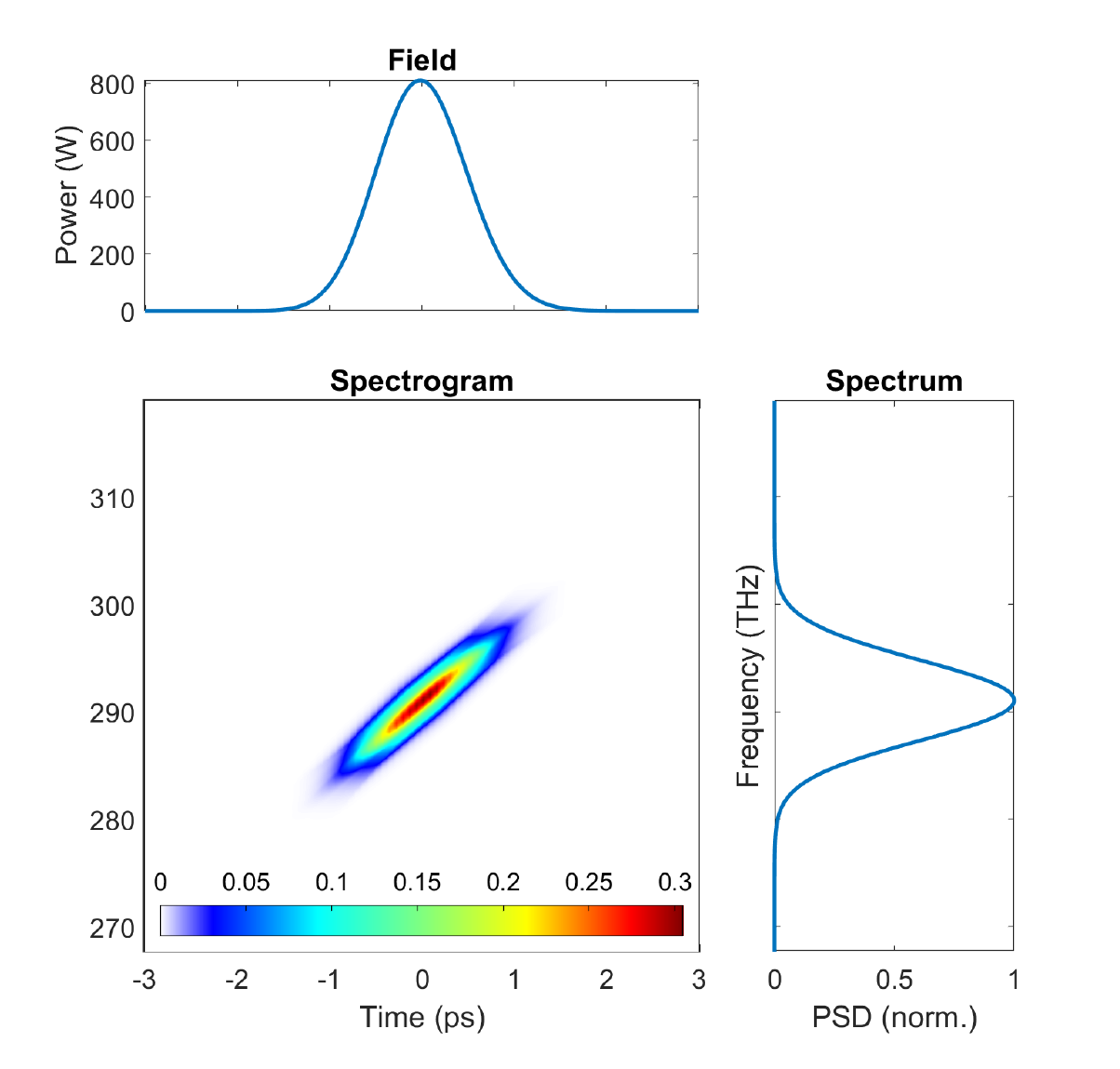}
\caption{Spectrogram of a positively-(linearly-)chirped signal [Fig.~\ref{fig:DFT_chirp}(c)]}
\label{fig:spectrogram_dispersion}
\end{figure}

\begin{figure}
\centering
\includegraphics[width=\linewidth]{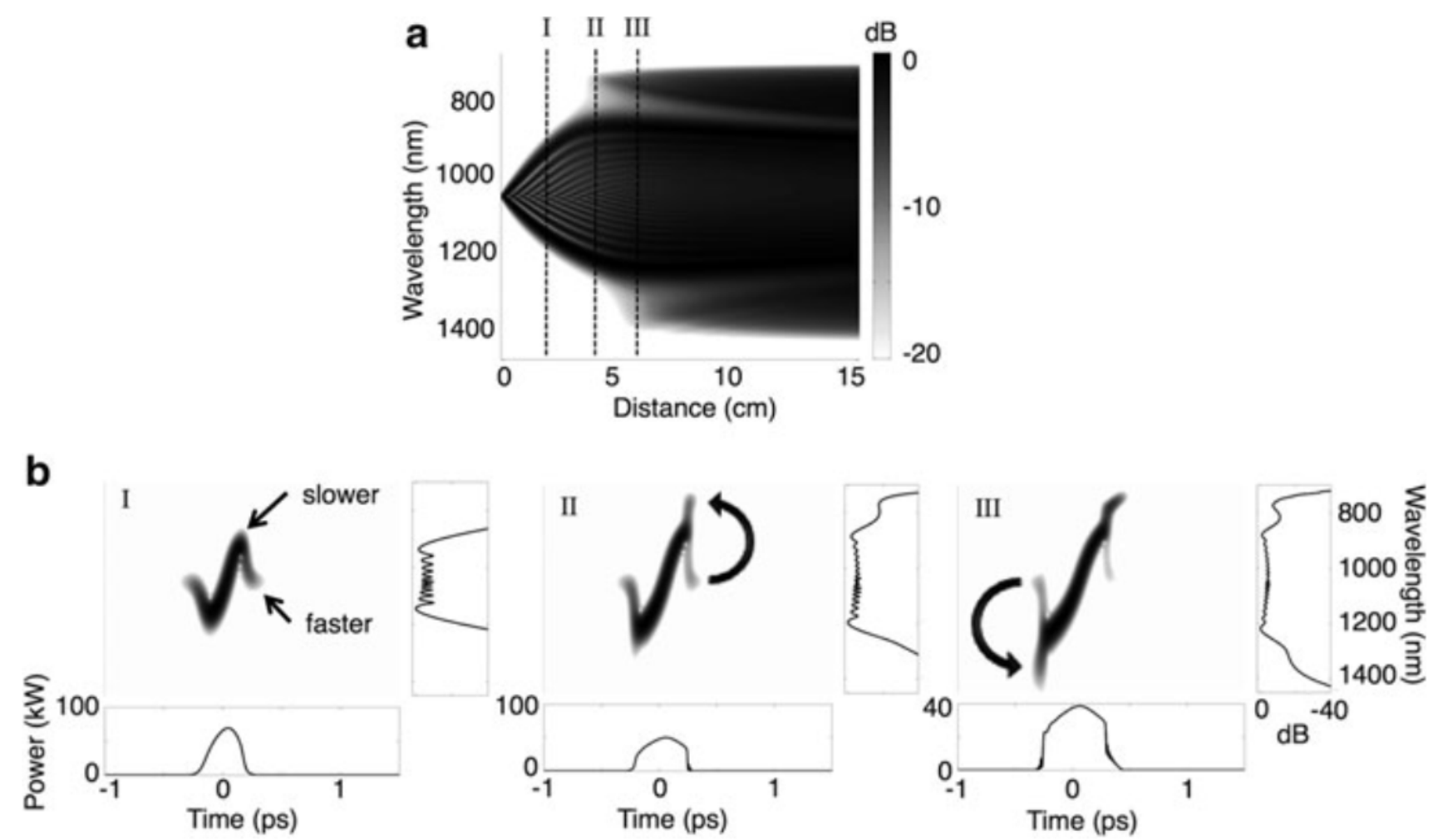}
\caption{Optical wave breaking. Figure is extracted from Fig.~6.5 of \cite{Heidt2016}. (a) Simulated spectral evolution of the supercontinuum generation process in the all-normal-dispersion photonic crystal fiber, pumped with a \SIadj{200}{\fs}, \SIadj{90}{\kW} pulse at \SI{1050}{\nm}. (b) Projected axes spectrograms of the supercontinuum pulse at \SI{2.0}{\cm}, \SI{3.7}{\cm}, and \SI{5.2}{\cm}. Arrows indicate energy transfer process due to optical wave breaking.}
\label{fig:optical_wave_breaking}
\end{figure}
\clearpage

\section{Hankel transform: 2D spatial Fourier transform of \texorpdfstring{$A(r)$}{A(r)}}\label{sec:Hankel_transform}
In scenarios that require a full-field, instead of mode-resolved, simulation, pulse propagation most likely exhibits radial symmetry, such as in a multipass cell \cite{Schulte2016,Hanna2017} and a multiplate compressor \cite{Centurion2006,Zhang2021,Wang2024}. Hankel transform is the core element of computations of such a radially-symmetric system, which we will introduce in this section.

\subsection{Introduction}\label{subsec:Hankel_intro}
During solving the full-field pulse propagation equation, the Hankel transform $\mathfrak{H}_0$ emerges from the two-dimensional ``spatial'' Fourier transform $\mathfrak{F}_{k_{\perp}}=\mathfrak{F}_x\mathfrak{F}_y$ of a radially-symmetric function:
\begin{align}
A(k_{\perp})=\mathfrak{F}_{k_{\perp}}[A(\vec{r}_{\perp}=r_{\perp})] & \equiv C_{\mathfrak{F}}^2\int^{\infty}_{-\infty}\int^{\infty}_{-\infty}A(r_{\perp})\Exp^{ic_s\left(k_xx+k_yy\right)}\diff x\diff y \nonumber \\
& =C_{\mathfrak{F}}^2\int^{2\pi}_{0}\int^{\infty}_{0}A(r_{\perp})\Exp^{ic_s\abs{\vec{k}_{\perp}}\abs{\vec{r}_{\perp}}\cos\theta}\abs{\vec{r}_{\perp}}\diff \abs{\vec{r}_{\perp}}\diff\theta\quad, \vec{k}_{\perp}\cdot\vec{r}_{\perp}=\abs{\vec{k}_{\perp}}\abs{\vec{r}_{\perp}}\cos\theta \nonumber \\
& =C_{\mathfrak{F}}^2\int^{\infty}_{0}A(r_{\perp})\abs{\vec{r}_{\perp}}\left(\int^{2\pi}_{0}\Exp^{ic_s\abs{\vec{k}_{\perp}}\abs{\vec{r}_{\perp}}\cos\theta}\diff\theta\right)\diff \abs{\vec{r}_{\perp}} \nonumber \\
& =2\pi C_{\mathfrak{F}}^2\int^{\infty}_{0}A(r_{\perp})J_0(\abs{c_s}k_{\perp}r_{\perp})r_{\perp}\diff r_{\perp} \nonumber \\
& \equiv2\pi C_{\mathfrak{F}}^2\frac{1}{C_{\mathfrak{H}}}\mathfrak{H}_0[A](k_{\perp})\quad\because\abs{c_s}=1,
\label{eq:tmp_FH}
\end{align}
where the relation of the Bessel function of order $0$, $J_0(z)=\frac{1}{2\pi}\int_0^{2\pi}\Exp^{iz\cos\theta}\diff\theta$, is employed. Notations are simplified with $k_{\perp}=\abs{\vec{k}_{\perp}}=\abs{(k_x,k_y)}$ and $r_{\perp}=\abs{\vec{r}_{\perp}}=\abs{(x,y)}$. Similarly, the inverse Fourier transform becomes
\begin{align}
A(r_{\perp})=\mathfrak{F}_{k_{\perp}}^{-1}[A(k_{\perp})] & \equiv C_{\mathfrak{IF}}^2\int^{\infty}_{-\infty}\int^{\infty}_{-\infty}A(k_{\perp})\Exp^{-ic_s\left(k_xx+k_yy\right)}\diff k_x\diff k_y \nonumber \\
& =2\pi C_{\mathfrak{IF}}^2\int^{\infty}_{0}A(k_{\perp})J_0(\abs{c_s}k_{\perp}r_{\perp})k_{\perp}\diff k_{\perp} \nonumber \\
& \equiv2\pi C_{\mathfrak{IF}}^2\frac{1}{C_{\mathfrak{IH}}}\mathfrak{H}_0^{-1}[A](k_{\perp}).
\end{align}
The minus sign in the exponent disappears due to the integral over a full $2\pi$ angle; as a result, the Hankel transform displays the same formulation, regardless of the Fourier-transform convention. In general, Hankel transform is defined for $\nu\geq-1/2$; however, because we are more interested in using it to solve radially-symmetric systems, discussion is limited only to Hankel transform of order $0$ (\ie $\nu=0$). The Hankel transform of order $0$ is defined as
\begin{equation}
\begin{aligned}
A_{H_0}(k_{\perp})=\mathfrak{H}_0[A(r_{\perp})] & \equiv C_{\mathfrak{H}}\int^{\infty}_0A(r_{\perp})J_0(k_{\perp}r_{\perp})r_{\perp}\diff r_{\perp} \\
A(r_{\perp})=\mathfrak{H}_0^{-1}[A_{H_0}(k_{\perp})] & \equiv C_{\mathfrak{IH}}\int^{\infty}_0A_{H_0}(k_{\perp})J_0(k_{\perp}r_{\perp})k_{\perp}\diff k_{\perp},
\end{aligned} \label{eq:Hankel_transform}
\end{equation}
where $C_{\mathfrak{H}}C_{\mathfrak{IH}}=1$. Generally, there are two Hankel-transform conventions: one is $C_{\mathfrak{H}}=C_{\mathfrak{IH}}=1$, which gives the same mathematical formulation for both the Hankel and inverse Hankel transforms; the other is $C_{\mathfrak{H}}=2\pi$, which absorbs the $(2\pi)$ factor in Eq.~(\ref{eq:tmp_FH}) such that $\mathfrak{F}_{k_{\perp}}=\mathfrak{H}_0$ if $C_{\mathfrak{F}}=1$.

With the orthogonality relation
\begin{equation}
\int_0^{\infty}J_{\nu}(kr)J_{\nu}(k'r)r\diff r=\frac{\delta(k-k')}{k},\quad k,k'>0,
\end{equation}
Eq.~(\ref{eq:Hankel_transform}) forms a Hankel-transform pair with $\mathfrak{H}_0\mathfrak{H}_0^{-1}=\mathfrak{H}_0^{-1}\mathfrak{H}_0=1$. This relation can also be verified with one-dimensional $\mathfrak{F}\mathfrak{F}^{-1}=\mathfrak{F}^{-1}\mathfrak{F}=1$ by noting that $C_{\mathfrak{F}}C_{\mathfrak{IF}}=\frac{1}{2\pi}$ with the following Fourier-transform representation:
\begin{subequations}
\begin{align}
\mathfrak{F}_{k_{\perp}} & =\frac{2\pi C_{\mathfrak{F}}^2}{C_{\mathfrak{H}}}\mathfrak{H}_0 \label{eq:FH1} \\
\mathfrak{F}_{k_{\perp}}^{-1} & =\frac{2\pi C_{\mathfrak{IF}}^2}{C_{\mathfrak{IH}}}\mathfrak{H}_0^{-1} \label{eq:FH2}.
\end{align} \label{eq:Fourier_Hankel}
\end{subequations}

Hankel transform satisfies a similar convolution theorem to the Fourier transform, which follows
\begin{subequations}
\begin{align}
\mathfrak{H}_0[A**B] & =\frac{2\pi}{C_{\mathfrak{H}}}\mathfrak{H}_0[A]\mathfrak{H}_0[B] \label{eq:cc1H} \\
\mathfrak{H}_0^{-1}[A**B] & =\frac{2\pi}{C_{\mathfrak{IH}}}\mathfrak{H}_0^{-1}[A]\mathfrak{H}_0^{-1}[B]. \label{eq:cc2H}
\end{align} \label{eq:Hankel_convolution}
\end{subequations}
Here, I use $**$ to represent the two-dimensional convolution:
\begin{equation}
A**B=\int_{-\infty}^{\infty}\int_{-\infty}^{\infty}A(x,y)B(X-x,Y-y)\diff x\diff y.
\end{equation}
Eq.~(\ref{eq:Hankel_convolution}) can be derived by realizing the fact that if $A$ and $B$ are radially symmetric, their two-dimensional convolution also exhibits radial symmetry as it can be rewritten as follows:
\begin{equation}
A**B==\int_0^{2\pi}\int_0^{\infty}A(r)B(\abs{\vec{R}-\vec{r}})r\diff r\diff\theta,
\end{equation}
which can be further simplified to depend only on $R$ (Fig.~\ref{fig:Hankel_convolution}). Due to its radial symmetry, the two-dimensional Fourier transform of $(A**B)$ [2D of Eq.~(\ref{eq:continuous_convolution_thm})]
\begin{equation}
\mathfrak{F}_{k_{\perp}}[A**B]=\frac{1}{C_{\mathfrak{F}}^2}\mathfrak{F}_{k_{\perp}}[A]\mathfrak{F}_{k_{\perp}}[B]
\end{equation}
can be reformulated as a Hankel transform. With Eq.~(\ref{eq:FH1}), it can be rewritten into the convolution-like theorem for the Hankel transform [Eq.~(\ref{eq:cc1H})]. A similar process can be applied to derive its inverse-Hankel-transform counterpart [Eq.~(\ref{eq:cc2H})].

\begin{figure}[!ht]
\centering
\includegraphics[width=0.5\linewidth]{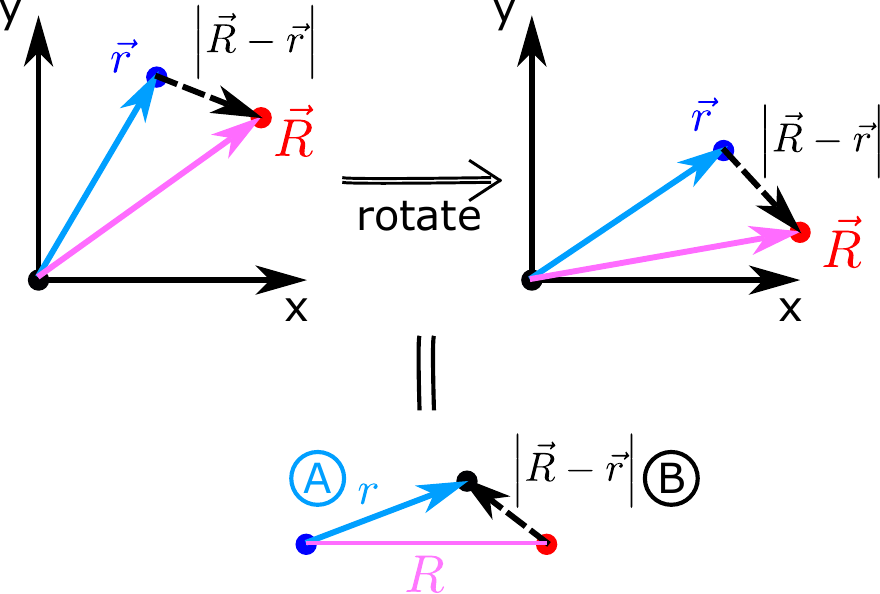}
\caption{Two-dimensional convolution $A**B$ of two radially-symmetric functions $A(r)$ and $B(r)$. Because $A$ depends only on $\vec{r}$ and $B$ only on $\abs{\vec{R}-\vec{r}}$, the convolution $\left(A**B\right)(R)=\int_0^{2\pi}\int_0^{\infty}A(r)B(\abs{\vec{R}-\vec{r}})r\diff r\diff\theta$ is also radially symmetric. This convolution can be treated as multiplying values depending on the black dot moving over the entire space (bottom figure), whose only independent variable is the distance $R$ between the blue and red dots.}
\label{fig:Hankel_convolution}
\end{figure}

Hankel transform satisfies the Parseval's theorem as well. The two-dimensional Parseval's theorem of the Fourier transform is
\begin{equation}
\frac{1}{C_{\mathfrak{IF}}^2}\int_{-\infty}^{\infty}\int_{-\infty}^{\infty}\abs{A(\vec{r}_{\perp})}^2\diff x\diff y=\frac{1}{C_{\mathfrak{F}}^2}\int_{-\infty}^{\infty}\int_{-\infty}^{\infty}\abs{A(\vec{k}_{\perp})}^2\diff k_x\diff k_y.
\label{eq:Parseval_2D}
\end{equation}
If $A$ is radially symmetric (and thus its 2D Fourier transform $A(\vec{k}_{\perp})$ is also radially symmetric [Eq.~(\ref{eq:tmp_FH})]). With Eq.~(\ref{eq:FH1}), it becomes
\begin{equation}
\frac{1}{C_{\mathfrak{IF}}^2}2\pi\int_0^{\infty}\abs{A(r_{\perp})}^2r_{\perp}\diff r_{\perp}=\frac{1}{C_{\mathfrak{F}}^2}2\pi\int_0^{\infty}\abs{\frac{2\pi C_{\mathfrak{F}}^2}{C_{\mathfrak{H}}}A_{H_0}(k_{\perp})}^2k_{\perp}\diff k_{\perp},
\end{equation}
which leads to the Parseval's theorem for the Hankel transform:
\begin{equation}
\frac{1}{C_{\mathfrak{IH}}}\int_0^{\infty}\abs{A(r_{\perp})}^2r_{\perp}\diff r_{\perp}=\frac{1}{C_{\mathfrak{H}}}\int_0^{\infty}\abs{A_{H_0}(k_{\perp})}^2k_{\perp}\diff k_{\perp}.
\label{eq:Hankel_Parseval}
\end{equation}

\subsection{Numerical computation: Fast Hankel transform of high accuracy (FHATHA)}\label{subsec:FHATHA}
In this section, we will discuss how to numerically compute the Hankel transform with the ``fast Hankel transform of high accuracy'' (FHATHA), proposed by \etal{Magni} \cite{Magni1992}. It is based on the quasi-fast Hankel transform (QFHT), independently developed by Siegman \cite{Siegman1977} and Talman \cite{Talman1978}, and was improved by Agrawal and Lax \cite{Agrawal1981}. QFHT relies on the well-established FFT that offers computational efficiency. FHATHA outperforms them by exhibiting two orders of magnitude lower error. However, it exhibits a problem with energy conservation and potential inaccuracy due to omission of the spatial center. The importance of spatial center was first pointed out by Agrawal and Lax to correct the original QFHT \cite{Agrawal1981} but not the FHATHA proposed later. Further improvements need to be included for optical modeling, which will be discussed. Although this document begins as a tutorial, rather than a research paper of a new topic, FHATHA's improvement steps are, for the first time, proposed here.

\subsubsection{Magni's approach}\label{subsubsec:Magni}
In FHATHA, both the fields in real $\vec{r}_{\perp}$-space and $\vec{k}_{\perp}$-space follow an exponential sampling strategy:
\begin{subequations}
\begin{align}
r_{\perp,n} & =r_{\perp}^{\max}\zeta_n \\
k_{\perp,n} & =k_{\perp}^{\max}\zeta_n,
\end{align}
\end{subequations}
where $\zeta_n$ represents the normalized coordinate that follows
\begin{equation}
\zeta_n=\zeta_0\Exp^{\alpha n},\quad n=0,1,\cdots,N_{\perp}-1
\label{eq:zeta}
\end{equation}
with $\zeta_0=\frac{1+\Exp^{\alpha}}{2}\Exp^{-\alpha N_{\perp}}$ such that $\zeta_n,\forall n\geq1$ are in the center of each interval $[\xi_n,\xi_{n+1}]$. $\xi_n$ is the normalized coordinate following
\begin{equation}
\xi_n=\begin{cases}
0, & n=0 \\
\xi'_0\Exp^{\alpha n}, & n=1,2,\cdots,N_{\perp}
\end{cases},
\label{eq:xi}
\end{equation}
where $\xi'_0=\Exp^{-\alpha N_{\perp}}$ (Fig.~\ref{fig:Hankel_sampling}). $\alpha$ is given such that the widths of the first and the last intervals are the same:
\begin{align}
& \xi_1-\xi_0=\xi_{N_{\perp}}-\xi_{N_{\perp}-1} \nonumber \\
& \Rightarrow \Exp^{-\alpha\left(N_{\perp}-1\right)}=1-\Exp^{-\alpha},
\end{align}
which can be easily solved numerically. During computing the Hankel transform, the function needs to be evaluated at the center of each interval. However, $\zeta_0$ is not at the center of $[\xi_0,\xi_1]$, as we redefine $\xi_0$ from $\xi'_0$ to $0$ [Eq.~(\ref{eq:xi})]. Therefore, $A(r_{\perp,0}=r_{\perp}^{\max}\zeta_0)$ needs to be replaced with $A(r'_{\perp,0}=r_{\perp}^{\max}\zeta'_0)$ (with $\zeta'_0=\xi_1/2$ being at the center of $[\xi_0=0,\xi_1]$), which is approximated by extrapolation of $A(r_{\perp,0})$ and $A(r_{\perp,1})$ with a parabola with zero derivative at the origin, a reasonable approximation curve for a radially-symmetric function. In FHATHA, an additional evaluation is required at $\zeta_{N_{\perp}}$, beyond the range of Eq.~(\ref{eq:zeta}), which is simply treated with $A(r_{\perp,N_{\perp}}=r_{\perp}^{\max}\zeta_{N_{\perp}})=0$. If we denote $A_n$ as $A(r_{\perp,n})$, eventually the $A$ that is used in FHATHA is
\begin{equation}
\tilde{A}_n=\begin{cases}
A(r'_{\perp,0})=\ell_0\left(A_0-A_1\right)+A_1, & n=0\quad\text{(from the parabola approximation)} \\
A_n, & n=1,2,\cdots,N_{\perp}-1 \\
0, & n=N_{\perp}
\end{cases} \label{eq:Hankel_An}
\end{equation}
where
\begin{equation}
\ell_0=\frac{\Exp^{\alpha}\left(2+\Exp^{\alpha}\right)}{\left(1+\Exp^{\alpha}\right)^2\left(1-\Exp^{-2\alpha}\right)}.
\label{eq:FHATHA_l0}
\end{equation}

\begin{figure}[!ht]
\centering
\includegraphics[width=0.7\linewidth]{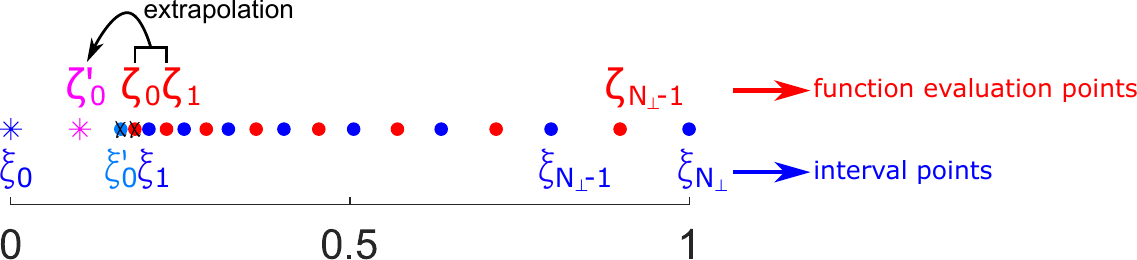}
\caption{Numerical sampling in FHATHA.}
\label{fig:Hankel_sampling}
\end{figure}

FHATHA solves the Hankel transform by summing over the Hankel-transform integral of each interval, whose function is evaluated at $(r_{\perp,n}=r_{\perp}^{\max}\zeta_n)$ [Eq.~(\ref{eq:zeta})] (with slight modification at $n=0$ [Eq.~(\ref{eq:Hankel_An})]) but interval is defined with $(r_{\perp}^{\max}\xi_n)$ [Eq.~(\ref{eq:xi})]:
\begin{align}
\int^{r_{\perp}^{\max}\xi_{n+1}}_{r_{\perp}^{\max}\xi_n}A(r_{\perp})J_0(k_{\perp}r_{\perp})r_{\perp}\diff r_{\perp} & \approx \tilde{A}_n\int^{r_{\perp}^{\max}\xi_{n+1}}_{r_{\perp}^{\max}\xi_n}J_0(k_{\perp}r_{\perp})r_{\perp}\diff r_{\perp} \nonumber \\
& =\tilde{A}_n\frac{r_{\perp}}{k_{\perp}}J_1(k_{\perp}r_{\perp})\big\rvert_{r_{\perp}^{\max}\xi_n}^{r_{\perp}^{\max}\xi_{n+1}} \nonumber \\
& =\frac{\tilde{A}_n}{k_{\perp}}\left[\left(r_{\perp}^{\max}\xi_{n+1}\right)J_1(k_{\perp}\left(r_{\perp}^{\max}\xi_{n+1}\right))-\left(r_{\perp}^{\max}\xi_n\right)J_1(k_{\perp}\left(r_{\perp}^{\max}\xi_n\right))\right]
\end{align}
by use of the relation $J_0(u)u=\dod{}{u}\left(uJ_1(u)\right)$. Summing over all $n=0,1,\cdots,(N_{\perp}-1)$ leads to
\begin{align}
A_{H_0,m}=A_{H_0}(k_{\perp,m}) & =C_{\mathfrak{H}}\frac{1}{k_{\perp,m}}\sum_{n=0}^{N_{\perp}-1}\ell_n\left(A_n-A_{n+1}\right)\left(r_{\perp}^{\max}\xi_{n+1}\right)J_1(k_{\perp,m}\left(r_{\perp}^{\max}\xi_{n+1}\right)) \nonumber \\
& =C_{\mathfrak{H}}\frac{r_{\perp}^{\max}}{k_{\perp,m}}\sum_{n=0}^{N_{\perp}-1}\left[\ell_n\left(A_n-A_{n+1}\right)\Exp^{\alpha\left(n+1-N_{\perp}\right)}\right]J_1(k_{\perp}^{\max}r_{\perp}^{\max}\zeta_0\Exp^{\alpha\left(m+n+1-N_{\perp}\right)}),
\label{eq:AHm_tmp}
\end{align}
where $\ell_n=1$ if $n\neq0$, and $J_1(\cdot)$ is the Bessel function of order $1$.

If we define the discrete cross correlation $\star_D$ as
\begin{equation}
\left(P\star_D Q\right)_m\equiv\sum_n\overline{P_n}Q_{m+n}.
\label{eq:discrete_cross_convolution}
\end{equation}
It satisfies [Eq.~(\ref{eq:ccc1D})]
\begin{align}
\mathfrak{F}_D[P\star_D Q] & =\frac{1}{C_{\mathfrak{F}_D}}\overline{\mathfrak{F}_D[P]}\mathfrak{F}_D[Q]=\frac{1}{C_{\mathfrak{F}_D}}\left(\frac{C_{\mathfrak{F}_D}}{C_{\mathfrak{IF}_D}}\mathfrak{F}_D^{-1}[\overline{P}]\right)\mathfrak{F}_D[Q] \nonumber \\
& =\frac{1}{C_{\mathfrak{IF}_D}}\mathfrak{F}_D^{-1}[\overline{P}]\mathfrak{F}_D[Q].
\label{eq:discrete_cross_convolution_conjugate}
\end{align}
Since DFT inherently exhibits periodicity (Fig.~\ref{fig:DFT}), zero-padding is required if $P$ or $Q$ does not display periodicity beyond the sampling window; otherwise, aliasing will occur by direct application of Eq.~(\ref{eq:discrete_cross_convolution_conjugate}).

If we let
\begin{subequations}
\begin{align}
P_n & =\overline{R}_n=\begin{cases}
\ell_n\left(\overline{A}_n-\overline{A}_{n+1}\right)\Exp^{\alpha\left(n+1-N_{\perp}\right)}, & n=0,1,\cdots,N_{\perp}-1 \\
0, & n=N_{\perp},N_{\perp}+1,\cdots,T^w_{\text{extended}}\quad\text{(zero-padding)}
\end{cases} \label{eq:FHATHA_P} \\
Q_n & =J_1(k_{\perp}^{\max}r_{\perp}^{\max}\zeta_0\Exp^{\alpha\left(n+1-N_{\perp}\right)}),\hspace{6.3em}n=0,1,\cdots,T^w_{\text{extended}} \label{eq:FHATHA_Q}
\end{align} \label{eq:FHATHA_PQ}
\end{subequations}
Eq.~(\ref{eq:AHm_tmp}) becomes
\begin{subequations}
\begin{align}
A_{H_0,m} & =C_{\mathfrak{H}}\frac{r_{\perp}^{\max}}{k_{\perp,m}}\frac{1}{C_{\mathfrak{IF}_D}}\mathfrak{F}_D^{-1}\left[\mathfrak{F}_D^{-1}[R]\mathfrak{F}_D[Q]\right] \\
& \xlongequal{\text{or also}}C_{\mathfrak{H}}\frac{r_{\perp}^{\max}}{k_{\perp,m}}\frac{1}{C_{\mathfrak{F}_D}}\mathfrak{F}_D\left[\mathfrak{F}_D[R]\mathfrak{F}^{-1}_D[Q]\right],\quad\text{derived from Eq.~(\ref{eq:ccc2D}) with a similar process}
\end{align} \label{eq:FHATHA}
\end{subequations}
which is meaningful only for $m=0,1,\cdots,(N_{\perp}-1)$ with others discarded after FHATHA. The zero-padding range is determined by $T^w_{\text{extended}}\geq2N_{\perp}-2$ [so that there are a total of $(2N_{\perp}-1)$ sampling points]. This zero-padding procedure is similar to the convolution operation with clipped extended signals (Fig.~\ref{fig:convolution_zeropadding}). However, since the summation in Eq.~(\ref{eq:AHm_tmp}) includes $Q_n$ up to $n=2N_{\perp}-2$, $Q_n$ must be analytically computed at least for $n=0,1,\cdots,(2N_{\perp}-2)$, rather than simple zero-padding as in the convolution theorem (Sec.~\ref{sec:DFT}\ref{subsec:DFT_convolution}). Note that $Q$, or further its $\mathfrak{F}_D[Q]$ or $\mathfrak{F}^{-1}_D[Q]$ in Eq.~(\ref{eq:FHATHA}), can be precomputed only once before the simulation. In the physical Fourier-transform convention ($C_{\mathfrak{F}_D}=\frac{1}{\mathfrak{N}}$ and thus $C_{\mathfrak{IF}_D}=1$), $A_{H_0,m}=C_{\mathfrak{H}}\dfrac{r_{\perp}^{\max}}{k_{\perp,m}}\mathfrak{F}_D^{-1}\left[\mathfrak{F}_D^{-1}[R]\mathfrak{F}_D[Q]\right]$, which is (in MATLAB syntax below)
\begin{lstlisting}[language=MATLAB]
A_H = r_max./k.*fft(fft(R).*ifft(Q));
\end{lstlisting}
if we assume the Hankel-transform convention $C_{\mathfrak{H}}=1$.

Since the Hankel transform and the inverse Hankel transform exhibit the same formulation [Eq.~(\ref{eq:Hankel_transform})], inverse FHATHA follows Eq.~(\ref{eq:FHATHA}) by swapping the symbols $r\leftrightarrow k$ and changing $C_{\mathfrak{H}}$ into $C_{\mathfrak{IH}}$. Note that $Q$ is independent of Hankel or inverse Hankel transforms due to its $(r_{\perp},k_{\perp})$ symmetry. Fig.~\ref{fig:Hankel_verification} shows two verification examples that demonstrate its excellent consistency with the analytical formula.

\begin{figure}[!ht]
\centering
\includegraphics[width=0.7\linewidth]{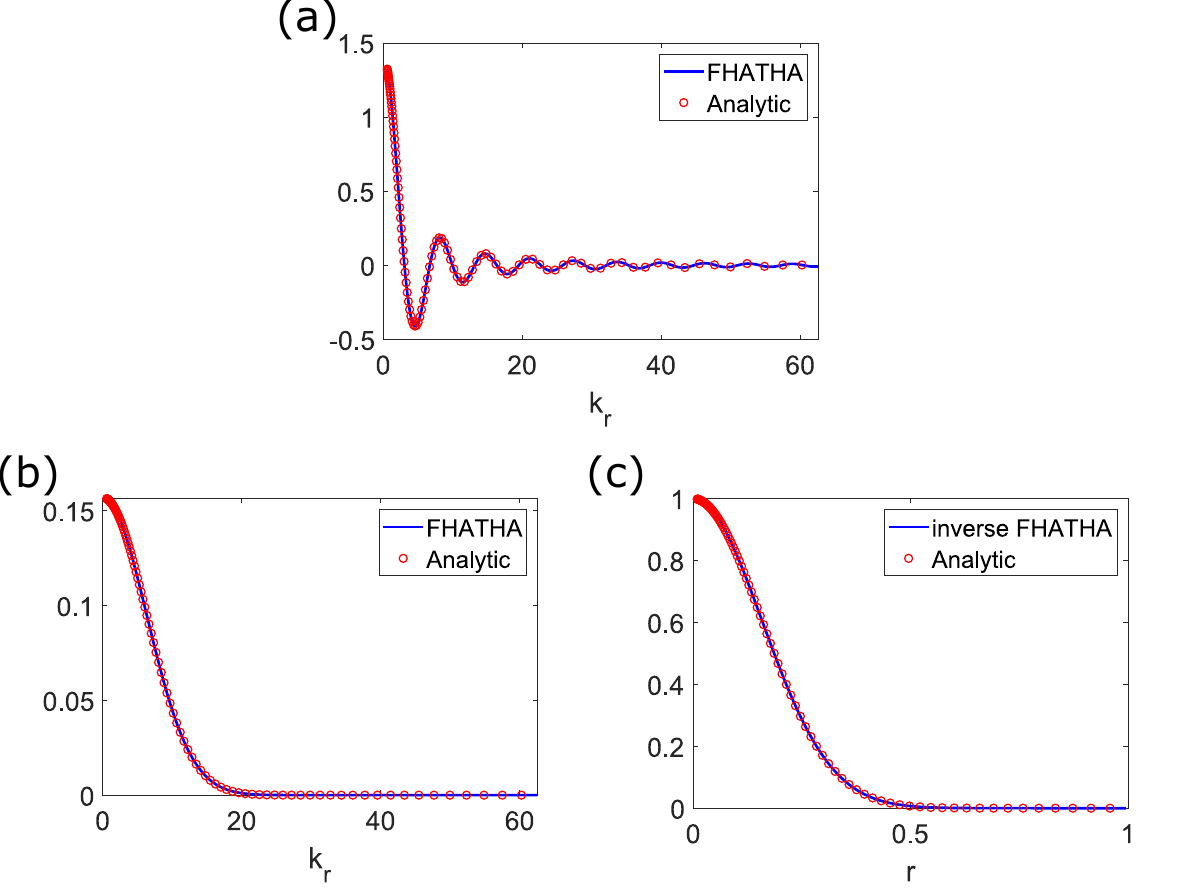}
\caption{Verifications of FHATHA. Zeroth-order Hankel transforms of (a) $f(r)=\sqrt{\frac{5}{2\pi}}r^2$ and (b) $f(r)=\Exp^{-20r^2}$, which are $\sqrt{10\pi}\frac{2\eta J_0(\eta)+(\eta^2-4)J_1(\eta)}{\eta^3}$ (with $\eta=r_{\max}k_r$) and $\frac{\pi}{20}\Exp^{-k_r^2/80}$, respectively. (c) Inverse zeroth-order Hankel transform of (b). This is used to further verify the operation of inverse Hankel transform and see whether the signal can be recovered.}
\label{fig:Hankel_verification}
\end{figure}

The exponential sampling strategy [Eqs.~(\ref{eq:zeta}) and (\ref{eq:xi})] of FHATHA introduces an advantage over typical uniform sampling. Mostly, the radially-symmetric field is the strongest at the spatial center. In particular, in cases such as self-focusing for a high-peak-power pulse (Fig.~\ref{fig:Hankel_sampling_benefit}), the spatial extent of the field decreases as it propagates. This can lead to undersampling of the field, which is resolved by FHATHA's exponential sampling that densely samples around the center.

\begin{figure}[!ht]
\centering
\includegraphics[width=0.7\linewidth]{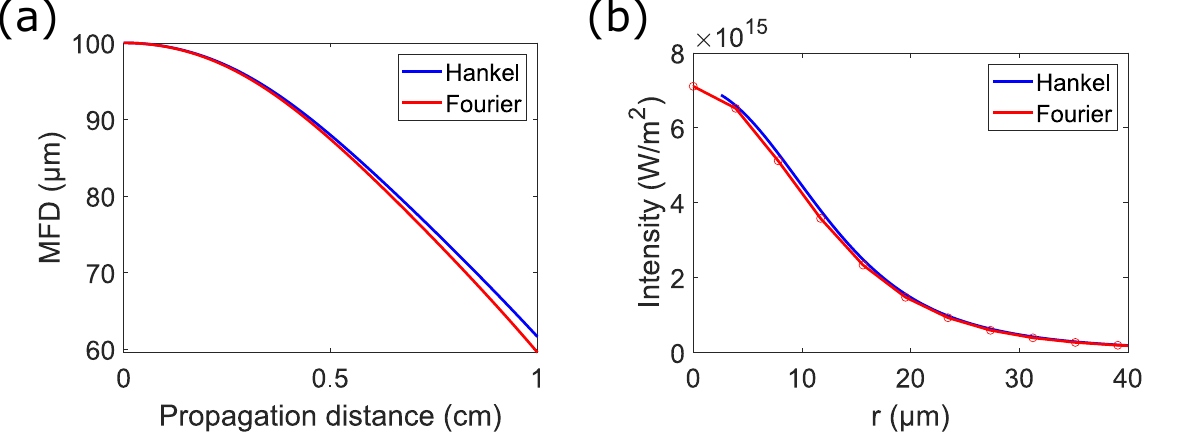}
\caption{(a) Mode-field-diameter (MFD) evolution of a pulse propagating in a \SIadj{1}{\cm}-long bulk silica, simulated with the Hankel transform or 2D spatial Fourier transform. The input pulse is \SIadj{1}{\ps}-long and has \SIadj{8}{\micro\joule} pulse energy, with a Gaussian spatial profile of \SIadj{100}{\micro\m} MFD. (b) The intensity profile of the pulse.}
\label{fig:Hankel_sampling_benefit}
\end{figure}

Unlike the frequency window in the Fourier transform that is determined by the temporal sampling period, the $k_{\perp}$ window can be chosen quite arbitrarily. Its value is determined by the signal's maximum $k_{\perp}$.

Lastly, although we discuss only the FHATHA with the zeroth order, it can be extended to an arbitrary integer order \cite{Zhang2002}. By use of the relation $u^{\nu+1}J_{\nu}(u)=\dod{}{u}\left(u^{\nu+1}J_{\nu+1}(u)\right)$, the Hankel-transform integral becomes
\begin{align}
\int^{r_{\perp}^{\max}\xi_{n+1}}_{r_{\perp}^{\max}\xi_n}A(r_{\perp})J_{\nu}(k_{\perp}r_{\perp})r_{\perp}\diff r_{\perp} & \approx \frac{\tilde{A}_n}{r_{\perp,n}^{\nu}}\int^{r_{\perp}^{\max}\xi_{n+1}}_{r_{\perp}^{\max}\xi_n}J_{\nu}(k_{\perp}r_{\perp})r_{\perp}^{\nu+1}\diff r_{\perp} \nonumber \\
& =\frac{\tilde{A}_n}{r_{\perp,n}^{\nu}}\frac{r_{\perp}^{\nu+1}}{k_{\perp}}J_{\nu+1}(k_{\perp}r_{\perp})\big\rvert_{r_{\perp}^{\max}\xi_n}^{r_{\perp}^{\max}\xi_{n+1}} \nonumber \\
& =\frac{\tilde{A}_n}{k_{\perp}r_{\perp,n}^{\nu}}\left[\left(r_{\perp}^{\max}\xi_{n+1}\right)^{\nu+1}J_{\nu+1}(k_{\perp}\left(r_{\perp}^{\max}\xi_{n+1}\right))-\left(r_{\perp}^{\max}\xi_n\right)^{\nu+1}J_{\nu+1}(k_{\perp}\left(r_{\perp}^{\max}\xi_n\right))\right]
\label{eq:Hankel_An_nu}
\end{align}
with $\dfrac{A(r_{\perp})}{r_{\perp}^{\nu}}$ evaluated at $r_{\perp,n}$. The Hankel transform of order $\nu$ eventually becomes
\begin{align}
A_{H_{\nu},m}=A_{H_{\nu}}(k_{\perp,m}) & =C_{\mathfrak{H}}\frac{1}{k_{\perp,m}}\sum_{n=0}^{N_{\perp}-1}\left(\frac{\tilde{A}_n}{r_{\perp,n}^{\nu}}-\frac{\tilde{A}_{n+1}}{r_{\perp,n+1}^{\nu}}\right)\left(r_{\perp}^{\max}\xi_{n+1}\right)^{\nu+1}J_{\nu+1}(k_{\perp,m}\left(r_{\perp}^{\max}\xi_{n+1}\right)) \nonumber \\
& =C_{\mathfrak{H}}\frac{r_{\perp}^{\max}}{k_{\perp,m}}\left(\frac{2}{1+\Exp^{\alpha}}\right)^{\nu}\sum_{n=0}^{N_{\perp}-1}\left[\left(\tilde{A}_n\Exp^{\alpha\nu}-\tilde{A}_{n+1}\right)\Exp^{\alpha\left(n+1-N_{\perp}\right)}\right]J_{\nu+1}(k_{\perp}^{\max}r_{\perp}^{\max}\zeta_0\Exp^{\alpha\left(m+n+1-N_{\perp}\right)})
\label{eq:AHm_tmp_nu}
\end{align}
The factor $\left[2/(1+\Exp^{\alpha})\right]$ results from $(\xi'_0/\zeta_0)$. Similarly, Eq.~(\ref{eq:AHm_tmp_nu}) can be solved with the FFT-based cross correlation. Note that there is a mistake in deriving Eq.~(\ref{eq:AHm_tmp_nu}) in \cite{Zhang2002} [their Eq.~(7)]. They naively added $\ell_n$ to Eq.~(\ref{eq:AHm_tmp_nu}) as in Eq.~(\ref{eq:AHm_tmp}). Such naive addition of $\ell_n$ represents that the parabola approximation is applied to $(A_n/r_{\perp,n}^{\nu})$, which is subsequently used to obtain the value of $A(r'_{\perp,0})$. However, for commonly-used Gaussian fields, the field $A(r_{\perp})\sim\Exp^{-ar_{\perp}^2}$ is parabola-like around the center $r_{\perp}=0$, not its $A(r_{\perp})/r_{\perp}^{\nu}\propto r_{\perp}^{2-\nu}$.

\clearpage
\subsubsection{Important steps to improve FHATHA for ultrafast pulse propagation}\label{subsubsec:Improved_FHATHA}
In fact, FHATHA proposed by \etal{Magni} requires modifications to be correctly employed in ultrafast pulse propagation. Because it does not satisfy any discrete Parseval's theorem as by the DFT [Eq.~(\ref{eq:discrete_Parseval})], the FHATHA-transformed signal ``numerically'' does not conserve energy. Although FHATHA approximates the continuous Hankel transform with a (continuous) Parseval's theorem [Eq.~(\ref{eq:Hankel_Parseval})], it inherently suffers from energy non-conservation, which can be visualized through comparisons of signals after repetitive operations of FHATHA and inverse-FHATHA operations [Figs.~\ref{fig:Hankel_improvements}(a) and (b) of both $N_{\perp}=100\text{ and }1000$].

To mitigate FHATHA's deviation due to a lack of energy conservation, energy restoration is added to the end of FHATHA:
\begin{lstlisting}[language=MATLAB]
% E0 and E_H are powers of the original and the transformed signals
% A is usually in "sqrt(W/m^2)", so E0 is in "W", the power.
E0 = 2*pi*trapz(r,abs(A).^2.*r);
E_H = 2*pi*trapz(kr,abs(A_H).^2.*kr);
% Energy restoration
A_H = A_H.*sqrt(E0./E_H);
\end{lstlisting}
The energy-restoration operation significantly improves FHATHA's accuracy when the number of sampling ($N_{\perp}$) is small [Figs.~\ref{fig:Hankel_improvements}(c) and (d) in the blue box]. However, at large $N_{\perp}$, the root-mean-squared error (RMSE) increases with the energy restoration, compared with the one without it [Figs.~\ref{fig:Hankel_improvements}(a--d) in the red box]. A reasonable speculation is that the simple energy-restoration operation proposed previously uniformly scales the transformed signal, which potentially modulates the weak (background) signal at the window edge and artificially induces an error. However, as shown in the simulation of a Gaussian field (Fig.~\ref{fig:Hankel_Galileo_telescope}), a lack of energy-restoration operation generates simulation results with varying energy.

Another reason for FHATHA's deviation is omission of signals at $r_{\perp}=0$ and $k_{\perp}=0$. In addition, the exponential sampling strategy [Eqs.~(\ref{eq:zeta}) and (\ref{eq:xi})] induces a huge gap between the first sampling point and $r_{\perp}=0$ (or $k_{\perp}=0$). \etal{Magni} verified FHATHA with a parabolic signal that has a higher amplitude at larger $r_{\perp}$. This signal justifies FHATHA's approach of approximating $A(r'_{\perp,0})$ with $A_0$ and $A_1$ [Eq.~(\ref{eq:Hankel_An})]. However, in practical ultrafast optical applications, the beam exhibits a spatially-confined profile, usually Gaussian. FHATHA's sampling gap and $A(r'_{\perp,0})$ approximation easily induce a significant error for such a signal that is the strongest around the spatial (or $\vec{k}_{\perp}$-space) center. To resolve this, the signal at the center needs to be taken into account. Assume that the signal to be transformed includes sampling at $r_{\perp}=0$, the numerical signal used in FHATHA [$\tilde{A}_n$ in Eq.~(\ref{eq:Hankel_An})] is modified to be
\begin{equation}
\tilde{A}_n=\begin{cases}
A(r'_{\perp,0})=\frac{1}{2}\left\{\left[\ell_0\left(A_0-A_1\right)+A_1\right]+\left(\frac{1}{1+\Exp^{\alpha}}A_{-1}+\frac{\Exp^{\alpha}}{1+\Exp^{\alpha}}A_0\right)\right\}, & n=0 \\
A_n, & n=1,2,\cdots,N_{\perp}-1 \\
0, & n=N_{\perp}
\end{cases} \label{eq:my_Hankel_An}
\end{equation}
which takes into account the signal at the spatial center, denoted as $A_{-1}$. $\left(\frac{1}{1+\Exp^{\alpha}}A_{-1}+\frac{\Exp^{\alpha}}{1+\Exp^{\alpha}}A_0\right)$ is the approximate signal at $r'_{\perp,0}$ with known signals at $r_{\perp}=0$ and $r_{\perp,0}$. With reasonably-fine sampling, $\alpha\approx0.01$ or less, and thus the approximate signal approaches $\left(A_{-1}+A_0\right)/2$. The modified signal considers, at $r'_{\perp,0}$, the mean of signals from the original FHATHA's parabolic approximation and from the linear interpolation of $A_{-1}$ and $A_0$. Thus, the Hankel-transform signal follows
\begin{align}
A_{H_0,m}=A_{H_0}(k_{\perp,m}) & =C_{\mathfrak{H}}\frac{1}{k_{\perp,m}}\sum_{n=0}^{N_{\perp}-1}\left(\tilde{A}_n-\tilde{A}_{n+1}\right)\left(r_{\perp}^{\max}\xi_{n+1}\right)J_1(k_{\perp,m}\left(r_{\perp}^{\max}\xi_{n+1}\right)) \nonumber \\
& =C_{\mathfrak{H}}\frac{r_{\perp}^{\max}}{k_{\perp,m}}\sum_{n=0}^{N_{\perp}-1}\left[\left(\tilde{A}_n-\tilde{A}_{n+1}\right)\Exp^{\alpha\left(n+1-N_{\perp}\right)}\right]J_1(k_{\perp}^{\max}r_{\perp}^{\max}\zeta_0\Exp^{\alpha\left(m+n+1-N_{\perp}\right)}),
\label{eq:my_AHm_tmp}
\end{align}
for $m=0,1,\cdots,(N_{\perp}-1)$. Eq.~(\ref{eq:my_AHm_tmp}) is subsequently solved with the FFT-based cross correlation [Eqs.~(\ref{eq:FHATHA_PQ}) and (\ref{eq:FHATHA})]. By employing the relation $\frac{r_{\perp}}{k_{\perp}}J_1(k_{\perp}r_{\perp})\approx\frac{r_{\perp}^2}{2}$ (due to $J_1(x)\approx\frac{x}{2}$ for $0\leq x\ll1$), we can find the transformed signal at $k_{\perp}=0$ through
\begin{align}
A_{H_0,-1}=A_{H_0}(k_{\perp,-1}=0) & =C_{\mathfrak{H}}\sum_{n=0}^{N_{\perp}-1}\left(\tilde{A}_n-\tilde{A}_{n+1}\right)\frac{r_{\perp,n+1}^2}{2} \nonumber \\
& =C_{\mathfrak{H}}\left(r_{\perp}^{\max}\right)^2\sum_{n=0}^{N_{\perp}-1}\left(\tilde{A}_n-\tilde{A}_{n+1}\right)\frac{\Exp^{2\alpha\left(n+1-N_{\perp}\right)}}{2}.
\label{eq:my_AHm_tmp_kr=0}
\end{align}
The comparisons of Figs.~\ref{fig:Hankel_improvements}(a,b) and (g,h) show that the inclusion of spatial and $\vec{k}_{\perp}$-space centers reduces FHATHA's error, with a significant amount at small $N_{\perp}$.

\begin{figure}[!ht]
\centering
\includegraphics[width=\linewidth]{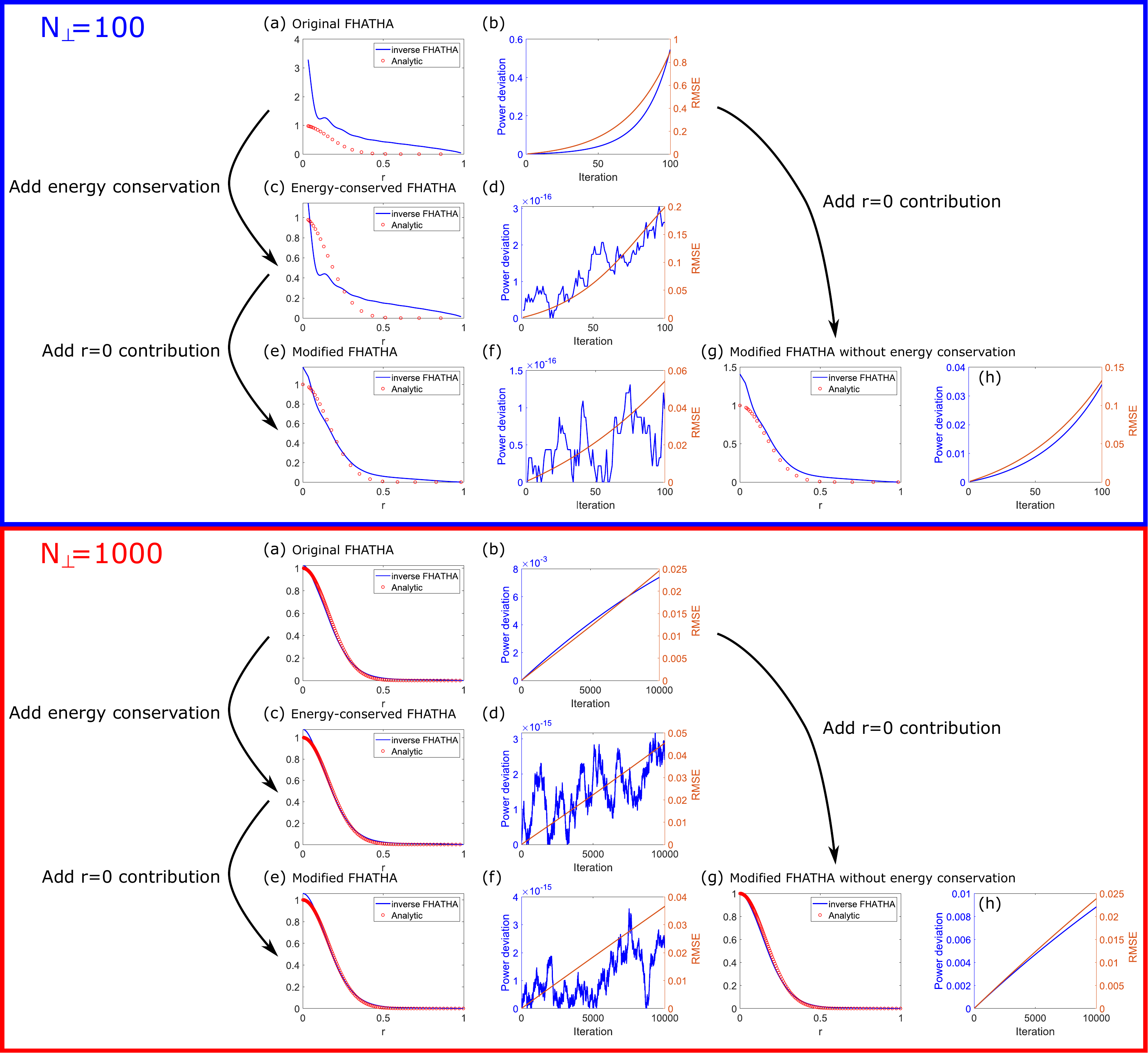}
\caption{Comparisons of FHATHA with different modifications. (a), (c), (e), and (g) are spatial profiles of the FHATHA-transformed signals after many operations of (FHATHA,inverse FHATHA) pair, and its analytic results that follow the Gaussian profile in Fig.~\ref{fig:Hankel_verification}(b). (d), (f), and (h) are power ($2\pi\int_0^{\infty}\abs{A}^2r\diff r$) deviations (blue) and root-mean-squared errors (orange) of two signals ($\sqrt{\frac{\sum_{n=1}^{N_{\perp}}\abs{A_{H_0,\text{analytic}}(n)-A_{H_0}(n)}^2}{N_{\perp}}}$) with increasing iterations. Top (those in the blue box) and bottom (in the red box) are results with different numbers of sampling points.}
\label{fig:Hankel_improvements}
\end{figure}

\begin{figure}[!ht]
\centering
\includegraphics[width=0.7\linewidth]{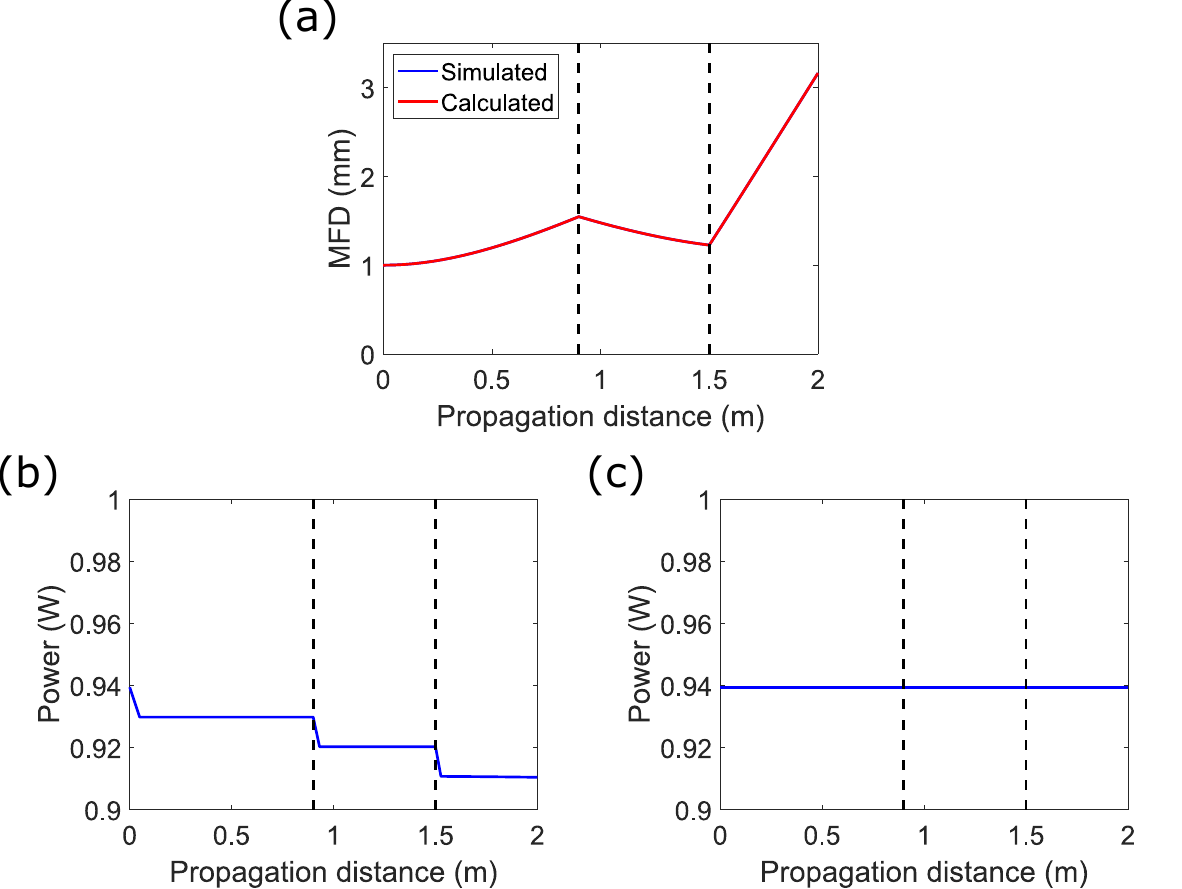}
\caption{Numerical simulation of a Gaussian field propagating through a Galileo telescope with the first ($f=\SI{0.9}{\m}$) and the second ($f=\SI{-0.3}{\m}$) lenses. This telescope increases the beam size by three times. (a) MFD evolution. The calculated red line is derived from typical ABCD matrices of a Gaussian field. Lines from the simulation (blue) and ABCD matrices (red) perfectly overlap. (b) and (c) are total powers of the field during propagation without and with energy restoration in FHATHA, respectively. This simulation involves three separate field propagations between lenses. Lens effect is considered by adding a quadratic spatial phase to the field.}
\label{fig:Hankel_Galileo_telescope}
\end{figure}

In conclusion, FHATHA requires additional steps of energy restoration and consideration of spatial and $\vec{k}_{\perp}$-space centers. Energy restoration (through a simple uniform scaling) can slightly distort the transformed signal but is a required step in simulations. On the other hand, consideration of spatial and $\vec{k}_{\perp}$-space centers only improves the result. In numerical pulse-propagation simulations, the field is usually solved in the $\vec{k}_{\perp}$ space through
\begin{equation}
A(k_{\perp},z+\diff z,\omega)=A(k_{\perp},z,\omega)+f(A(k_{\perp},z,\omega),k_{\perp},\omega)\diff z,
\end{equation}
where $\left(f\diff z\right)$ is effectively solved with various numerical stepping schemes, such as the Runge-Kutta. Solving $f$ usually involves a few Hankel and inverse Hankel transforms. Since $\abs{f\diff z}$ is generally small, FHATHA's error in energy does not significantly affect the propagating field, as seen in Fig.~\ref{fig:Hankel_Galileo_telescope}. The energy deviation occurs only in the first and last field transformations that transform the field to the $\vec{k}_{\perp}$ space to initiate the propagation and back to the $\vec{r}_{\perp}$ space when finished. As a result, energy-conserved FHATHA must be applied in the first and last transformations, while the main optical propagation can employ FHATHA, with or without energy restoration depending on simulation conditions (Fig.~\ref{fig:Hankel_propagation_scheme}). If the propagation distance is long or a simulation involves multiple propagations, energy restoration is recommended to avoid error accumulation; otherwise, it can be disabled for an improved computational performance and potentially higher transformation accuracy as described previously. In the meantime, space centers are always considered in FHATHA.

\begin{figure}[!ht]
\centering
\includegraphics[width=\linewidth]{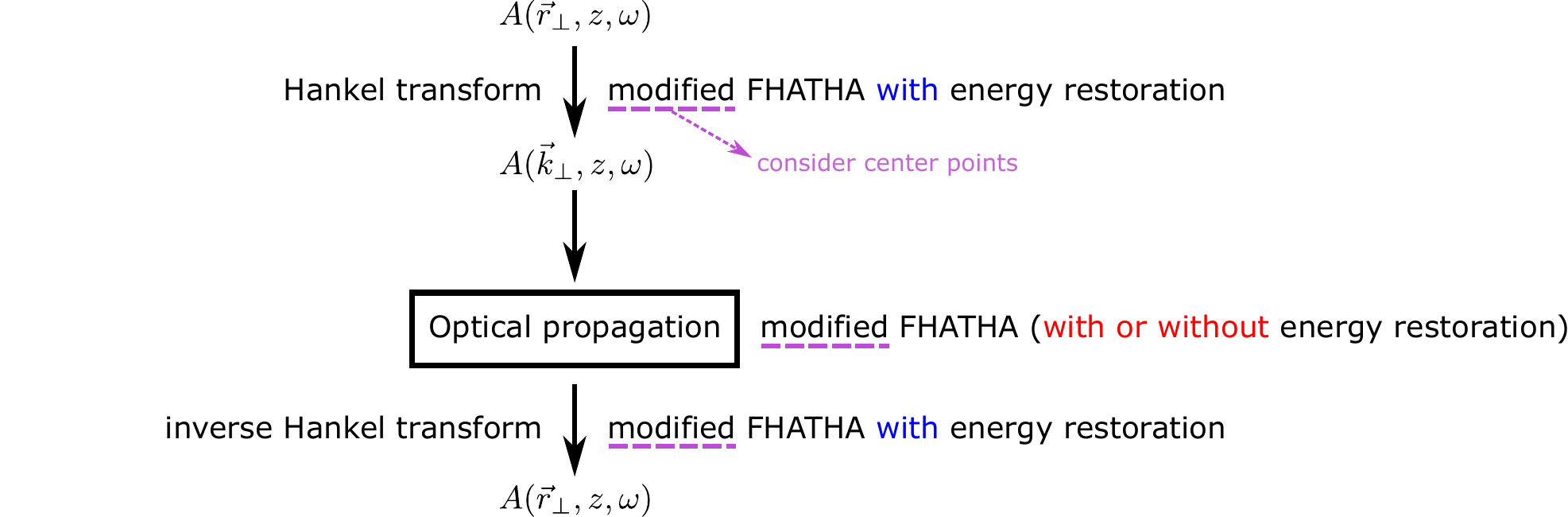}
\caption{Optical propagation simulation scheme with the FHATHA-based Hankel transform. During optical propagation, whether to impose FHATHA's energy-restoration operation depends on the amount of potential error accumulation.}
\label{fig:Hankel_propagation_scheme}
\end{figure}

\pagebreak
\subsection{Discrete Hankel transform (DHT)}\label{subsec:DHT}
DHT should be a transformation that exhibits its own rules \cite{Baddour2015,Baddour2017}, as DFT. In this perspective, Hankel transforms that simply approximate the continuous Hankel transform are not DHT, but rather quasi-HT.\footnote{In this perspective, Siegman’s and Talman's QFHT should be renamed to ``fast quasi-HT (FQHT)'' \cite{Siegman1977}.} The theory of DHT has been first proposed by \etal{Yu} for the zeroth-order Hankel transform \cite{Yu1998} and then by Guizar-Sicairos and Guti\'{e}rrez-Vega for higher orders \cite{GuizarSicairos2004}. Most importantly, DHT satisfies a discrete Parseval's theorem such that the transformation conserves the discrete energy or power, making it a perfect transformation for any physical simulations. However, DHT investigated to date suffers from a large computational complexity, compared to DFT that harnesses the FFT algorithm. Its complicated transformation kernel lacks a simple symmetry as the exponential kernel of DFT, and thus a similar divide-and-conquer method of FFT [to achieve a smaller $\mathcal{O}(N_{\perp}\log_2N_{\perp})$ computational complexity] is hindered. With the newly-proposed simulation scheme that utilizes quasi-HT (Fig.~\ref{fig:Hankel_propagation_scheme}), fast FFT-based quasi-HT, such as FHATHA, outperforms naive DHT with matrix multiplication, as it not only reasonably conserves the energy/power as DHT but also exhibits significant computational efficiency.

For completeness as a tutorial document, here I briefly introduce DHT. For a complete discussion of DHT, readers are referred to \cite{Baddour2015}, as well as \cite{Baddour2017} for numerical discussions.

The discrete version of the continuous Hankel-transform pair, which is the counterpart of Eq.~(\ref{eq:discrete_Fourier_transform_c}), follows
\begin{equation}
\begin{aligned}
A_{H_{\nu,D_c}}(k_{\nu,m})=\mathfrak{H}_{\nu,D_c}[A(r)] & \equiv C_{\mathfrak{H}}\frac{r_{\max}^2}{j_{\nu,N_{\perp}}}\sum_{n=1}^{N_{\perp}-1}Y^{\nu,N_{\perp}}_{mn}A(r_{\nu,n}) \\
A(r_{\nu,n})=\mathfrak{H}^{-1}_{\nu,D_c}[A_{H_{\nu,D_c}}(k)] & \equiv C_{\mathfrak{IH}}\frac{k_{\max}^2}{j_{\nu,N_{\perp}}}\sum_{m=1}^{N_{\perp}-1}Y^{\nu,N_{\perp}}_{nm}A_{H_{\nu,D_c}}(k_{\nu,m})=C_{\mathfrak{IH}}\frac{j_{\nu,N_{\perp}}}{r_{\max}^2}\sum_{m=1}^{N_{\perp}-1}Y^{\nu,N_{\perp}}_{nm}A_{H_{\nu,D_c}}(k_{\nu,m})
\end{aligned},
\label{eq:DcHT}
\end{equation}
where the transformation kernel is
\begin{equation}
[\mathbf{Y}^{\nu,N_{\perp}}]_{mn}=Y^{\nu,N_{\perp}}_{mn}=\frac{2}{j_{\nu,N_{\perp}}J_{\nu+1}^2(j_{\nu,n})}J_{\nu}(\frac{j_{\nu,m}j_{\nu,n}}{j_{\nu,N_{\perp}}}),\quad1\leq m,n\leq N_{\perp}-1,
\end{equation}
where $J_{\nu}$ is the Bessel function of order $\nu$. Based on Eq.~(\ref{eq:DcHT}), we can derive DHT as
\begin{equation}
\begin{aligned}
A_{H_{\nu,D}}(k_{\nu,m})=\mathfrak{H}_{\nu,D}[A(r)] & \equiv C_{\mathfrak{H}_D}\sum_{n=1}^{N_{\perp}-1}Y^{\nu,N_{\perp}}_{mn}A(r_{\nu,n}) \\
A(r_{\nu,n})=\mathfrak{H}^{-1}_{\nu,D}[A_{H_{\nu,D}}(k)] & \equiv C_{\mathfrak{IH}_D}\sum_{m=1}^{N_{\perp}-1}Y^{\nu,N_{\perp}}_{nm}A_{H_{\nu,D}}(k_{\nu,m})
\end{aligned}.
\label{eq:DHT}
\end{equation}
Unlike DFT which has $C_{\mathfrak{F}_D}C_{\mathfrak{IF}_D}=\frac{1}{\mathfrak{N}}$, $C_{\mathfrak{H}_D}C_{\mathfrak{IH}_D}=1$; as a result, we always assume the convention of $C_{\mathfrak{H}_D}=C_{\mathfrak{IH}_D}=1$ and ignore them in Eq.~(\ref{eq:DHT}). The sampling needs to follow
\begin{subequations}
\begin{align}
r_{\nu,n} & =\frac{j_{\nu,n}}{j_{\nu,N_{\perp}}}r_{\max} \\
k_{\nu,n} & =\frac{j_{\nu,n}}{j_{\nu,N_{\perp}}}k_{\max},
\end{align}
\end{subequations}
for $n=1,2,\cdots,N_{\perp}$. $j_{\nu,n}$ is the $n$th positive (non-zero) Bessel-zero of the Bessel function of order $\nu$ ($j_{\nu,1}\neq0$) such that $r_{\nu,1}$ and $k_{\nu,1}$ are not zero. Because $Y^{\nu,N_{\perp}}_{N_{\perp}n}=Y^{\nu,N_{\perp}}_{mN_{\perp}}=0$, the field sampled at the boundaries must be zero: $A(r_{\nu,N_{\perp}}=r_{\max})=A_{H_{\nu,D_c}}(k_{\nu,N_{\perp}}=k_{\max})=0$. Therefore, calculations are required only for the other $(N_{\perp}-1)$ points [Eq.~(\ref{eq:DHT})]. Because the rows and columns of $\mathbf{Y}^{\nu,N_{\perp}}$ are orthonormal, $\mathbf{Y}^{\nu,N_{\perp}}\mathbf{Y}^{\nu,N_{\perp}}=\mathbf{I}$. Discrete version of the continuous Hankel transform [Eq.~(\ref{eq:DcHT})] satisfies the discrete Parseval's theorem:
\begin{equation}
\frac{r_{\max}^2}{C_{\mathfrak{IH}}}\sum_{n=1}^{(N_{\perp}-1)\text{ or }N_{\perp}}\abs{\frac{A(r_{\nu,n})}{J_{\nu+1}(j_{\nu,n})}}^2=\frac{k_{\max}^2}{C_{\mathfrak{H}}}\sum_{m=1}^{(N_{\perp}-1)\text{ or }N_{\perp}}\abs{\frac{A_{H_{\nu,D_c}}(k_{\nu,m})}{J_{\nu+1}(j_{\nu,m})}}^2,
\end{equation}
from which we can derive DHT's [Eq.~(\ref{eq:DHT})] Parseval's theorem:
\begin{equation}
\frac{1}{C_{\mathfrak{IH}_D}}\sum_{n=1}^{(N_{\perp}-1)\text{ or }N_{\perp}}\abs{\frac{A(r_{\nu,n})}{J_{\nu+1}(j_{\nu,n})}}^2=\frac{1}{C_{\mathfrak{H}_D}}\sum_{m=1}^{(N_{\perp}-1)\text{ or }N_{\perp}}\abs{\frac{A_{H_{\nu,D}}(k_{\nu,m})}{J_{\nu+1}(j_{\nu,m})}}^2.
\end{equation}
With fine sampling ($N_{\perp}\gg1$), $\mathfrak{H}_{\nu,D_c}\approx\mathfrak{H}_{\nu}$ and $\mathfrak{H}^{-1}_{\nu,D_c}\approx\mathfrak{H}_{\nu}^{-1}$. Similar to the window relation $\nu^wT^w=\frac{1}{\triangle t}\left(\mathfrak{N}\triangle t\right)=\mathfrak{N}$ of the DFT, DHT exhibits
\begin{equation}
k_{\max}r_{\max}=j_{\nu,N_{\perp}}.
\end{equation}

It is worth noting that due to the denominator in the Parseval's theorem, it is customary to define another DHT as
\begin{equation}
\begin{aligned}
A_{H_{\nu,D'}}(k_{\nu,m})=\mathfrak{H}_{\nu,D'}[A(r)] & \equiv C_{\mathfrak{H}_D}\sum_{n=1}^{N_{\perp}-1}T^{\nu,N_{\perp}}_{mn}A(r_{\nu,n}) \\
A(r_{\nu,n})=\mathfrak{H}^{-1}_{\nu,D'}[A_{H_{\nu,D'}}(k)] & \equiv C_{\mathfrak{IH}_D}\sum_{m=1}^{N_{\perp}-1}T^{\nu,N_{\perp}}_{nm}A_{H_{\nu,D'}}(k_{\nu,m})
\end{aligned},
\label{eq:DcHT2}
\end{equation}
where a new transformation kernel $\mathbf{T}^{\nu,N_{\perp}}$ follows
\begin{equation}
T^{\nu,N_{\perp}}_{mn}=Y^{\nu,N_{\perp}}_{mn}\frac{J_{\nu+1}(j_{\nu,n})}{J_{\nu+1}(j_{\nu,m})}.
\end{equation}
Its Parseval's theorem is simply
\begin{equation}
\frac{1}{C_{\mathfrak{IH}_D}}\sum_{n=1}^{(N_{\perp}-1)\text{ or }N_{\perp}}\abs{A(r_{\nu,n})}^2=\frac{1}{C_{\mathfrak{H}_D}}\sum_{m=1}^{(N_{\perp}-1)\text{ or }N_{\perp}}\abs{A_{H_{\nu,D'}}(k_{\nu,m})}^2.
\end{equation}
that bears a strong resemblance to the DFT's [Eq.~(\ref{eq:discrete_Parseval})]. This is the convention employed by \etal{Yu} in the first DHT derivation \cite{Yu1998}. However, Eq.~(\ref{eq:DHT}) with $\mathbf{Y}^{\nu,N_{\perp}}$ is preferred because its $A_{H_{\nu,D_c}}/A_{H_{\nu,D}}$ directly approximates the continuous Hankel transform, as DFT that directly approximates the continuous Fourier transform after adding the additional $\triangle t$ factor.

There is one distinction between DFT and DHT due to the boundary value. Since DHT requires the signals at the pre-defined boundaries to be zero, $A(r_{\max})=A_{H_{\nu,D}}(k_{\max})=0$, the naive discrete generalization of the continuous Hankel transform in Eq.~(\ref{eq:DcHT}) is, in fact, the continuous Hankel transform with an additional zero-signal assumption at the boundary. On the other hand, DFT applies a different assumption that adds periodicity to the applied signal.

\clearpage
\pagebreak
\bibliography{reference.bib}

\begin{thebibliography}{10}
\newcommand{\enquote}[1]{``#1''}

\bibitem{Boyd2008}
R.~W. Boyd, \emph{Nonlinear Optics} (Academic Press, Burlington, 2008), 3rd ed.

\bibitem{Conforti2013}
M.~Conforti, A.~Marini, D.~Faccio, and F.~Biancalana, \enquote{Negative frequencies get real: a missing puzzle piece in nonlinear optics,} {\protect\JournalTitle{arXiv preprint arXiv:1305.5264}}  (2013).

\bibitem{Chen2007}
Y.-H. Chen, S.~Varma, A.~York, and H.~M. Milchberg, \enquote{Single-shot, space- and time-resolved measurement of rotational wavepacket revivals in {$\mathrm{H}_{2}$}, {$\mathrm{D}_{2}$}, {$\mathrm{N}_{2}$}, {$\mathrm{O}_{2}$}, and {$\mathrm{N}_{2}\mathrm{O}$},} {\protect\JournalTitle{Opt. Express}} \textbf{15}, 11341--11357 (2007).

\bibitem{Chen2024}
Y.-H. Chen and F.~Wise, \enquote{Unified and vector theory of {Raman} scattering in gas-filled hollow-core fiber across temporal regimes,} {\protect\JournalTitle{APL Photonics}} \textbf{9}, 030902 (2024).

\bibitem{Konyashchenko2008}
A.~V. Konyashchenko, L.~L. Losev, V.~S. Pazyuk, and S.~Y. Tenyakov, \enquote{Frequency shifting of sub-100 fs laser pulses by stimulated {Raman} scattering in a capillary filled with pressurized gas,} {\protect\JournalTitle{Appl. Phys. B}} \textbf{93}, 455--461 (2008).

\bibitem{Chen2023}
Y.-H. Chen, J.~Moses, and F.~Wise, \enquote{Femtosecond long-wave-infrared generation in hydrogen-filled hollow-core fiber,} {\protect\JournalTitle{J. Opt. Soc. Am. B}} \textbf{40}, 796--806 (2023).

\bibitem{Loree1976}
T.~R. Loree, C.~D. Cantrell, and D.~L. Barker, \enquote{Stimulated {Raman} emission at {9.2 $\mu$m} from hydrogen gas,} {\protect\JournalTitle{Opt. Commun.}} \textbf{17}, 160--162 (1976).

\bibitem{Grasyuk1984}
A.~Z. Grasyuk, L.~L. Losev, D.~N. Nikogosyan, and A.~A. Oraevski{\u{\i}}, \enquote{Generation of single picosecond pulses of up to 0.6 {mJ} energy and of 9.2 {$\mu$m} wavelength by stimulated {Raman} scattering,} {\protect\JournalTitle{Sov. J. Quantum Electron.}} \textbf{14}, 1257--1258 (1984).

\bibitem{Kane1993}
D.~J. Kane and R.~Trebino, \enquote{Single-shot measurement of the intensity and phase of an arbitrary ultrashort pulse by using frequency-resolved optical gating,} {\protect\JournalTitle{Opt. Lett.}} \textbf{18}, 823--825 (1993).

\bibitem{Miranda2012}
M.~Miranda, T.~Fordell, C.~Arnold, A.~L'Huillier, and H.~Crespo, \enquote{Simultaneous compression and characterization of ultrashort laser pulses using chirped mirrors and glass wedges,} {\protect\JournalTitle{Opt. Express}} \textbf{20}, 688--697 (2012).

\bibitem{Miranda2012a}
M.~Miranda, C.~L. Arnold, T.~Fordell, F.~Silva, B.~Alonso, R.~Weigand, A.~L'Huillier, and H.~Crespo, \enquote{Characterization of broadband few-cycle laser pulses with the d-scan technique,} {\protect\JournalTitle{Opt. Express}} \textbf{20}, 18732--18743 (2012).

\bibitem{Chen2024a}
Y.-H. Chen and F.~Wise, \enquote{A simple accurate way to model noise-seeded ultrafast nonlinear processes,} {\protect\JournalTitle{arXiv preprint arXiv: 2410.20567}}  (2024).

\bibitem{Eggleston1980}
J.~Eggleston and R.~Byer, \enquote{Steady-state stimulated {Raman} scattering by a multimode laser,} {\protect\JournalTitle{IEEE J. Quantum Electron.}} \textbf{16}, 850--853 (1980).

\bibitem{Dudley2006}
J.~M. Dudley, G.~Genty, and S.~Coen, \enquote{Supercontinuum generation in photonic crystal fiber,} {\protect\JournalTitle{Rev. Mod. Phys.}} \textbf{78}, 1135--1184 (2006).

\bibitem{Frosz2010}
M.~H. Frosz, \enquote{Validation of input-noise model for simulations of supercontinuum generation and rogue waves,} {\protect\JournalTitle{Opt. Express}} \textbf{18}, 14778--14787 (2010).

\bibitem{Genier2019}
E.~Genier, P.~Bowen, T.~Sylvestre, J.~M. Dudley, P.~Moselund, and O.~Bang, \enquote{Amplitude noise and coherence degradation of femtosecond supercontinuum generation in all-normal-dispersion fibers,} {\protect\JournalTitle{J. Opt. Soc. Am. B}} \textbf{36}, A161--A167 (2019).

\bibitem{Strickland1985}
D.~Strickland and G.~Mourou, \enquote{Compression of amplified chirped optical pulses,} {\protect\JournalTitle{Opt. Commun.}} \textbf{55}, 447--449 (1985).

\bibitem{Najmi1994}
A.-H.~Y. Najmi, \enquote{{THE WIGNER DISTRIBUTION: A TIME-FREQUENCYANALYSIS TOOL},} {\protect\JournalTitle{Johns Hopkins APL Technical Digest}} \textbf{15}, 298--305 (1994).

\bibitem{Cohen1989}
L.~Cohen, \enquote{Time-frequency distributions-a review,} {\protect\JournalTitle{Proc. IEEE}} \textbf{77}, 941--981 (1989).

\bibitem{MATLAB_stft}
Mathworks, \enquote{stft,} https://www.mathworks.com/help/signal/ref/stft.html.

\bibitem{Khan2011}
N.~A. Khan, M.~N. Jafri, and S.~A. Qazi, \enquote{Improved resolution short time {Fourier} transform,} in \emph{2011 7th International Conference on Emerging Technologies,}  (2011), pp. 1--3.

\bibitem{Anderson1992}
D.~Anderson, M.~Desaix, M.~Lisak, and M.~L. Quiroga-Teixeiro, \enquote{Wave breaking in nonlinear-optical fibers,} {\protect\JournalTitle{J. Opt. Soc. Am. B}} \textbf{9}, 1358--1361 (1992).

\bibitem{Heidt2016}
A.~M. Heidt, A.~Hartung, and H.~Bartelt, \emph{{Generation of Ultrashort and Coherent Supercontinuum Light Pulses in All-Normal Dispersion Fibers}} (Springer New York, New York, NY, 2016), pp. 247--280.

\bibitem{Finot2018}
C.~Finot, F.~Chaussard, and S.~Boscolo, \enquote{Simple guidelines to predict self-phase modulation patterns,} {\protect\JournalTitle{J. Opt. Soc. Am. B}} \textbf{35}, 3143--3152 (2018).

\bibitem{Schulte2016}
J.~Schulte, T.~Sartorius, J.~Weitenberg, A.~Vernaleken, and P.~Russbueldt, \enquote{Nonlinear pulse compression in a multi-pass cell,} {\protect\JournalTitle{Opt. Lett.}} \textbf{41}, 4511--4514 (2016).

\bibitem{Hanna2017}
M.~Hanna, X.~D\'{e}len, L.~Lavenu, F.~Guichard, Y.~Zaouter, F.~Druon, and P.~Georges, \enquote{Nonlinear temporal compression in multipass cells: theory,} {\protect\JournalTitle{J. Opt. Soc. Am. B}} \textbf{34}, 1340--1347 (2017).

\bibitem{Centurion2006}
M.~Centurion, M.~A. Porter, Y.~Pu, P.~G. Kevrekidis, D.~J. Frantzeskakis, and D.~Psaltis, \enquote{{Modulational Instability in a Layered Kerr Medium: Theory and Experiment},} {\protect\JournalTitle{Phys. Rev. Lett.}} \textbf{97}, 234101 (2006).

\bibitem{Zhang2021}
S.~Zhang, Z.~Fu, B.~Zhu, G.~Fan, Y.~Chen, S.~Wang, Y.~Liu, A.~Baltuska, C.~Jin, C.~Tian, and Z.~Tao, \enquote{Solitary beam propagation in periodic layered {Kerr} media enables high-efficiency pulse compression and mode self-cleaning,} {\protect\JournalTitle{Light Sci. Appl.}} \textbf{10}, 53 (2021).

\bibitem{Wang2024}
W.~Wang, Y.~Eisenberg, Y.-H. Chen, C.~Xu, and F.~Wise, \enquote{Efficient temporal compression of {10-$\mu$J} pulses in periodic layered {Kerr} media,} {\protect\JournalTitle{Opt. Lett.}} \textbf{49}, 5787--5790 (2024).

\bibitem{Magni1992}
V.~Magni, G.~Cerullo, and S.~D. Silvestri, \enquote{High-accuracy fast {Hankel} transform for optical beam propagation,} {\protect\JournalTitle{J. Opt. Soc. Am. A}} \textbf{9}, 2031--2033 (1992).

\bibitem{Siegman1977}
A.~E. Siegman, \enquote{Quasi fast {Hankel} transform,} {\protect\JournalTitle{Opt. Lett.}} \textbf{1}, 13--15 (1977).

\bibitem{Talman1978}
J.~D. Talman, \enquote{Numerical {Fourier} and {Bessel} transforms in logarithmic variables,} {\protect\JournalTitle{J. Comput. Phys.}} \textbf{29}, 35--48 (1978).

\bibitem{Agrawal1981}
G.~P. Agrawal and M.~Lax, \enquote{End correction in the quasi-fast {Hankel} transform for optical-propagation problems,} {\protect\JournalTitle{Opt. Lett.}} \textbf{6}, 171--173 (1981).

\bibitem{Zhang2002}
D.~W. Zhang, X.-C. Yuan, N.~Q. Ngo, and P.~Shum, \enquote{Fast {Hankel} transform and its application for studying the propagation of cylindrical electromagnetic fields,} {\protect\JournalTitle{Opt. Express}} \textbf{10}, 521--525 (2002).

\bibitem{Baddour2015}
N.~Baddour and U.~Chouinard, \enquote{Theory and operational rules for the discrete {Hankel} transform,} {\protect\JournalTitle{J. Opt. Soc. Am. A}} \textbf{32}, 611--622 (2015).

\bibitem{Baddour2017}
N.~Baddour and U.~Chouinard, \enquote{{Matlab Code for the Discrete Hankel Transform},} {\protect\JournalTitle{Journal of Open Research Software}}  (2017).

\bibitem{Yu1998}
L.~Yu, M.~Huang, M.~Chen, W.~Chen, W.~Huang, and Z.~Zhu, \enquote{Quasi-discrete {Hankel} transform,} {\protect\JournalTitle{Opt. Lett.}} \textbf{23}, 409--411 (1998).

\bibitem{GuizarSicairos2004}
M.~Guizar-Sicairos and J.~C. Guti\'{e}rrez-Vega, \enquote{Computation of quasi-discrete {Hankel} transforms of integer order for propagating optical wave fields,} {\protect\JournalTitle{J. Opt. Soc. Am. A}} \textbf{21}, 53--58 (2004).

\end{thebibliography}

\end{document}